%% file: paper.tex
\documentclass[a4paper]{article}
\usepackage{etex}
\usepackage{a4wide}
\usepackage{graphicx}
\usepackage{amsbsy,amssymb,amsgen,amsfonts}
\usepackage{amsmath,amsfonts,color,latexsym}
\usepackage[breaklinks=true,bookmarks=true,colorlinks=true,linkcolor=blue,citecolor=blue,allcolors=blue]{hyperref}
\usepackage{natbib}
\shortcites{amoura:10,caporaloni:75,eames:11,gerashenko:08,glowinski:99,lsvezzo:10,mordant:02,qureshi:08,qureshi:07,tencate:04,calzavarini:09,villalba:12,tencate:04,kidanemariam:13,Silbert2001,constantinides:08,Aussillous2013,Loiseleux2005a,Charru2007a,Ouriemi2007,Lobkovsky2008,Joseph2001,TenCate2002,Foerster1994a,BrandledeMotta2013,Kidanemariam2013,Glowinski1999} 
\usepackage[title]{appendix}
\usepackage{natbib}

\usepackage{doi}
\newcommand{\ben} {\begin{equation}}
\newcommand{\een} {\end{equation}}
\newcommand{\be} [1] {\begin{equation} \label{#1}}
\newcommand{\ee} {\end{equation}}
\newcommand{\bse} [1] {\begin{subequations} \label{#1}}
\newcommand{\ese} {\end{subequations}}
\newcommand{\ban} {\begin{eqnarray*} }
\newcommand{\ean} {\end{eqnarray*} }
\newcommand{\bea} {\begin{eqnarray}}
\newcommand{\eea} {\end{eqnarray}}

\input{my_line_styles}

\input{my_abbreviations}
\newcommand{\revision}[2]{#2}
\newcommand{\revisions}[1]{}

\newcommand{\figpap}[2]{#1#2}
\usepackage{sectsty}
\allsectionsfont{\sffamily}
\let\LaTeXmaketitle\maketitle
\renewcommand{\maketitle}{{\sf\LaTeXmaketitle}}

\begin{document}
\epstopdfsetup{suffix=} 
%
  \title{%
    Interface-resolved direct numerical simulation of 
    the erosion of a sediment bed sheared by laminar
    \revision{}{channel} flow%
  }

%
%
%
%
\author{Aman G.\
  Kidanemariam\footnote{\href{mailto:aman.kidanemariam@kit.edu}{aman.kidanemariam@kit.edu}}
  \hspace*{1ex}and 
  Markus
  Uhlmann\footnote{\href{mailto:markus.uhlmann@kit.edu}{markus.uhlmann@kit.edu}} 
  \\[1ex]
  {\small 
    Institute for Hydromechanics, Karlsruhe Institute of
    Technology}\\ 
  {\small 
    76131 Karlsruhe, Germany
  }
}
\date{\small DOI:
  \href{http://dx.doi.org/10.1016/j.ijmultiphaseflow.2014.08.008}{10.1016/j.ijmultiphaseflow.2014.08.008}} 
\maketitle
\input{abstract}
%
%
%
\input{introduction}
%
\input{numerical_method}

%
\input{single_sphere_bouncing}
%
\input{lam_bdld_tras_setup}
%
\input{lam_bdld_tras_results}
\input{conclusion}
%
\input{acknowledgement}
%
\appendix
\input{appendix_suppl_mat.tex}
\input{appendix_averaging_operations}

\clearpage
\bibliographystyle{../style/model2-names}
\addcontentsline{toc}{section}{Bibliography}
%

\input{paper.bbl}
%
%
%
\end{document}

%% file: my_line_styles.tex
\usepackage{relsize}
\def\solidthick{\protect\rule[2pt]{10.pt}{.5pt}}

\newcommand{\opencircle}
           {$\mathlarger{\mathlarger{\mathlarger{\circ}}}$}
\newcommand{\opentriup}{$\vartriangle$}
\newcommand{\opensquare}{$\mathsmaller{\square}$}
\newcommand{\solidcircle}
           {$\mathlarger{\mathlarger{\mathlarger{\bullet}}}$}
\newcommand{\solidtriup}{$\blacktriangle$}

\newcommand{\soliddiamond}{$\blacklozenge$}      
\newcommand{\solidsquare}{$\mathsmaller{\blacksquare}$}
\definecolor{mygray}{rgb}{0.7,0.7,0.7}         

%% file: my_abbreviations.tex
\newcommand\Lx{\ensuremath{L_x}}
\newcommand\Ly{\ensuremath{L_y}}
\newcommand\Lz{\ensuremath{L_z}}
\newcommand\Npt{\ensuremath{N_p}}          
\newcommand\dryrest{\ensuremath{\varepsilon_d}}   
\newcommand\efferest{\ensuremath{\varepsilon}}   
\newcommand\impactVel{\ensuremath{v_{pT}}}   
\newcommand\reboundVel{\ensuremath{v_{pR}}}   
\newcommand\stokes{\ensuremath{St}}   
\newcommand\Reb{\ensuremath{Re}}   
\newcommand\Ga{\ensuremath{Ga}}    
\newcommand\qfmean{\ensuremath{q_f}} 
\newcommand\qpmean{\ensuremath{\langle q_p\rangle}} 
\newcommand\qfstar{\ensuremath{\qfmean/\qrefhf}}
\newcommand\qrefv{\ensuremath{q_{visc,D}}}    
\newcommand\qrefhf{\ensuremath{q_{visc,h}}} 
\newcommand\shields{\ensuremath{\Theta_{Pois}}} 
\newcommand\shieldsgen{\ensuremath{\Theta}} 
\newcommand\shieldscrit{\ensuremath{\Theta_{Pois}^{(c)}}} 
\newcommand\shieldsgencrit{\ensuremath{\Theta^{(c)}}} 
\newcommand\dratio{\ensuremath{\rho_p/\rho_f}} 
\newcommand\hfluid{\ensuremath{h_f}} 
\newcommand\pileparam{\ensuremath{H_b}}
\newcommand\hmobile{\ensuremath{h_m}} 
\newcommand\hufmax{\ensuremath{h^*}} 
\newcommand\yref{\ensuremath{y_0}} 

\newcommand\ufmean{\ensuremath{\langle u_f \rangle}}
\newcommand\ufmax{\ensuremath{U_f^m}}
\newcommand\upmean{\ensuremath{\langle u_p \rangle}}
\newcommand\phimean{\ensuremath{\langle \phi_s \rangle}}
\newcommand\phimeansmooth{\ensuremath{\langle \phi_p \rangle}}
\newcommand\usettling{\ensuremath{U_{s}}}

\newcommand\solidvolfracbed{\ensuremath{\Phi_{bed}}}


%% file: abstract.tex
%
\begin{abstract}
A numerical method based upon the immersed boundary technique for the
fluid-solid coupling and on a soft-sphere approach for solid-solid
contact is used to perform direct numerical simulation of the
flow-induced motion of a thick bed of spherical particles in a
horizontal plane channel.  
The collision model features a normal force component with a spring
and a damper, as well as a damping tangential component, limited by
a Coulomb friction law. 
The standard test case of a single particle colliding perpendicularly
with a horizontal wall in a viscous fluid is simulated over a broad
range of Stokes numbers, yielding values of the effective restitution
coefficient in close agreement with experimental data. 
The case of bedload particle transport by laminar channel flow is
simulated for 24 different parameter values covering a broad range of
the Shields number.  
Comparison of the present results with reference data from the
experiment of \citet{Aussillous2013} yields excellent agreement. 
It is confirmed that the particle flow rate varies with the third
power of the Shields number once the known threshold value is
exceeded. 
The present data suggests that the thickness of the mobile particle
layer (normalized with the height of the clear fluid region) increases
with the square of the normalized fluid flow rate. 
\end{abstract}

%% file: introduction.tex
\section{Introduction}
\label{introduction}
Subaqueous sediment transport is a dense particulate flow problem,
which involves the erosion, entrainment, transport and 
deposition of sediment particles as a result of the 
net effect of hydrodynamic forces, gravity forces as well as forces
arising from inter-particle contacts. 
Systems which are significantly affected by the sediment transport
process involve many fields of engineering, 
in particular civil and environmental engineering  
(e.g.\ river morphology and dune formation).
Therefore, an improved understanding of the mechanisms leading to
fluid-induced transport of sediment and to its accurate prediction is
highly desirable.  

A considerable amount of experimental and theoretical studies 
have been carried out in the past, 
leading to a number of 
(semi-) empirical predictive models for engineering purposes. 
For instance, there exist several
algebraic expressions 
for the particle flux as a function of the local bed 
shear stress both in the laminar and turbulent
flow regimes \citep[see e.g.][]{Garcaia2008,Ouriemi2009}. 
Critically assessing  the 
validity of the proposed models
is a challenging task due to
the complex interaction between the flow and the mobile sediment bed,
and due to the dependence
on multiple parameters. 
For similar reasons, available experimental data is widely dispersed 
\revision{}{\citep{Ouriemi2009}.}

%
In order to investigate the fundamental aspects of granular transport, 
the problem was simplified in some studies  
by considering the erosion of a sediment bed consisting 
of mono-dispersed spherical particles under laminar shear flows
\citep[see e.g.][]{Charru2004,Loiseleux2005a,Charru2007a,Ouriemi2007,
           Lobkovsky2008,Mouilleron2009,Ouriemi2009,Aussillous2013}.
%
%
There is a general consensus  that the onset of particle motion and
bedload transport is controlled by the Shields number \shieldsgen,
which 
is proportional to the wall shear stress times the cross-sectional
area of a particle, divided by its apparent weight 
\citep{Shields1936}. 
Below a critical value \shieldsgencrit, almost independent of the
particle Reynolds number in the laminar regime, no erosion of
sediment is observed. 
\citet[][]{Ouriemi2007}
have performed an experimental investigation of the cessation of
motion (which indirectly yields the threshold for the onset of motion) 
of spherical beads in laminar pipe flow. 
They inferred that the critical Shields number has a value 
$\shieldsgencrit=0.12\pm 0.03$. 
This value has also been reported by other authors 
\citep{Charru2004,Loiseleux2005a}.
Note that, \shieldsgencrit\ is different from and larger in value
than another critical Shields number 
which
corresponds to the initiation of the motion of individual particles in
an initially loosely packed granular bed. 
Experiments show that particles, when they are initially 
set in motion, move in an erratic manner by rolling over other 
particles, temporarily halting in troughs and then starting to 
move again, most of the time impacting other particles and 
possibly setting them in motion. After a sufficient duration, 
the particles rearrange and the sediment bed gets compacted.
\citet{Charru2004} refer to this phenomenon as the `armoring
effect of the bed'; they explain the observed initiation of
motion of particles at $\shieldsgen \approx 0.04$ in an initially 
loosely packed sediment bed and describe 
the gradual increase of the critical shear number towards
\shieldsgencrit. 

At super-critical values of the Shields number, 
the resulting sediment flux is usually expressed as a function 
of the local bed shear number (or the excess 
shear number $\shieldsgen - \shieldsgencrit$).  
\citet{Charru2002}, applying the viscous 
resuspension model of \citet{Leighton1986},
found that the particle flux varies cubically with the Shields number.
\citet{Ouriemi2009}, considering
an alternative continuum description of bedload transport and
assuming a frictional rheology of the mobile granular layer,
proposed an expression for the dimensionless particle flux 
which likewise predicts a cubic variation with the Shields number for
$\shieldsgen \gg \shieldsgencrit$. 

In an attempt to complement experiments, 
a number of numerical simulations of the 
transport of particles as bedload 
have been performed in the granular flow community, 
albeit without 
properly resolving the near-field around the particles 
\citep[see e.g.][]{Schmeeckle2003,Heald2004a}.
These simulations are based on the discrete element model 
(DEM) by which the trajectory of all individual particles,
 which constitute the sediment bed, is accounted for.
The main feature of DEM is the modelling of the
inter-particle collisions. Various
collision models have been proposed which are usually based 
either on the hard-sphere or the soft-sphere approach.
In the hard-sphere approach particles are assumed to 
be rigid and to exchange momentum during 
instantaneous binary collision events
\citep[see e.g.][]{Foerster1994a}. 
On the other hand,
in the soft-sphere approach the 
deformation of particles during contact is indirectly 
considered by 
\revision{admitting}{allowing} 
them to overlap.
The contact forces, which are assumed 
to be functions of the overlap thickness and/or the relative 
particle velocities, 
are computed based on mechanical models such as springs, dash-pots and sliders
\citep{Cundall1979}.

Recently, a number of studies has emerged in which 
the fluid flow even in the near vicinity of individual grains is being
fully resolved while at the same time a realistic contact model is
employed
\citep{Yang2008,Wachs2009,Li2011,Simeonov2012,Kempe2012a,BrandledeMotta2013}. 
For the purpose of validation of the 
coupling between the fluid-solid solver and the contact model, 
most of these studies have considered benchmark cases where a single
spherical particle collides with a plane wall or with
another particle in a viscous fluid. 
Experimental studies of this configuration have shown that when a
particle freely approaches and collides with another particle or a 
wall, in addition to energy dissipation from the solid-solid contact, 
it loses energy as a result of the work done to 
squeeze out the viscous fluid from the gap between the contacting edges,
thereby decelerating it prior to contact
\citep[][]{Joseph2001,Gondret2002,TenCate2002,Joseph2004a,Yang2006}.
Similarly, additional fluid-induced losses occur during the 
rebound phase. 
The effect of the viscous fluid on the bouncing behavior 
of the particle is classically characterized by an effective 
coefficient of restitution \efferest\ which is the ratio of
the particle's pre- and post-collision normal velocities. 
Thus \efferest\ accounts for the total 
energy dissipation both from viscous fluid resistance as
well as from the actual solid-solid contact,
in contrast to the dry coefficient of restitution defined 
in the same way but for collisions happening in vacuum.
It is well established that the Stokes number, defined as
$\stokes = (\dratio)Re_p/9$ where $Re_p$ is the particles Reynolds number
based on its diameter and 
\revision{it}{its}
 velocity before impact,
is the relevant parameter which determines the degree of 
viscous influence on the bouncing behavior of a particle.
At large values of the Stokes number (above $\stokes=1000$ say), 
the effect of the fluid on the collision becomes negligible and 
\efferest\ approaches the dry coefficient of restitution.
On the other hand, at small values of the Stokes number
($\stokes\lesssim 10$), viscous damping is so large
that no rebound of the particle is observed.
%
Various sets of experimental data are available for the
bouncing sphere 
\citep[see e.g.][]{Joseph2001,Gondret2002}
and it has become a standard
benchmark for the validation of numerical approaches which model
the collision dynamics of finite-size objects fully immersed in
a viscous fluid.

One of the goals 
of the present work 
is to investigate the formation of patterns from an initially flat bed
of erodible sediment particles. 
A precursor stage to this process is bedload transport,
i.e.\ the featureless motion of sediment particles which involves
multiple contacts, sliding, rolling and saltation of particles.
For this latter case, detailed experimental data are available from
the recent experiments of \citet{Aussillous2013}. 
Using an index-matching technique these authors were able to determine 
the velocity profiles of both the fluid and the particulate phase in
pressure driven flow through a rectangular duct. 
As will be shown in the present paper, the experimental conditions and
the range of the main control parameter (the Shields number) covered
therein is accessible to interface-resolved numerical simulation. 
For these reasons, the case of bedload transport in horizontal
wall-bounded flow is an attractive benchmark configuration for the
purpose of validation of numerical approaches to the transport of
dense sediment. 

%
%
In the present contribution we first present an extension of the
immersed boundary method of \citet{Uhlmann2005a} to include
solid-solid contact forces by means of a soft-sphere model similar to
the approach of \citet{Wachs2009}. 
This coupled DNS-DEM technique is then validated in
\S~\ref{sec:collision-sphere-wall-viscous-fluid} through simulations
of the standard test case of a single sphere colliding with a plane
wall. 
In \S~\ref{sec:erosion-of-granular-bed-sheared-by-laminar-flow}
we present simulation results of bedload transport in laminar
plane channel flow over a broad range of parameters, comparing them to
data from the reference experiment.  
This second test serves to validate the coupled DNS-DEM approach in a
case with many particles simultaneously interacting. As such it
provides an important step towards the simulation of sediment pattern 
formation. 
Furthermore, the present results contribute new data to the ongoing
discussion of scaling laws in bedload transport. 
The paper closes with conclusions in \S~\ref{sec-conclusion}. 
%
%
%

%% file: numerical_method.tex
\section{Numerical method}
\label{sec:numerical-method}
\subsection{Immersed boundary method}
\label{subsec:immersed-boundary-method}
The numerical method employed is a variant of the immersed
boundary method as proposed by \citet[][]{Uhlmann2005a}.  
The incompressible Navier-Stokes equations are solved throughout the
entire computational domain 
$\Omega$ comprising the fluid domain $\Omega_{\mathrm{f}}$ and the
space occupied by the suspended particles $\Omega_{\mathrm{s}}$. 
For this purpose a force term is added to the right-hand side of the
momentum equation which serves to impose the no-slip condition at the
fluid-solid interfaces. 
The direct numerical simulation (DNS) code has been
validated on a whole range of benchmark problems
\citep[][]{Uhlmann2004,Uhlmann2005a,Uhlmann2005b,Uhlmann2006,uhlmann:13a},
and has been previously employed for the simulation of various
particulate flow configurations
\citep[][]{Uhlmann2008,Chan-braun2011,Garcia-villalba2012,
  Kidanemariam2013}.
\subsection{Inter-particle collision model}
\label{subsec:inter-particle-collision-model}
In direct numerical simulations of systems with a low solid volume fraction 
the particle-particle or particle-wall encounters are often 
treated with the aid of an artificial repulsion force model
\citep[such as the one proposed by][]{Glowinski1999}.
This technique prevents the occurrence of 
non-physical overlap of particles in the simulation 
while frictional losses during the inter-particle or wall-particle
contact are typically not accounted for
\citep{Uhlmann2008,Garcia-villalba2012}. 
However, in dense systems, such as sediment transport in wall-bounded shear
flow, 
inter-particle contact forces are expected to be significant in
addition to the hydrodynamic forces acting on the particles.
For this purpose, in the present work we resort to a discrete element
model (DEM) in order to describe the collision dynamics between the
submerged solid objects instead of using an artificial repulsion model.
%
\begin{figure}
  \figpap{
  \centering
  \begin{minipage}{.5\linewidth}
    \includegraphics[width=\linewidth,clip=true,
    viewport=80 450 360 700]
    {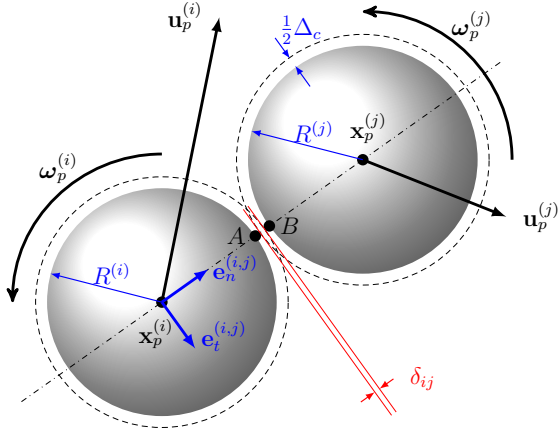}
  \end{minipage}
  }{
  \caption{
    Schematic diagram showing two spherical particles with indices
    $i$ and $j$ colliding in the context of a soft-sphere collision
    model.
  \protect\label{fig:collision-particle-schematic}
  }
  }
\end{figure}
%

%
The presently employed DEM uses a standard
soft-sphere approach which is based on a linear
mass-spring-damper system.
More specifically, 
let us consider a pair of
particles with indices $i$ and $j$ belonging to a system with a total 
number of \Npt\ particles  
%
(for an illustration of the geometrical relations 
please refer to figure~\ref{fig:collision-particle-schematic}). 
We define an overlap length $\delta_{ij}(t)$ as follows 
\begin{equation}
     \delta_{ij}(t) =  R^{(i)}+R^{(j)}+\Delta_{\rm c}
                 - |\mathbf{x}^{(j)}_p(t) - \mathbf{x}^{(i)}_p(t)|\;,
     \label{eq:admitted-ovelap}
\end{equation}
where $R^{(i)}$ denotes the $i$th particle's radius and
$\mathbf{x}^{(i)}_p(t)$ is its center position at time $t$. 
The length $\Delta_{\rm c}$ in (\ref{eq:admitted-ovelap}) is an
admitted gap termed `force range' by \citet{patankar:01c} in the
context of a point-particle description coupled to a soft-sphere model. 
It accounts for the distance over which the actual fluid-solid
interfaces are `smeared out' in the discrete formulation of the fluid
flow problem. Therefore, $\Delta_{c}$ is related to the support of the
regularized delta function used in the context of the present immersed
boundary method \citep{Uhlmann2005a}; its magnitude is of the order of
the mesh width $\Delta x$ (precise values will be stated for each flow
case below). 
%
%

We can now determine whether a given particle pair with indices $i$,
$j$ ($i\neq j$) is in contact at time $t$ by introducing the following
function $I_c^{(i,j)}$: 
\begin{equation}
  I_c^{(i,j)}(t) =
    \left\{
       \begin{array}{lll}
        1&\mbox{if}&
        %
        \delta_{ij}(t) \ge 0
        \;,
         \\
        0&\mbox{else}\;.&
        \end{array}
    \right.
 \label{eq:particle-contact-indicator-function}
\end{equation}
The force $\mathbf{f}_{c}^{(i)}(t)$ acting on particle $i$ as a result
of its contact with neighboring particles is then defined as 
\begin{equation}
   \mathbf{f}_{c}^{(i)}(t)
   =\sum_{\substack{j=1\\j\neq i}}^{N_p}
    \Big(\mathbf{f}_{el}^{(i,j)}(t) + 
         \mathbf{f}_{d}^{(i,j)}(t) + 
         \mathbf{f}_{t}^{(i,j)}(t) 
     \Big)I_c^{(i,j)}(t)\;.
  \label{eq:total-contact-force}
\end{equation}
The three individual force contributions entering the 
contact force in \eqref{eq:total-contact-force} are modeled
as follows. 
The elastic part $\mathbf{f}_{el}^{(i,j)}(t)$ of the normal force
component is  a linear function of the overlap between each 
particle pair. Its value (acting on the $i$th particle) is given by:
%
%
\begin{equation}
     \mathbf{f}_{el}^{(i,j)}(t) = 
     -k_n\,\delta_{ij}(t)\,\mathbf{e}_n^{(i,j)}(t)\;,
     \label{eq:normal-replusive-force}
\end{equation}
where  $k_n$ is a constant stiffness coefficient, 
and $\mathbf{e}_n^{(i,j)}$ is the unit normal vector along the line
connecting the two particle centers pointing from 
$\mathbf{x}^{(i)}_p$ to $\mathbf{x}^{(j)}_p$:
\begin{equation}
  \mathbf{e}_n^{(i,j)}(t) 
  = \frac{\mathbf{x}^{(j)}_p(t) - \mathbf{x}^{(i)}_p(t)}
         {|\mathbf{x}^{(j)}_p(t) - \mathbf{x}^{(i)}_p(t)|}\;.
 \label{eq:unit-normal-vector}
\end{equation}
The dissipative part $\mathbf{f}_{d}^{(i,j)}(t)$ of the normal force
acting on the $i$th particle is defined as:
\begin{equation}
   \mathbf{f}_d^{(i,j)}(t) =
    - c_{dn}\, \mathbf{u}_{r,n}^{(i,j)}(t)\;,
 \label{eq:normal-dissipative-force}
\end{equation}
where  $c_{dn}$ is a constant damping coefficient, and
$\mathbf{u}_{r,n}^{(i,j)}$ is the normal component of
the relative velocity of particle $i$ with respect to particle $j$
at the contacting points (cf.\ points $A$ and $B$ in figure
\ref{fig:collision-particle-schematic}). This relative velocity is
defined as follows:
\begin{eqnarray}
  \mathbf{u}_{r}^{(i,j)}(t) &=& 
      \mathbf{u}_p^{(i)}(t) -\mathbf{u}_p^{(j)}(t) + 
      \boldsymbol{\omega}_p^{(i)}(t)\times R^{(i)}\mathbf{e}_n^{(i,j)}(t)
      +
      \boldsymbol{\omega}_p^{(j)}(t)\times R^{(j)}\mathbf{e}_n^{(i,j)}(t)\;,
 \label{eq:particle-relative-velocity-j-i}\\
 \mathbf{u}_{r,n}^{(i,j)}(t) &=&
     \Big(\mathbf{e}_n^{(i,j)}(t)\cdot\mathbf{u}_{r}^{(i,j)}(t)\Big)
      \mathbf{e}_n^{(i,j)}(t).
  \label{eq:particle-relative-velocity-j-i-normal}
\end{eqnarray} 
%
%
The force acting tangentially (on the $i$th particle) applied
at the contact points between particles $i$ and $j$ is
computed according to the following expression:
\begin{equation}
 \mathbf{f}_t^{(i,j)}(t) = 
  \left\{
  \begin{array}{lll}
     -\Big[
     \min\big\{\mu_{\rm c}\big|\mathbf{f}_{el}^{(i,j)}(t)
                         +\mathbf{f}_d^{(i,j)}(t)\big|
               \,,\,
               c_{dt}\big|\mathbf{u}_{r,t}^{(i,j)}(t)\big|\big\}
      \Big]\mathbf{e}_t^{(i,j)}(t)
      &\mbox{if}& |\mathbf{u}_{r,t}^{(i,j)}(t)| \neq 0\;,\\[20pt]
      0 &\mbox{else}\;.&
   \end{array}
   \right.
 \label{eq:tangential-dissipative-force}
\end{equation} 
Relation \eqref{eq:tangential-dissipative-force} expresses
the fact that the tangential frictional force with 
damping coefficient $c_{dt}$ is proportional to the tangential
component of the relative velocity
at the contact point
(cf.\ below); it is, however, limited by the Coulomb friction, 
which is in turn proportional to the normal force acting at the same
contact point, multiplied by a friction coefficient $\mu_c$.
The tangential component of the relative velocity 
$\mathbf{u}_{r,t}^{(i,j)}(t)$ is given by the following relation:
\begin{equation}
   \mathbf{u}_{r,t}^{(i,j)}(t) = 
            \mathbf{u}_{r}^{(i,j)}(t)-
              \mathbf{u}_{r,n}^{(i,j)}(t)\;,
   \label{eq:tangential-component-of-relative-velocity}
\end{equation} 
while the tangential unit vector  $\mathbf{e}_t^{(i,j)}$ is defined as:
\begin{equation}
 \mathbf{e}_t^{(i,j)}(t) = \frac{\mathbf{u}_{r,t}^{(i,j)}(t)}
                     {|\mathbf{u}_{r,t}^{(i,j)}(t)|}, \quad
          \forall\quad |\mathbf{u}_{r,t}^{(i,j)}(t)| \neq 0\;.
 \label{eq:unit-tangential-vector}
\end{equation} 
It is important to note that the tangential component of the 
collision force generates a torque. The net torque $\mathbf{t}_c^{(i)}$ 
acting on particle $i$ due to all binary collisions at time $t$ is
given by:
\begin{equation}
 \mathbf{t}_c^{(i)}(t) = R^{(i)}\mathbf{e}_n^{i,j}\times
    \sum_{j=1}^{N_{\rm p}}
      \mathbf{f}_c^{(i,j)}(t) 
      I_c^{(i,j)}(t)\;.
 \label{eq:frictional-torque}
\end{equation} 

The model described in 
(\ref{eq:total-contact-force}-\ref{eq:frictional-torque}) introduces 
four parameters affecting the collision process namely: 
$k_{n}$, $c_{dn}$, $c_{dt}$, $\mu_{c}$ as well as the above-mentioned
force range $\Delta_c$. 
From an analytical solution of the linear mass-spring-damper system in
an idealized configuration (considering a binary normal collision in
vacuum and in the absence of external forces), 
a relation between the normal stiffness coefficient $k_n$ and the
normal damping coefficient $c_{dn}$ can be found.
For this purpose one can define a dry restitution coefficient
$\efferest_d$ as the ratio of the post-collision to pre-collision 
normal relative velocities in vacuum as
\begin{equation}
\efferest_d = -\frac{|\mathbf{u}_{r,n}^{(i,j)}|_{post}}
                    {|\mathbf{u}_{r,n}^{(i,j)}|_{pre}}\;.
\label{eq:dry-restitution-coefficient}
\end{equation}
With this definition, it can be shown that the following relation
holds between the normal damping coefficient and the stiffness
coefficient 
\revision{}{\citep{crowe:98}:}
\begin{equation}
  c_{dn} = -2\sqrt{M_{ij}k_{n}} \frac{\ln{\efferest_d}}
  {\sqrt{\pi^2+\ln^2\efferest_d}}\;,
  \label{damping-coefficient-and-stiffness-coefficient}
\end{equation}
where $M^{(l)}$ is the mass of the $l$th particle and
\begin{equation}\label{equ-def-reduced-mass}
  M_{ij} = M^{(i)}M^{(j)}/( M^{(i)}+M^{(j)})
\end{equation}
is the reduced mass of the particle pair $i,j$. 
\revision{\citep{crowe:98}.}{}
A crucial parameter of the collision model described above is
the duration of a generic collision event, since it determines 
the time step required to numerically integrate the Newton 
equations of solid body motion. 
It can again be shown that the duration of a collision for the 
ideal configuration mentioned above is given by the following formula 
\revision{:}{\citep{crowe:98}:}
\begin{equation}
  T_c = \frac{2\pi M_{ij}}{\sqrt{4M_{ij}k_n - c_{dn}^2}},
  \label{eq:collision-duration}
\end{equation}
which provides a useful reference value.

%
The collision between a particle with index $i$ and a plane wall is
treated in a similar way as the inter-particle collisions. 
However, in the former case we set $\mathbf{x}^{(j)}_p(t)$ equal to
the position of the wall-contact point, 
set the overlap length to 
$\delta_{ij}(t) =  R^{(i)}+\Delta_{\rm c}
- |\mathbf{x}^{(j)}_p(t) - \mathbf{x}^{(i)}_p(t)|$ instead of
(\ref{eq:admitted-ovelap}), 
use $M_{ij}=M^{(i)}$ instead of (\ref{equ-def-reduced-mass}), 
and replace the $j$th particle's velocity by the wall velocity in
(\ref{eq:particle-relative-velocity-j-i}).  
%

Similar to \cite{Wachs2009} we have kept the solid-solid 
contact model as simple as possible. For this purpose, we do not
include any additional force into the formulation, such as a
lubrication correction \citep[e.g.][]{nguyen:02}.  
%
\subsection{Time discretization of the equations of particle motion}
\label{subsec:time-discretization-of-newton-equations}
Due to the comparatively large stiffness of typical solid materials, 
the collision time scale $T_c$ is often much smaller than the typical
time step of the fluid solver $\Delta t$. 
It is straightforward to show for the configurations considered in the
present work that these two scales are typically separated by at least
one order of magnitude. 
Taking into account the fact that ${\cal O}(10)$ time steps are
typically required to accurately describe a generic collision event
\citep{Cleary2004}, it turns out that the ratio between fluid and
solid time steps easily reaches ${\cal O}(100)$. 
In order to avoid this restriction, one approach is to apply
a sub-stepping strategy. In this framework the Newton equations for  
particle motion are solved with a smaller time step $\Delta t_{sub}$
than the one used for solving the Navier-Stokes equations $\Delta t$
\citep[see e.g.][]{Wachs2009}.  
Recently
\citet{Kempe2012a} have proposed a strategy of artificial collision
time stretching, matching the time scale of the collision to that of
the fluid.  
In the present work,
we have adopted the former strategy of separately resolving
the two time scales. 
Sub-stepping amounts to keeping the hydrodynamic contribution to force
and torque acting on the particles constant for the
duration of a number of time steps $N_{sub}$ (where
$N_{sub}=\Delta t/\Delta t_{sub}$) between two consecutive updates of
the fluid phase.
%
It should be noted that the particle-related computational load
is not critically limited by the collision treatment in our
computational implementation. This allows us
to choose a conservatively small value for $\Delta t_{sub}$. 
In the cases treated in the present work, the number of sub-steps
was in the range of $100$-$240$.

%% file: single_sphere_bouncing.tex
%
\begin{table}\scriptsize
   \centering
   \begin{tabular}{lcccccc}
    \hline 
    Case & \dratio\ & \Ga\ & $Re_p$& \stokes\ &$k_n^*$& domain\\
    \input{headon_collision_relevant_paramters_table}
    \hline 
   \end{tabular}
   \caption{Physical parameters of the different simulations 
            corresponding
            to the vertically-oriented normal collision of a single
            particle with a wall in a viscous fluid.
            \dratio\ is the particle-to-fluid density ratio;
            \Ga\ is the Galileo number;
            $Re_{p}$ and \stokes\ are the particle's 
            Reynolds number and the Stokes number based on the
            particle's terminal  velocity; the normalized stiffness
            constant in the normal collision force is defined as 
            $k_n^\ast=k_n/((\rho_p/\rho_f-1)|\mathbf{g}|V_p/D)$. 
            The computational domain size and numerical
            parameters corresponding to the abbreviation in column 7
            are listed in
            table~\ref{tab:numerical_parameters_of_single_particle_collision}. 
           }
    \label{tab:physical_parameters_of_single_particle_collision}
\end{table}
%
%
%
%
%
\begin{table}\scriptsize
   \centering
   \begin{tabular}{*{5}{c}}
   \hline
   Domain & $ [\Lx \times \Ly\times \Lz]/D $ &
   $N_x \times N_y\times N_z$ & $D/\Delta x$ & 
   $\Delta_c/\Delta x$ \\
   D1 & 
   $6.4 \times 25.6 \times 6.4$ & 
   $128 \times 513 \times 128$ & 
   20 &  
   2 \\
   D2 & 
   $6.4 \times 51.2 \times 6.4$ & 
   $128 \times 1025 \times 128$ & 
   20 &  
   2 \\
   D2$^b$ & 
   $6.4 \times 51.2 \times 6.4$ & 
   $128 \times 1025 \times 128$ & 
   20 &  
   1 \\
   D2$^c$ & 
   $6.4 \times 51.2 \times 6.4$ & 
   $192 \times 1537 \times 192$ & 
   30 &  
   2 \\
   D2$^d$ & 
   $6.4 \times 51.2 \times 6.4$ & 
   $192 \times 1537 \times 192$ & 
   30 &  
   1 \\
   D3 & 
   $6.4 \times 102.4 \times 6.4$ & 
   $128 \times 2049 \times 128$ & 
   20 &  
   2 \\
   \hline
   \end{tabular}
   \caption{Numerical parameters of the particle-wall 
           rebound simulations.
           $L_i$ and $N_i$ are the computational domain length 
           and number of grid points in the
           $i$th coordinate direction, respectively; 
           $\Delta x$ is the uniform grid spacing 
           and $\Delta_{\rm c}$ is the force range. 
         }
   \label{tab:numerical_parameters_of_single_particle_collision}
\end{table}
%
\section{Collision of a sphere with a wall
            in a viscous fluid}
\label{sec:collision-sphere-wall-viscous-fluid}
\subsection{Computational setup and parameter values}
\label{subsec:collision-sphere-wall-viscous-fluid-setup}
For validation purposes we have simulated the case of an isolated
sphere settling on a straight vertical path under the action of
gravity before colliding with a plane horizontal wall. 
The chosen setup is similar to the experiment of \citet{Gondret2002}. 
Our computational domain has periodic boundary conditions 
in the wall-parallel directions $x$ and $z$, and a 
no-slip condition is imposed upon the fluid at the domain boundaries
in the wall-normal direction $y$. 
Gravity is directed in the negative $y$ direction. 
A spherical particle of diameter $D$ is 
released from a position near the upper wall
located at a wall-normal distance \Ly\ from the bottom wall. 
From dimensional considerations this problem is characterized  by two
dimensionless numbers, e.g.\ the solid-to-fluid density ratio
$\rho_p/\rho_f$ and the Galileo number $Ga=u_g D/\nu$, 
where $u_g=((\rho_p/\rho_f-1)|\mathbf{g}|D)^{1/2}$, $\mathbf{g}$ being
the vector of gravitational acceleration and $\nu$ the kinematic
viscosity. 
However, the data for the particle motion under normal collision with a
wall is well described as a function of the Stokes number $St$ alone
\citep{Joseph2001,Gondret2002}. 
%
%
The Stokes number can be defined from the terminal settling
velocity $v_{pT}$ of the sphere as 
\begin{equation}\label{equ-def-stokes}
  St=\frac{\rho_p}{\rho_f}\frac{Re_T}{9}
  \,,
\end{equation}
where $Re_T=v_{pT}D/\nu$. 
%

%
In the present work we have simulated $24$ different combinations of
values for the density ratio and the Galileo number. These values are 
given in
table~\ref{tab:physical_parameters_of_single_particle_collision} along
with the corresponding values of the terminal Reynolds number
and of the Stokes number. The present simulations cover the range of 
$St=4.7-1880$. 
The size of the computational domain as well as the grid resolution
(expressed as the ratio between the particle diameter and the mesh
width, $D/\Delta x$) are given in
table~\ref{tab:numerical_parameters_of_single_particle_collision}. 
It should be pointed out that, although the flow remains axisymmetric
at all times, the simulations were done with a full three-dimensional 
Cartesian grid with uniform mesh size. 
Three additional simulations (termed C08$^b$, C08$^c$, C08$^d$) with
the same physical parameter values as in case C08 have been run in
order to check the sensitivity of the results with respect to the
spatial resolution and the choice of the value for the force range
$\Delta_c$ (cf.\
table~\ref{tab:numerical_parameters_of_single_particle_collision}). 

Concerning the parameters for the solid-solid contact model, the
following values were chosen. 
The stiffness parameter $k_n$, normalized by the submerged weight of
the particle divided by its diameter, varies in the range of
$2\cdot10^4-2\cdot10^5$, cf.\
table~\ref{tab:physical_parameters_of_single_particle_collision}. 
With this choice the maximum penetration distance recorded in the
different simulations measures only a few percent of the force range
$\Delta_c$ (i.e.\ $\max(\delta_{ij}(t))\leq0.05\Delta_c$). 
In two of the cases (C08$^b$ and C08$^d$ the
maximum penetration reaches a value of approximately $0.2\Delta_c$,
which still only corresponds to one percent of the respective particle
diameter. 
The dry coefficient of restitution was set to a value
$\dryrest=0.97$ corresponding e.g.\ to dry collisions of steel or
glass spheres on a glass wall \citep{Gondret2002}. 
The normal damping coefficient $c_n$ was determined according
to relation \eqref{damping-coefficient-and-stiffness-coefficient}. 
%
\subsection{Results}
\label{sec:collision-sphere-wall-viscous-fluid-results}
\begin{figure}
  \figpap{
   \centering
        \begin{minipage}{2ex}
          \rotatebox{90}
          {\small $(y_p-D/2)/D$}
        \end{minipage}
        \begin{minipage}{.45\linewidth}
          \begin{minipage}{\linewidth}
            \centerline{(\textit{a})}
            \includegraphics[width=\linewidth]
            {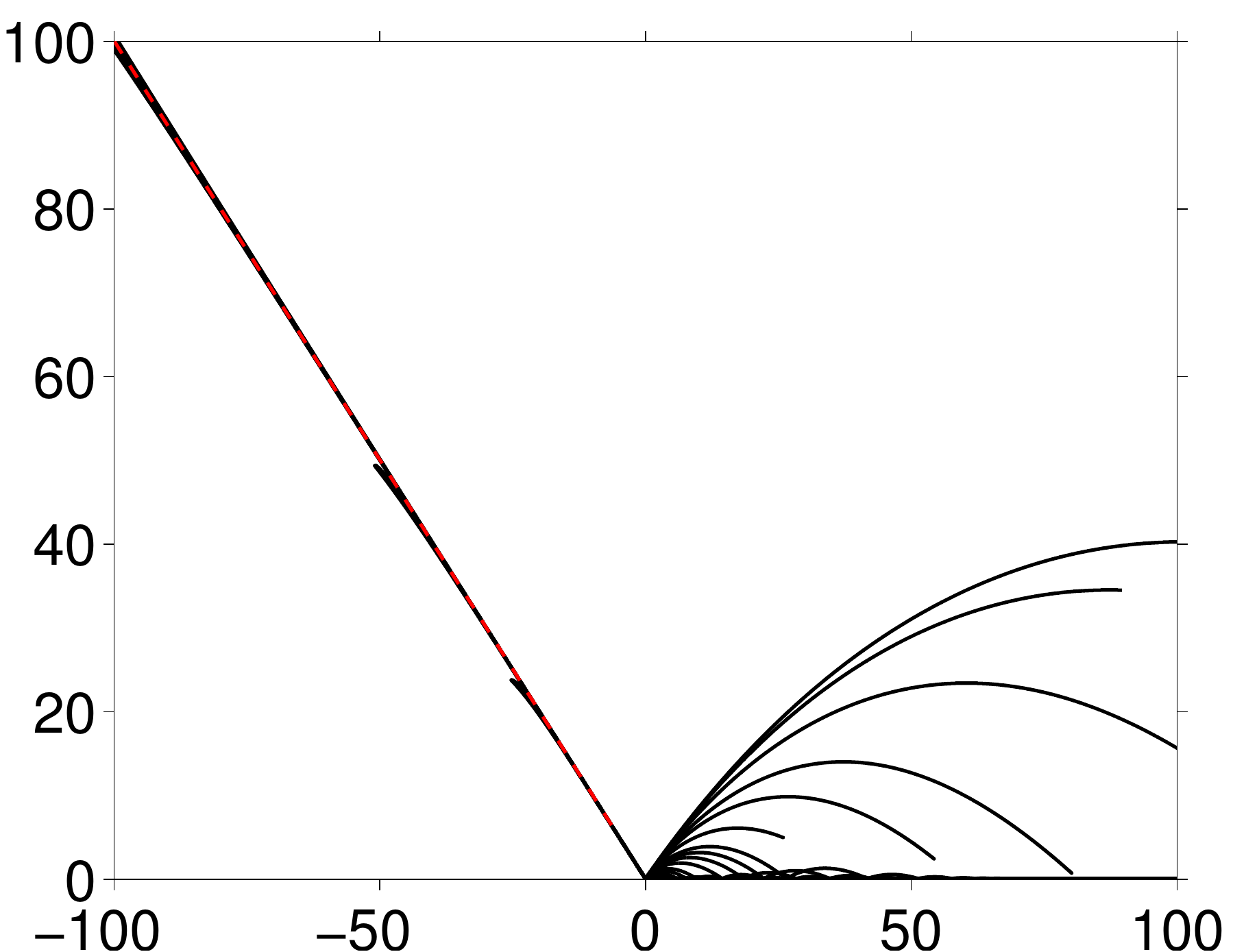}
            \centerline{\small $(t-t_1)v_{pT}/D$}
          \end{minipage}
          %
          \hspace{-.6\linewidth}
          \begin{minipage}{.39\linewidth}
             \vspace{-.8\linewidth}
            \includegraphics[width=\linewidth]
            {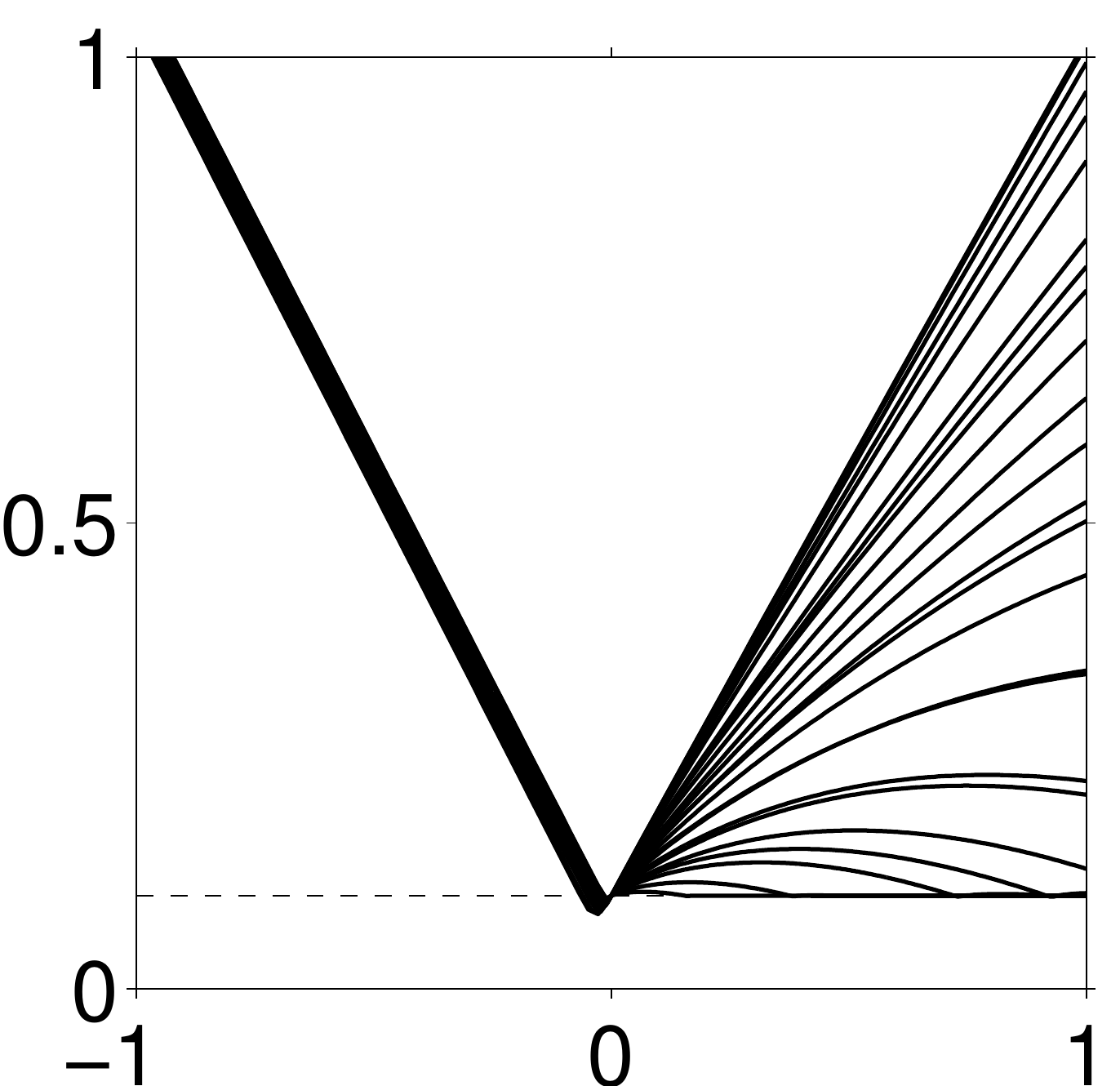}
           \end{minipage}
         \end{minipage}
         \hspace{1ex}
         \begin{minipage}{2ex}
           \rotatebox{90}{\small $v_p/v_{pT}$}
         \end{minipage}
         \begin{minipage}{.45\linewidth}
          \begin{minipage}{\linewidth}
            \centerline{(\textit{b})}
            \includegraphics[width=\linewidth]
            {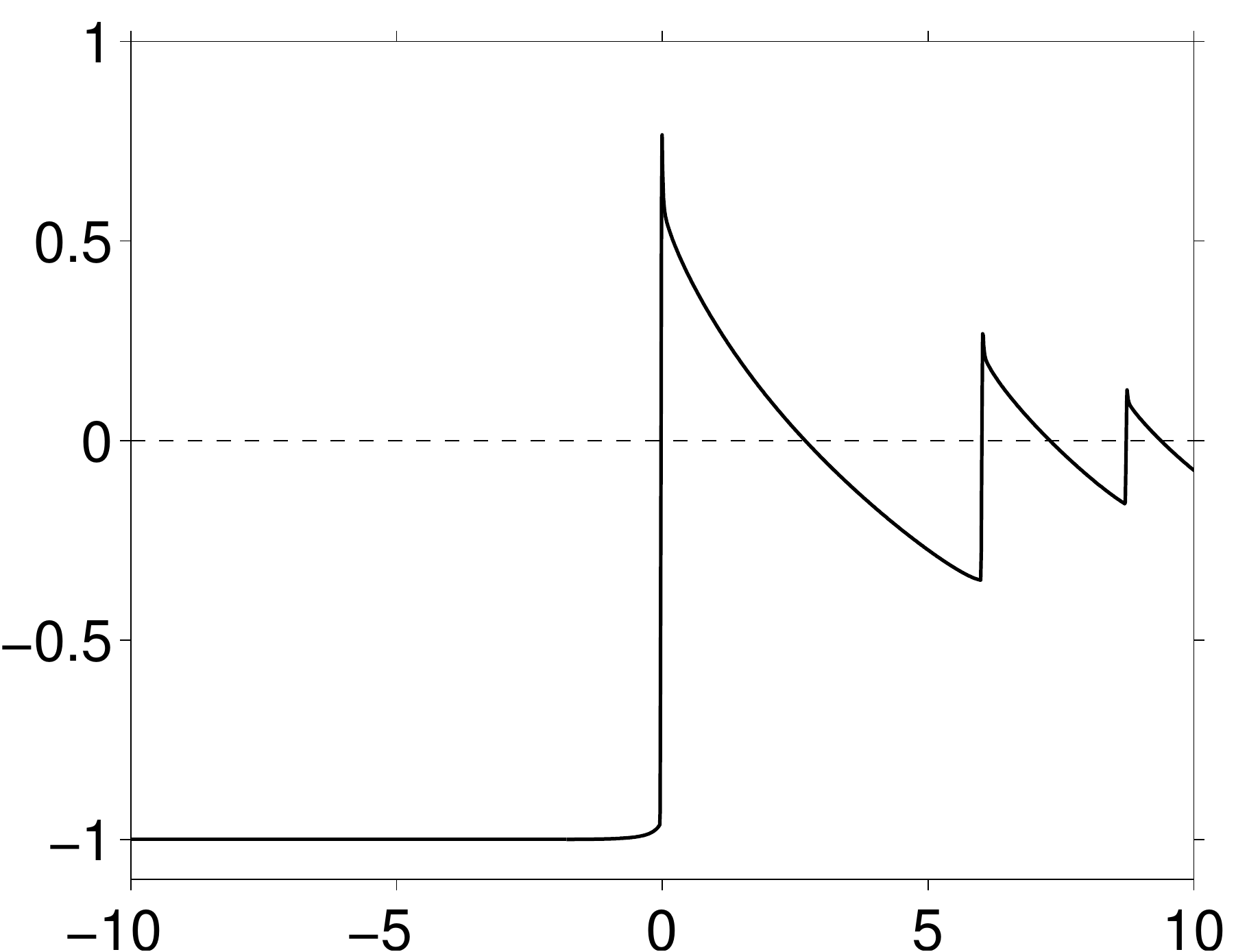}
            \centerline{\small $(t-t_1)v_{pT}/D$}
         \end{minipage}
         %
         \hspace{-.92\linewidth}
         \raisebox{-1ex}{
           \begin{minipage}[t]{.4\linewidth}
             \includegraphics[width=\linewidth]
             {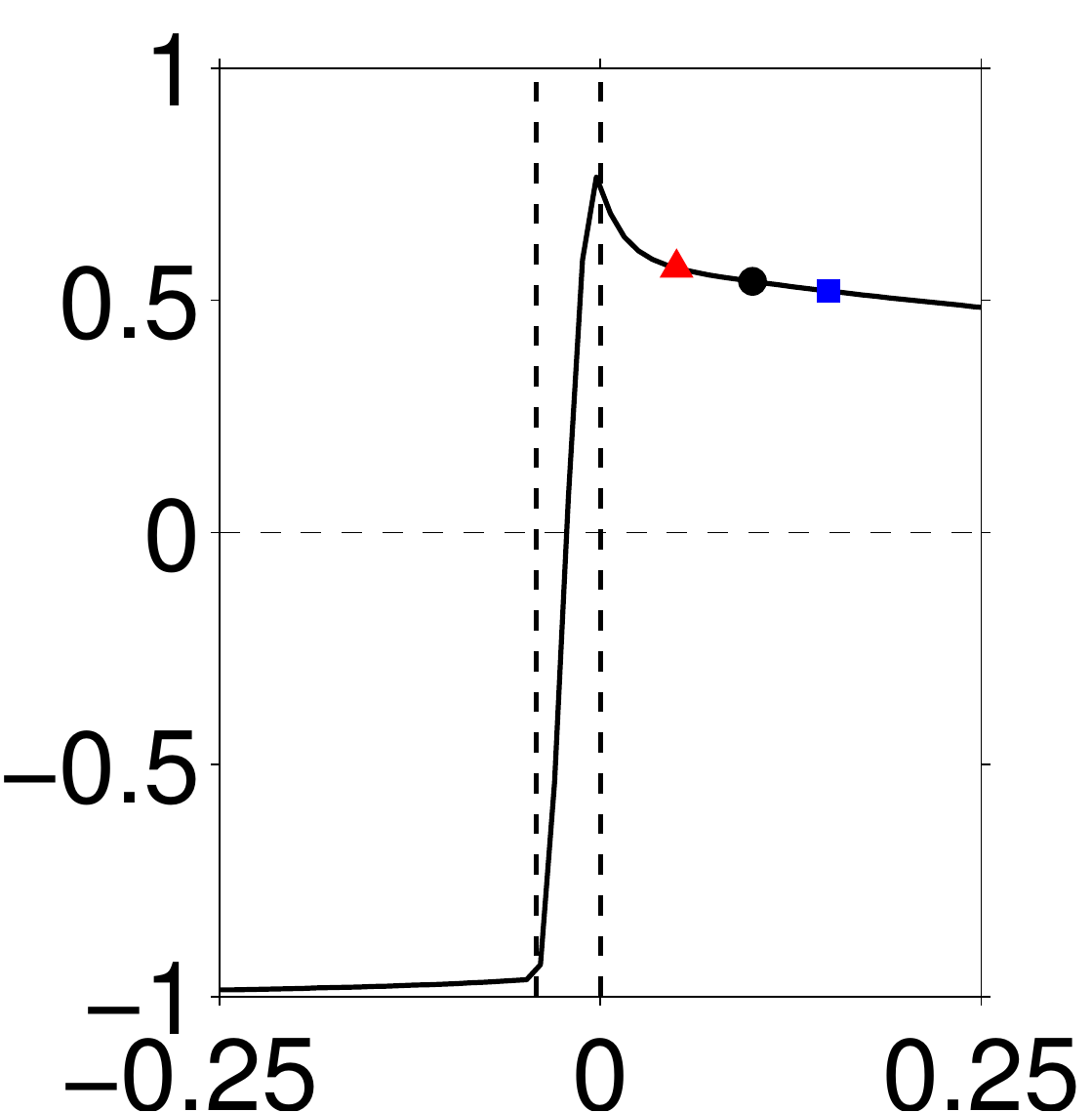}
           \end{minipage}
         }
       \end{minipage}
  }{
    \caption{(\textit{a}) Wall-normal position
      of the lowest point of the sphere $y_p-D/2$ as a
      function of time. 
      In each case the time coordinate is shifted
      such that all particle impacts
      coincide with the origin of the graph. 
      The red dotted line has
      a slope of $-1$, corresponding to motion at the
      terminal velocity. 
      The horizontal dashed line in the inset 
      indicates the $\Delta_c$ offset which is
      the extent of the collision force. 
      (\textit{b})
      Wall-normal velocity of the particle in 
      case C08 as a function of time. 
      The symbols mark the particle rebound
      velocity for different values of the time delay: 
      ({\color{red}$\blacktriangle$}) $t_R=0.05 D/v_{pT}$; 
      ({\color{black}\solidcircle}) $t_R=0.1 D/v_{pT}$; 
      ({\color{blue}$\blacksquare$}) $t_R=0.15 D/v_{pT}$. 
      The vertical dashed lines mark
      the interval during which the collision force is
      non-zero.            
    \protect\label{fig:particle-position-and-velocity-evolution}
    }
    }
\end{figure}

Figure~\ref{fig:particle-position-and-velocity-evolution}\textit{a} 
shows the wall-normal particle position (scaled with the
particle diameter) as a function of time (normalized with the
reference time $D/\impactVel$). 
It is seen that in all simulated cases the particle has attained its
terminal velocity before feeling the hydrodynamic influence of the
solid wall.   
After rebound the maximum distance between the wall and the closest
point on the particle surface 
varies from a negligibly small value ($0.004D$) for case C01
up to approximately $40D$ for case C24.
In three cases (C01, C02, C03) the maximum rebound height was below
$0.04$ times the particle diameter such that the results can be
qualified as `no rebound'; 
the corresponding Stokes numbers are $St=4.7$, $7.9$, $11.6$,
respectively, which is in agreement with the general observation that
the bouncing transition occurs at $St_c\approx10$ 
\citep{Gondret2002}. 

A typical time evolution of the wall-normal particle velocity  
is shown in figure~\ref{fig:particle-position-and-velocity-evolution}\textit{b}.
The figure shows the steady settling regime before impact and the 
subsequent rapid deceleration phase which is followed by an
acceleration phase as a result of the action of the contact force. 
In figure~\ref{fig:particle-position-and-velocity-evolution}\textit{b}
one can also observe the effect of the hydrodynamic forces the
particle experiences just before and after the collision event.  
%

%
One of the principal quantities of interest in this case is the
effective coefficient of restitution $\efferest$, defined as the
ratio of the particle velocity values before and after the collision,
viz: 
\begin{equation}\label{equ-def-efffective-coeff-restitution}
  \efferest=-\frac{\reboundVel}{\impactVel}
  \,.
\end{equation}
In collision experiments in a viscous fluid the determination of the
rebound velocity \reboundVel\ is a somewhat delicate issue, and no
unique definition seems to exist in the literature. 
%
%
In laboratory experiments this quantity is typically computed from
measured particle position data, involving 
some kind of gradient computation based on the records in the direct
vicinity of the wall. Therefore, the result may be sensitive to the
temporal resolution of the measuring device. 
In the experiment of \citet{Gondret2002} the time interval between
successive images was $2\,ms$ which means the first data point after
impact was recorded approximately at a time of $0.3D/\impactVel$ after 
the particle loses contact with the wall (under typical conditions of
their experiment). 
In the experiment of \cite{Joseph2001}, which featured a pendulum
set-up with a horizontal impact on a vertical wall, the temporal
resolution was comparably higher, corresponding to approximately
$0.03D/\impactVel$.  
In both cases the temporal resolution is significantly larger than the
duration of the actual collision. As a consequence, the measured
rebound velocity \reboundVel\ (and therefore the effective coefficient
of restitution $\efferest$) takes into account to some extent the
fluid-induced damping after the particle-wall contact. 
%
\begin{figure}
  \figpap{
  \centering
  \begin{minipage}{2ex}
    \rotatebox{90}
    {\small $\efferest/\dryrest$}
  \end{minipage}
  \begin{minipage}{.45\linewidth}
    \includegraphics[width=\linewidth]
    {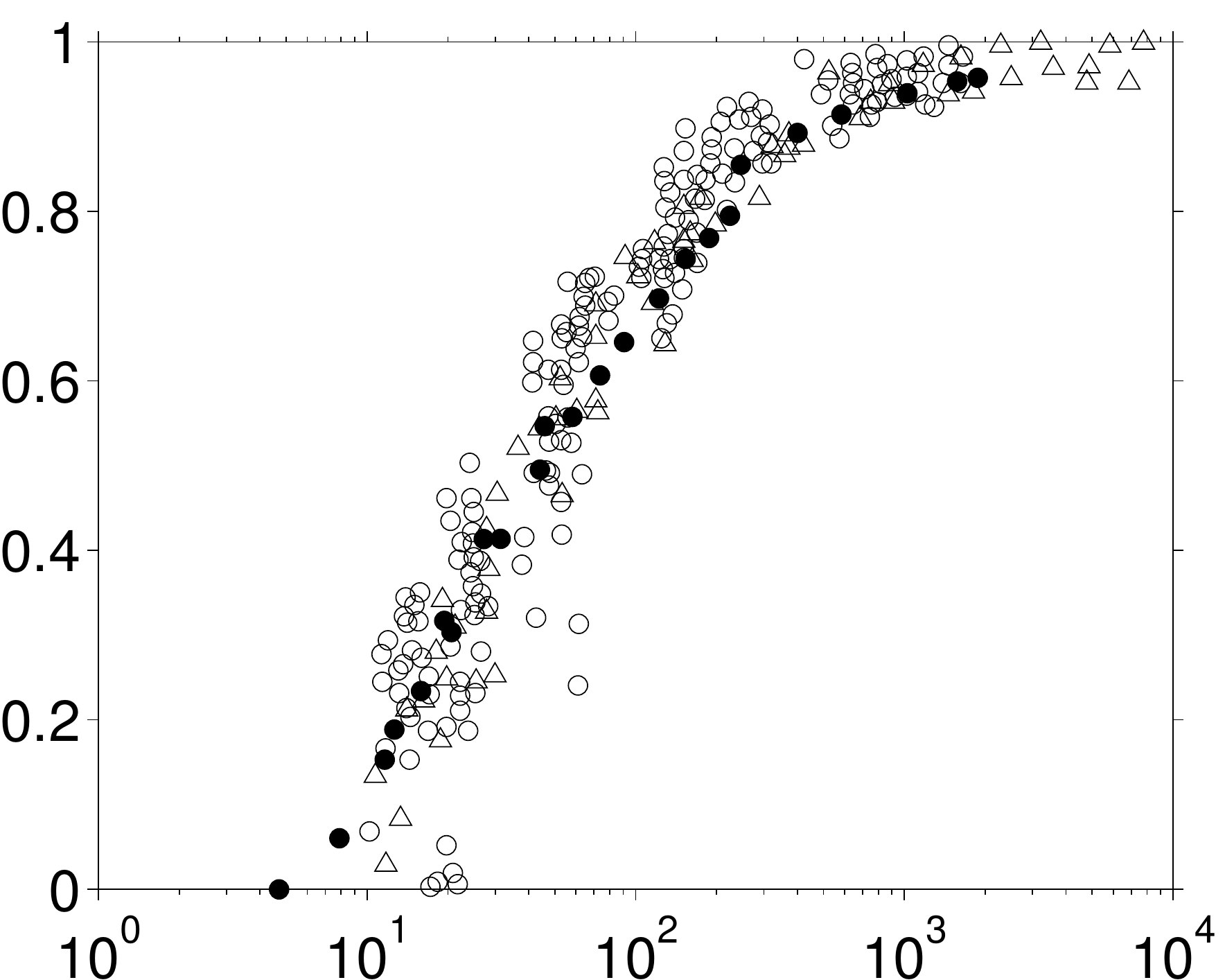}
    \centerline{\small $St$}
  \end{minipage}
  }{
  \caption{%
    Ratio between the effective coefficient of restitution \efferest\ 
    and the dry value \dryrest\ as a function of the Stokes number.
    The present results are indicated by the symbols
    ({\color{black}\solidcircle}), 
    and the rebound velocity has been computed from
    (\ref{equ-define-rebound-vel}) with $t_R=0.1D/\impactVel$.  
    Experimental data of 
    \citet{Joseph2001} and 
    \citet{Gondret2002} are marked by the symbols
    (\opencircle) and (\opentriup) respectively. 
  \protect\label{fig:restitution-coefficient-vs-stokes-number}
  }
  }
\end{figure}
%
\begin{figure}
  \figpap{
  \begin{minipage}{2ex}
    \rotatebox{90}
    {\small $\efferest/\dryrest$}
  \end{minipage}
  \begin{minipage}{.45\linewidth}
    \centerline{(\textit{a})}
    \includegraphics[width=\linewidth]
    {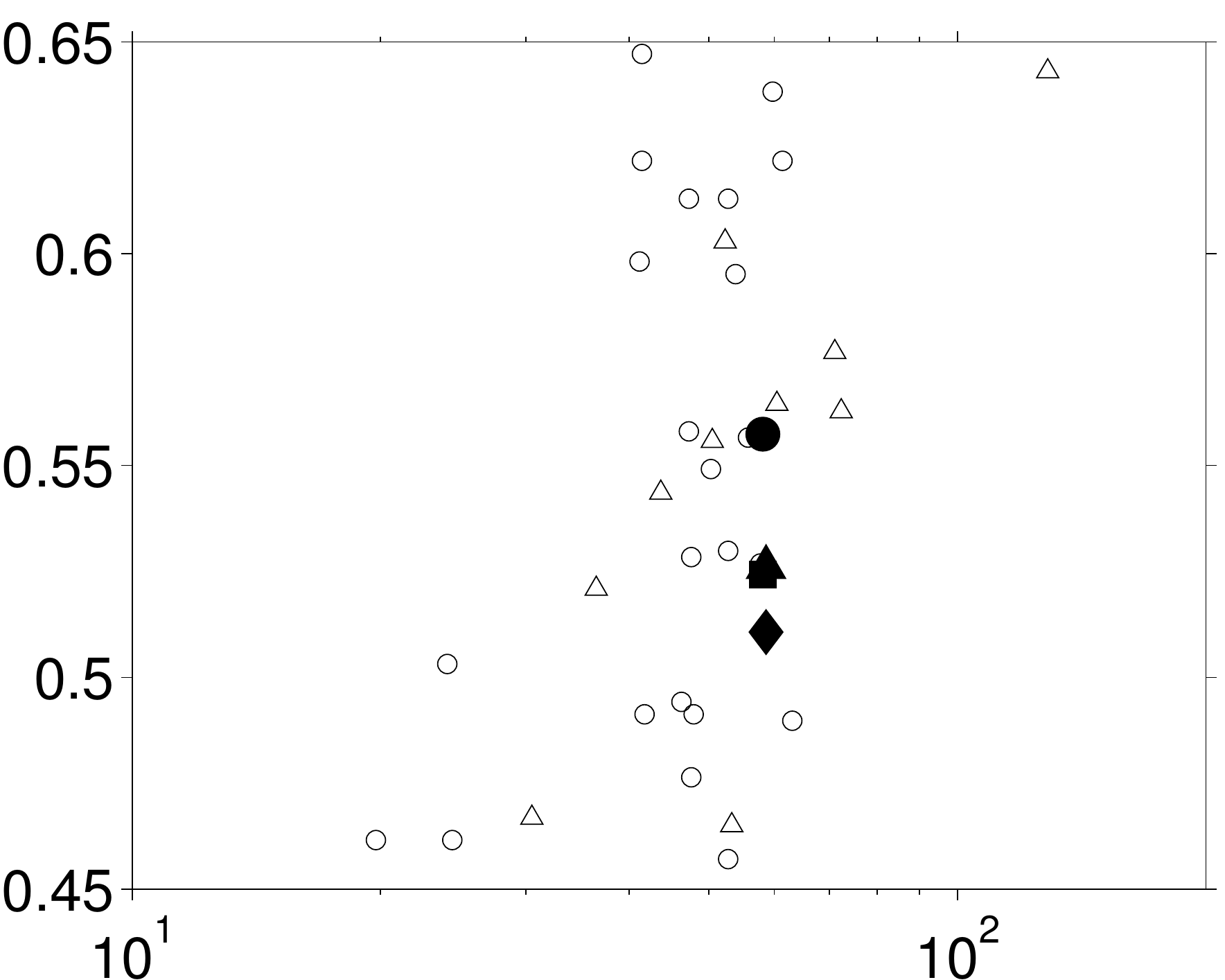}
    \centerline{\small $St$}
  \end{minipage}
  \hfill
  \begin{minipage}{2ex}
    \rotatebox{90}
    {\small $\efferest/\dryrest$}
  \end{minipage}
  \begin{minipage}{.45\linewidth}
    \centerline{(\textit{b})}
    \includegraphics[width=\linewidth]
    {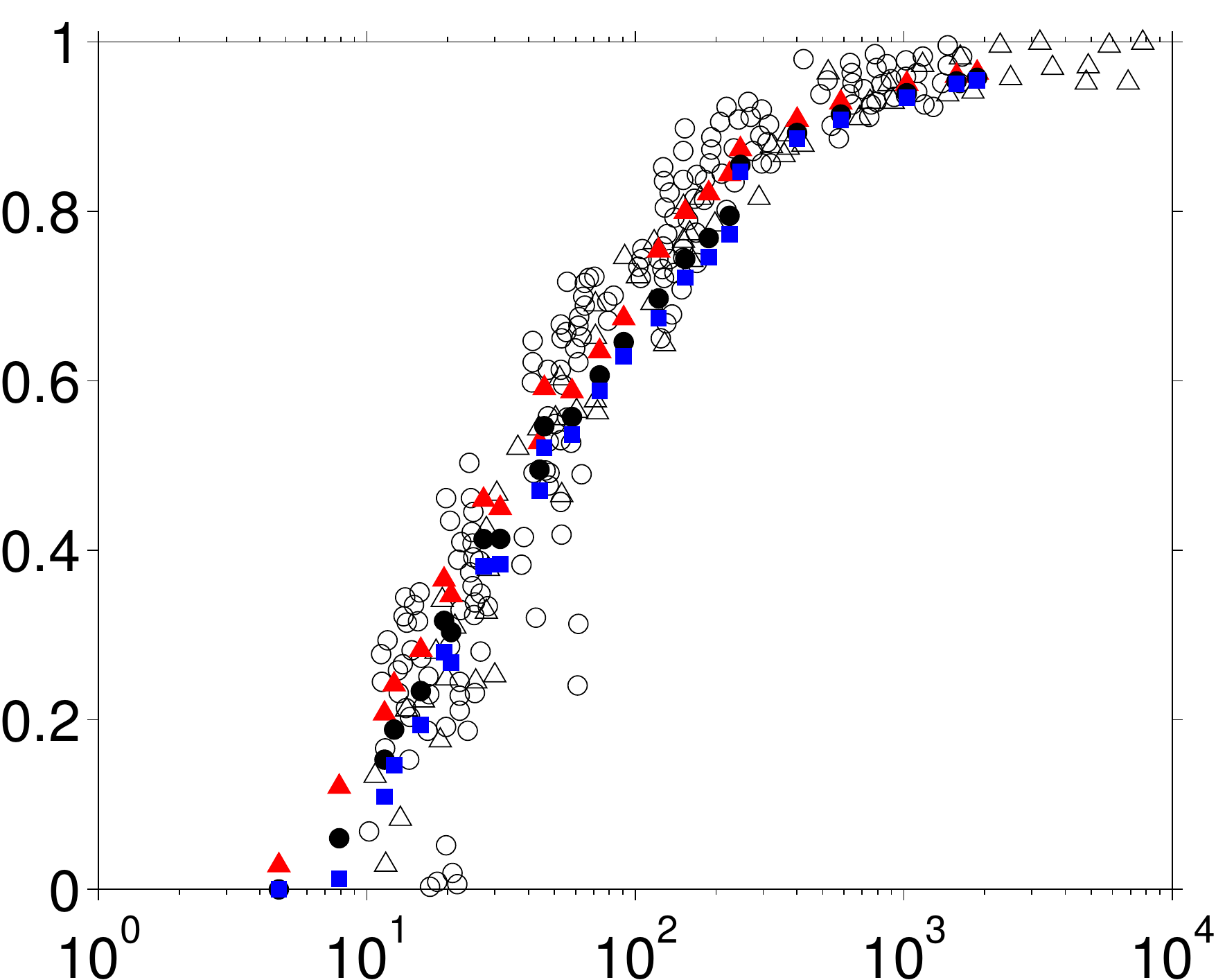}
    \centerline{\small $St$}
  \end{minipage}
  }{
  \caption{%
    (\textit{a}) 
    Sensitivity of the computed effective coefficient of restitution
    with respect to the numerical parameters for $\dratio=7$ and
    $\Ga=75.6$: 
    ({\color{black}\solidcircle}) case~C08 ($D/\Delta x=20$,
    $\Delta_c/\Delta x=2$); 
    ({\color{black}\solidsquare}) case~C08$^b$ ($D/\Delta x=20$,
    $\Delta_c/\Delta x=1$); 
    ({\color{black}\solidtriup}) case~C08$^c$ ($D/\Delta x=30$,
    $\Delta_c/\Delta x=2$); 
    ({\color{black}\soliddiamond}) case~C08$^d$ ($D/\Delta x=30$,
    $\Delta_c/\Delta x=1$). 
    (\textit{b}) 
    Sensitivity of $\efferest$ with respect of the chosen value for
    the delay time $t_R$ for all simulations: 
    ({\color{red}\solidtriup}) $t_R=0.05 D/v_{pT}$; 
    ({\color{black}\solidcircle}) $t_R=0.1 D/v_{pT}$; 
    ({\color{blue}\solidsquare}) $t_R=0.15 D/v_{pT}$. 
    In both graphs the symbols (\opencircle) and (\opentriup)
    correspond to the experimental data of \citet{Joseph2001} and 
    \citet{Gondret2002}, respectively. 
  \label{fig:restitution-coefficient-vs-stokes-number-add-1}
  }
  }
\end{figure}

%
In the present work we measure the rebound velocity at a predefined
time after the particle loses contact with the wall. More precisely,
we define
\begin{equation}\label{equ-define-rebound-vel}
  \reboundVel=v_{p}(t_1+t_R)
  \,,
\end{equation}
where $t_1$ is the instant in time when the contact force first
becomes zero after the particle-wall collision, and $t_R$ is a
prescribed delay which was set to the value of $t_R=0.1D/\impactVel$.   
Note that the chosen value for the delay is comparable to the one used
in the work of \citet{Gondret2002}. 
The insert in
figure~\ref{fig:particle-position-and-velocity-evolution}\textit{b}
shows the time evolution of the particle velocity around the collision
interval in the present case C08, where the rebound velocity obtained
from the definition in (\ref{equ-define-rebound-vel}) is marked by a
symbol. Also shown in that figure are two alternative choices of delay
times ($t_R\,\impactVel/D=0.05$ and $0.15$) 
which will be further discussed below. 

The effective coefficient of restitution computed according to 
(\ref{equ-def-efffective-coeff-restitution}) is shown as a function of
the Stokes number in
figure~\ref{fig:restitution-coefficient-vs-stokes-number}.  
A very good match with the data points provided by the experimental
measurements of \citet{Gondret2002} and those of \cite{Joseph2001}
can be observed over the whole parameter range, with an exponential
increase in $\efferest$ beyond $St_c\approx10$. 
This result demonstrates that the comparatively simple collision model
employed in the present work is capable of accurately reproducing the
fluid-mediated impact of a spherical particle on a solid wall in the
framework of an immersed boundary technique. 

The results obtained in the additional simulations C08$^b$, C08$^c$,
C08$^d$ are shown in
figure~\ref{fig:restitution-coefficient-vs-stokes-number-add-1}$(a)$
in comparison to the original case C08 (all four simulations are for
$\dratio=7$ and $\Ga=75.6$). In simulation C08$^b$ the grid resolution
is kept the same as in case C08 ($D/\Delta x=20$), but the force range
$\Delta_c$ is divided by two. The result is a reduction of the effective
coefficient of restitution by approximately 6\%. Nearly the same effect
is obtained in case~C08$^c$ where the force range is maintained
constant (in multiples of the mesh width), but the spatial resolution
is increased to $D/\Delta x=30$. When setting the smaller value for
the force range ($\Delta_c/\Delta x=1$) and simultaneously choosing
the decreased mesh width ($D/\Delta x=30$) as in case~C08$^d$, a
reduction of the value of \efferest\ by approximately 9\% is
obtained. 
In comparison to the available experimental data, this sensitivity
analysis shows that the value of the force range is not a crucial 
quantity. It also demonstrates that a mesh width of $D/\Delta x=20$
is adequate at the present parameter point. 

Finally, let us consider the sensitivity of our results with respect
to the choice of the delay time
$t_R$. 
For all present simulations
figure~\ref{fig:restitution-coefficient-vs-stokes-number-add-1}$(b)$ 
shows the results for the effective coefficient of restitution
computed with three different values of the delay time as indicated in
figure~\ref{fig:particle-position-and-velocity-evolution}$(b)$: 
$t_R\,v_{pT}/D=0.05$, $0.1$ and $0.15$. 
Choosing a smaller delay time has the effect of systematically
increasing the coefficient of restitution, since the fluid-mediated
damping after the particle-wall contact has less time to act. 
However, it can be observed that over the whole range of values the
simulation results match the experimental data very well. 

%% file: headon_collision_relevant_paramters_table.tex
%
%
C01&2.0&30.9&21.2&4.7&40053&D1\\
C02&2.5&37.8&28.4&7.9&33377&D1\\
C03&3.0&43.7&34.9&11.6&30039&D1\\
C04&3.5&48.8&40.9&15.9&28037&D1\\
C05&4.0&53.5&46.4&20.6&26702&D1\\
C06&5.0&61.8&56.6&31.4&25033&D1\\
C07&6.0&69.1&66.0&44.0&24032&D2\\
C08&7.0&75.6&74.7&58.1&23364&D2\\
C09&8.0&81.7&82.9&73.7&22887&D2\\
C10&9.0&87.3&90.6&90.6&22530&D2\\
C11&4.0&37.8&28.4&12.6&26702&D2\\
C12&5.0&43.7&34.9&19.4&25033&D2\\
C13&6.0&48.8&40.8&27.2&23969&D2\\
C14&8.0&57.8&51.6&45.9&22843&D2\\
C15&14.0&78.7&78.5&122.2&215427&D3\\
C16&16.0&84.6&86.4&153.6&213406&D3\\
C17&18.0&90.0&93.9&187.7&211675&D3\\
C18&20.0&95.2&101.0&224.5&210474&D3\\
C19&30.0&75.1&73.9&246.3&206629&D3\\
C20&40.0&87.1&90.2&400.7&204997&D3\\
C21&50.0&97.7&105.0&583.3&204031&D3\\
C22&70.0&115.9&132.0&1026.7&202939&D3\\
C23&90.0&131.6&157.3&1573.2&202162&D3\\
C24&100.0&138.8&169.1&1878.3&201970&D3\\
C08$^b$&7.0&75.6&74.7&58.1&23364&D2$^b$\\
C08$^c$&7.0&75.6&75.4&58.6&23364&D2$^c$\\
C08$^d$&7.0&75.6&75.4&58.6&23364&D2$^d$\\
%
%

%% file: lam_bdld_tras_setup.tex
\section{Erosion of granular bed sheared by laminar flow}
\label{sec:erosion-of-granular-bed-sheared-by-laminar-flow}
\subsection{Computational setup}
\label{subsec:computational-setup}
\subsubsection{Flow configuration and parameter values}
\label{subsubsec:flow-configuration-and-parameter-values}
\begin{figure}
  \figpap{
      
  \begin{minipage}{0.8\linewidth}
    \centerline{$(a)$\hspace*{30ex}$(b)$}
    \includegraphics[width=\linewidth,
    clip=true,viewport=93 420 555 675]
    {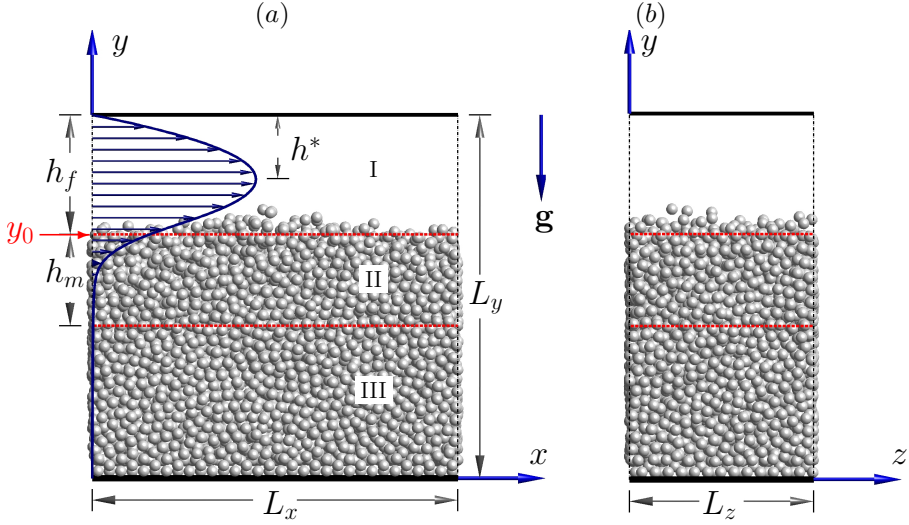}
  \end{minipage}
  }{
  \caption{Schematic diagram showing the configuration of
    the bedload transport simulations in
    \S~\ref{sec:erosion-of-granular-bed-sheared-by-laminar-flow}. 
    Periodicity in the $x$- and $z$-directions is assumed. 
    $(a)$ shows the streamwise/wall-normal plane; in $(b)$ the view is
    into the streamwise direction.  
    \protect\label{fig:bedload-schematic-diagram}
  }
  }
\end{figure}
In the present section we are considering the motion of spherical
particles induced by laminar flow in a horizontal plane channel, as
sketched in figure~\ref{fig:bedload-schematic-diagram}. 
The computational set-up features a number of particles (\Npt) forming
a sediment bed which takes up a large fraction of the channel height. 
As figure~\ref{fig:bedload-schematic-diagram} shows, the Cartesian
coordinates $x$, $y$, and $z$ correspond to the streamwise,
wall-normal and spanwise directions, respectively. The cuboidal
computational domain is of size $L_x$, $L_y$, $L_z$ in the respective
coordinate directions; periodicity is imposed in $x$ and $z$, while a
no-slip condition is applied at the two wall planes. 
The flow is driven by a streamwise pressure gradient 
adjusted at each time step such as to maintain
a constant flow rate \qfmean\ (note that the actual volumetric flow
rate is divided by the spanwise domain size, i.e.\ \qfmean\
corresponds to a flow rate per unit span with units of velocity times
length).  
The particulate flow problem features 10 relevant quantities
($\rho_p$, $\rho_f$, \qfmean, $|\mathbf{g}|$, $D$, $\nu$, $L_x$,
$L_y$, $L_z$, \Npt) which means that it is fully described by 7
non-dimensional parameters. 
These can be chosen as follows:
the particle-to-fluid density ratio \dratio; 
the Galileo number \Ga\ 
(defined in
\S~\ref{subsec:collision-sphere-wall-viscous-fluid-setup});  
the bulk Reynolds number given by
\begin{equation}
  Re 
  = \frac{\qfmean}{\nu}
  \,;
  \label{reynolds-number}
\end{equation}
the global solid volume fraction defined as
\begin{equation}\label{equ-def-solid-volume-fraction}
  \Phi_s=
  \frac{\Npt V_p}{L_xL_yL_z}
  \,,
\end{equation}
where $V_p=(\pi/6)D^3$ is the particle volume; 
the three relative length scales $L_x/D$, $L_y/D$, $L_z/D$. 
As an alternative to one of the latter length scale ratios one can
e.g.\ choose an imposed pile height parameter given by
\citep{Silbert2001}
\begin{equation}\label{equ-define-pile-height-param}
  \pileparam
  =
  \Npt D^2/(\Lx\,\Lz)
  \,.
\end{equation}
%
\begin{table}\scriptsize
  \centering
  \begin{tabular}{lcccccccccc}
    \hline
    Case & 
    \Reb\ & 
    \Ga\ & 
    $k_n^*$& 
    \pileparam\ &
    $\Phi_s$&
    \Npt\ &
    $\hfluid/D$ &
    \shields\ & 
    $\Phi_{bed}$ &
    domain 
    \\
    \hline
    \input{bdld_relevant_paramters_table_A}
    \hline
  \end{tabular}
  \caption{Physical parameters of the bedload transport simulations
    in \S~\ref{sec:erosion-of-granular-bed-sheared-by-laminar-flow}: 
    bulk Reynolds number \Reb, 
    Galileo number \Ga,   
    normalized solid stiffness coefficient
    $k_n^\ast=k_nD/((\rho_p/\rho_f-1)V_p|\mathbf{g}|)$, 
    pile height parameter \pileparam\ (given in \ref{equ-define-pile-height-param}), 
    global solid volume fraction $\Phi_s$ (defined in
    \ref{equ-def-solid-volume-fraction}), 
    number of particles \Npt, 
    fluid height \hfluid\ (as defined in 
    \S~\ref{subsubsec:determination-of-fluid-height}), 
    Shields number \shields\ (defined in
    \ref{equ-def-shields-poiseuille})  
    and solid volume fraction inside the bulk of the bed $\Phi_{bed}$ 
    (as defined in \ref{equ-def-mean-solid-vol-fract-in-bulk}). 
    The particle-to-fluid density ratio was kept at $\dratio=2.5$ in
    all cases. 
    The computational domain size and numerical
    parameters corresponding to the abbreviation in column 11 
    are listed in table~\ref{tab:numerical_parameters}. 
    %
  }
  \label{tab:physical_parameters}
\end{table}
%
%
%
\begin{table}\scriptsize
   \centering
   \begin{tabular}{*{6}{c}}
   \hline
   Domain & $ [\Lx \times \Ly\times \Lz]/D $ &
   $N_x \times N_y\times N_z$ & $D/\Delta x$ & 
   $\Delta_c/\Delta x$ & symbol \\
   D4 & 
   $32 \times 32 \times 16$ & 
   $512 \times 513 \times 256$ & 
   16 &  
   2 &
   {\color{black}\solidthick},{\color{black}\solidcircle}\\
   D5 & 
   $32 \times 36 \times 16$ & 
   $512 \times 577 \times 256$ & 
   16 &  
   2 &
   {\color{black}\solidthick},{\color{black}\solidcircle}\\
   D6 & 
   $32 \times 16 \times 16$ & 
   $512 \times 257 \times 256$ & 
   16 &  
   2 &
   {\color{black}\solidthick},{\color{black}\solidcircle}\\
   D7 & 
   $38.4 \times 25.6 \times 12.8$ & 
   $384 \times 257 \times 128$ & 
   10 &  
   1  &
   {\color{black}\solidthick},{\color{red}\solidsquare}\\
   D8 & 
   $38.4 \times 51.2 \times 12.8$ & 
   $384 \times 513 \times 128$ & 
   10 &  
   1 &
   {\color{black}\solidthick},{\color{red}\solidsquare}\\
   D8$^b$ & 
   $38.4 \times 51.2 \times 12.8$ & 
   $768 \times 1025 \times 256$ & 
   20 &  
   2 &
   {\color{blue}\solidthick},{\color{blue}\solidtriup}\\
   D9 & 
   $32 \times 36 \times 16$ & 
   $512 \times 577 \times 256$ & 
   16 &  
   1 &
   {\color{magenta}\solidthick},{\color{magenta}\soliddiamond}\\
   D10 & 
   $32 \times 32 \times 16$ & 
   $512 \times 513 \times 256$ & 
   16 &  
   1 &
   {\color{magenta}\solidthick},{\color{magenta}\soliddiamond}\\
    \hline
   \end{tabular}
   \caption{Numerical parameters used in 
           the simulations of bedload transport 
           (for the notation cf.\
           table~\ref{tab:numerical_parameters_of_single_particle_collision}).  
           The color and symbol coding given in the last column 
           correspond to a chosen pair of numerical parameter values  
           ($D/\Delta x$,$\Delta_c/\Delta x$) and will be used
           in subsequent plots.        
           %
         }
   \label{tab:numerical_parameters}
\end{table}
%
Further global parameters of interest 
are computed a posteriori from the observed height \hfluid\
of the fluid-bed interface (defined in
\S~\ref{subsubsec:determination-of-fluid-height}) as well as from the
characteristic shear stress. 
The Shields number is defined as 
\begin{equation}\label{equ-def-shields-general-case}
  \shieldsgen
  =
  \frac{u_\tau^2}{\left(\rho_p/\rho_f-1\right)|\mathbf{g}|D}
    \,,
\end{equation}
where $u_\tau=(\tau_w/\rho_f)^{1/2}$ is the friction velocity (with $\tau_w$
denoting the wall shear stress). In laminar plane Poiseuille flow with
smooth walls, the Shields number can be written as (using the present
notation):
\begin{equation}\label{equ-def-shields-poiseuille}
  \shields
  =
  \frac{6\,Re}{Ga^2}\,\left(\frac{D}{\hfluid}\right)^2
  \,.
\end{equation}
Although in a region near the rough fluid-bed surface the mean
velocity profile is not identical to a parabolic Poiseuille profile
(cf.\ discussion in
\S~\ref{subsec:mean-fluid-and-particle-velocities}),  
the Shields number based upon the definition
(\ref{equ-def-shields-poiseuille}) is often used as a parameter in
experimental studies. Therefore, its value is included in
table~\ref{tab:physical_parameters} for each simulated case. 
%
%

Out of the total number of \Npt\ particles, one layer adjacent to the
wall 
\revision{(in dense hexagonal arrangement and with small-amplitude 
  modulation of the vertical position)}{%
  (in dense hexagonal arrangement and with a vertical displacement by
  one particle radius applied to every second particle in that layer)} 
is kept fixed during the
entire simulation in order to form a rough bottom wall surface. 
The chosen parameter values in the 26 independent simulations which we
have performed are listed in table~\ref{tab:physical_parameters}. 
Note that 24 different physical parameter combinations have been
chosen. The remaining two simulations (B20$^b$ and B21$^b$) were
conducted at one half the mesh width in order to verify the adequacy
of the spatial resolution. 

%
%
%
Let us now turn to the parameter values of the solid contact
model. Analogous to the strategy employed in
\S~\ref{subsec:collision-sphere-wall-viscous-fluid-setup} we have set
the stiffness parameter $k_n$ of the elastic normal force component
such that the maximum penetration length is kept below a few percent
of the force range (i.e.\ $\max(\delta_{ij}(t))\leq0.05\Delta_c$). 
This criterion is fulfilled for values of
$k_n^\ast=k_nD/((\rho_p/\rho_f-1)V_p|\mathbf{g}|)$ in the interval of
$5400-17000$ (for the precise choice per flow case cf.\
table~\ref{tab:physical_parameters}).  
The dry restitution coefficient was set to a value of $\dryrest=0.3$,
which is smaller than the material property commonly reported for
glass ($\dryrest=0.97$, cf.\
\S~\ref{subsec:collision-sphere-wall-viscous-fluid-setup}) or   
plexiglass \citep[PMMA, $\dryrest=0.8$, cf.\ ][]{constantinides:08}. 
%
%
%
With this choice the normal damping coefficient $c_{dn}$ can be
computed from relation
(\ref{damping-coefficient-and-stiffness-coefficient}).  
The tangential damping coefficient was set to the same value, i.e.\ 
$c_{dt}=c_{dn}$ \citep{Tsuji1993}. 
The value for the Coulomb friction coefficient was fixed at
$\mu_c=0.4$ 
which is similar to but somewhat larger than values reported for wet
contact between glass-like materials \citep{Foerster1994a,Joseph2004a}. 
%
%
%
\begin{figure}
  \figpap{
        \begin{minipage}{2ex}
          \rotatebox{90}
          {\small $y/D$}
        \end{minipage}
        \begin{minipage}{0.4\linewidth}
          \centerline{$(a)$}
          \includegraphics[width=\linewidth,clip=true,
          viewport=260 60 2200 1710]
          {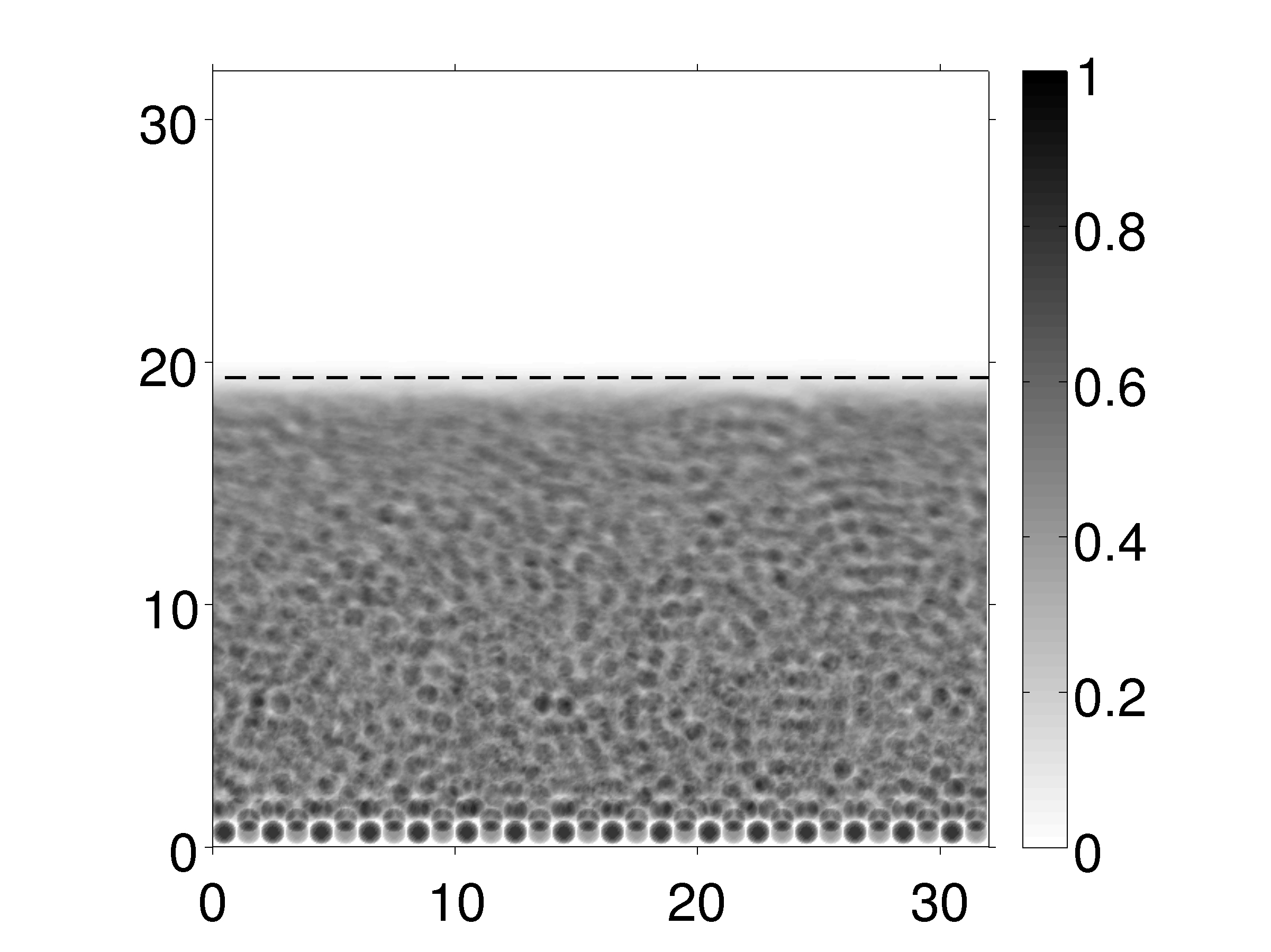}
          \centerline{\small $x/D$ }
        \end{minipage}
        \hfill
        \begin{minipage}{2ex}
          \rotatebox{90}
          {\small $(y-\yref)/\hfluid$}
        \end{minipage}
        \begin{minipage}{0.45\linewidth}
          \centerline{$(b)$}
          \includegraphics[width=\linewidth]
          {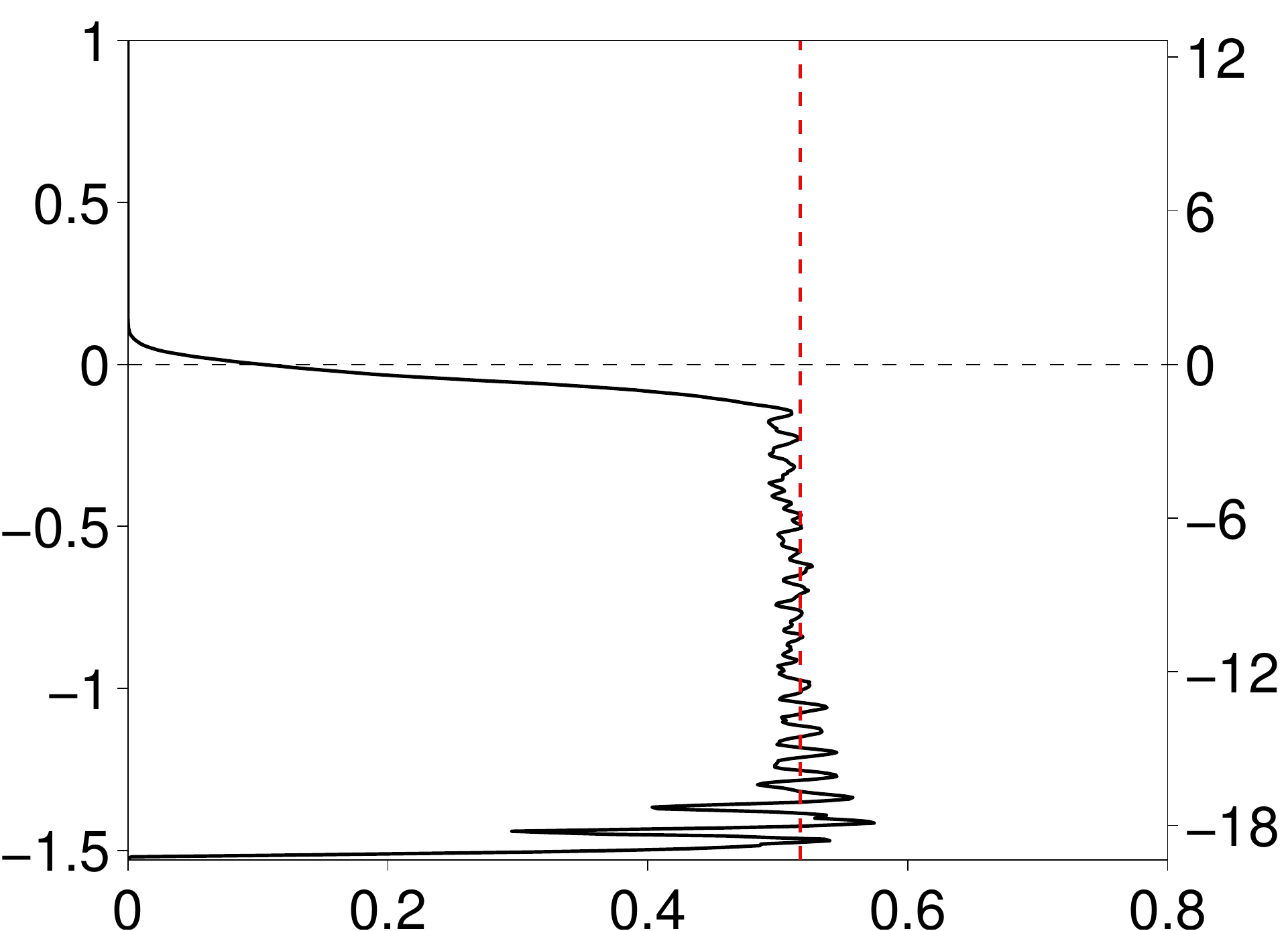}
          \centerline{\small $\phimeansmooth$ }
        \end{minipage}
        \begin{minipage}{2ex}
          \rotatebox{90}
          {\small $(y-\yref)/D$}
        \end{minipage}
  }{
        \caption{(\textit{a})
         Solid volume fraction averaged in time and over the spanwise
         direction, $\phimeansmooth_{zt}(x,y)$, in case BL24. 
         The horizontal dashed line represents
         the mean fluid-bed interface location \yref\ defined in
         (\ref{eq:fluid-height-and-bed-height}).
         (\textit{b})
         The corresponding profile of the mean solid volume fraction
         $\phimeansmooth(y)$. 
         The mean value in the bulk of the bed, $\Phi_{bed}$, is
         indicated by a vertical dashed line (in red color).
         \protect\label{fig:particle_bed_grayscale}
       }
     }
\end{figure}
\subsubsection{Simulation start-up and initial transient}
\label{subsubsec:simulation-start-up-and-initial-transient}
\begin{figure}
  \figpap{
        %
        \begin{minipage}{2ex}
          \rotatebox{90}
          {\small $q_p/\qpmean$}
        \end{minipage}
        \begin{minipage}{.45\linewidth}
         \centerline{$(a)$}
          \includegraphics[width=\linewidth]
          {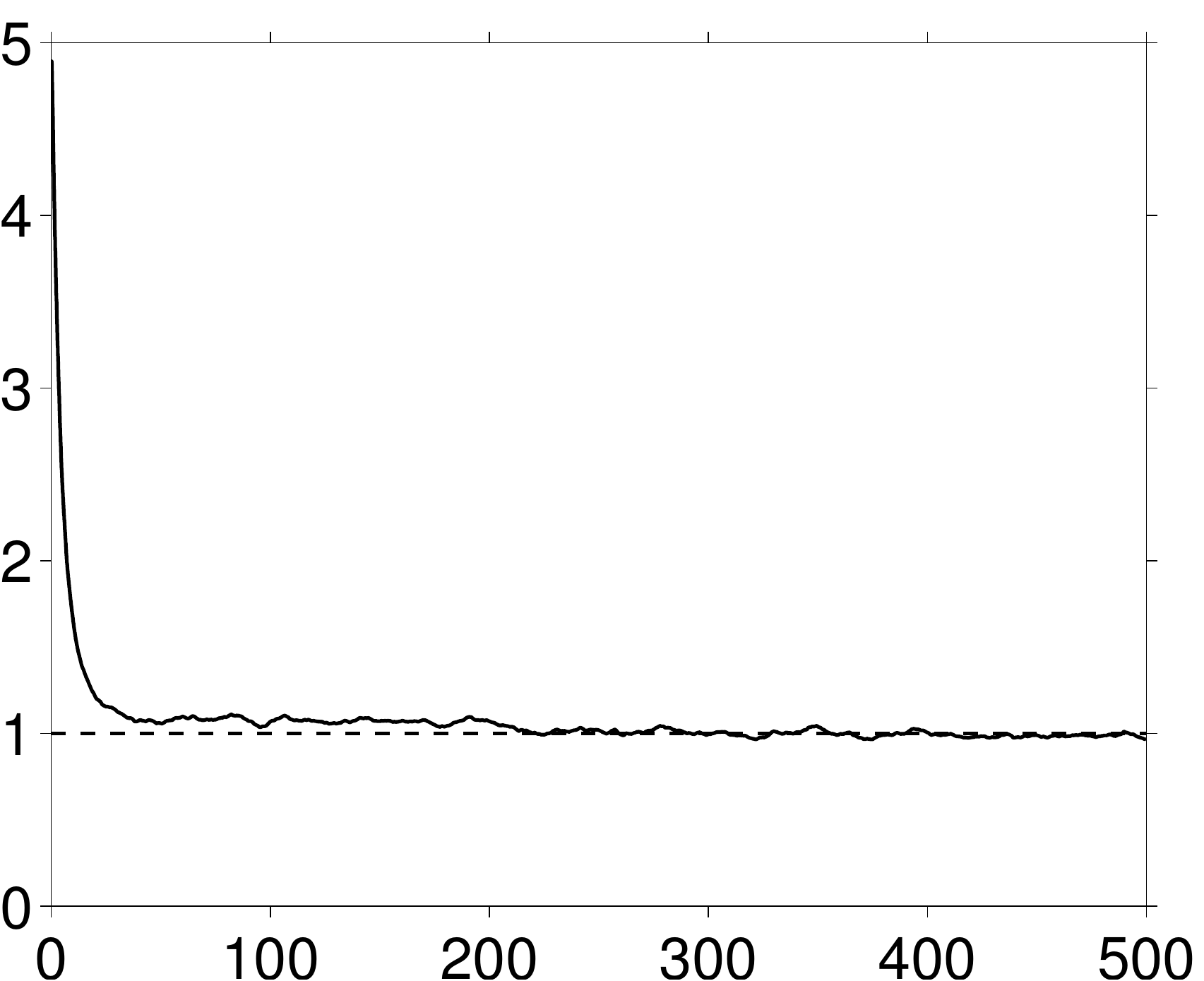}
          \centerline{\small 
            $t\, q_{\rm f}/h_{\rm f}^2$}
        \end{minipage}
        \hfill
        \begin{minipage}{2ex}
          \rotatebox{90}
          {\small $q_p/\qpmean$}
        \end{minipage}
        \begin{minipage}{.45\linewidth}
         \centerline{$(b)$}
          \includegraphics[width=\linewidth]
          {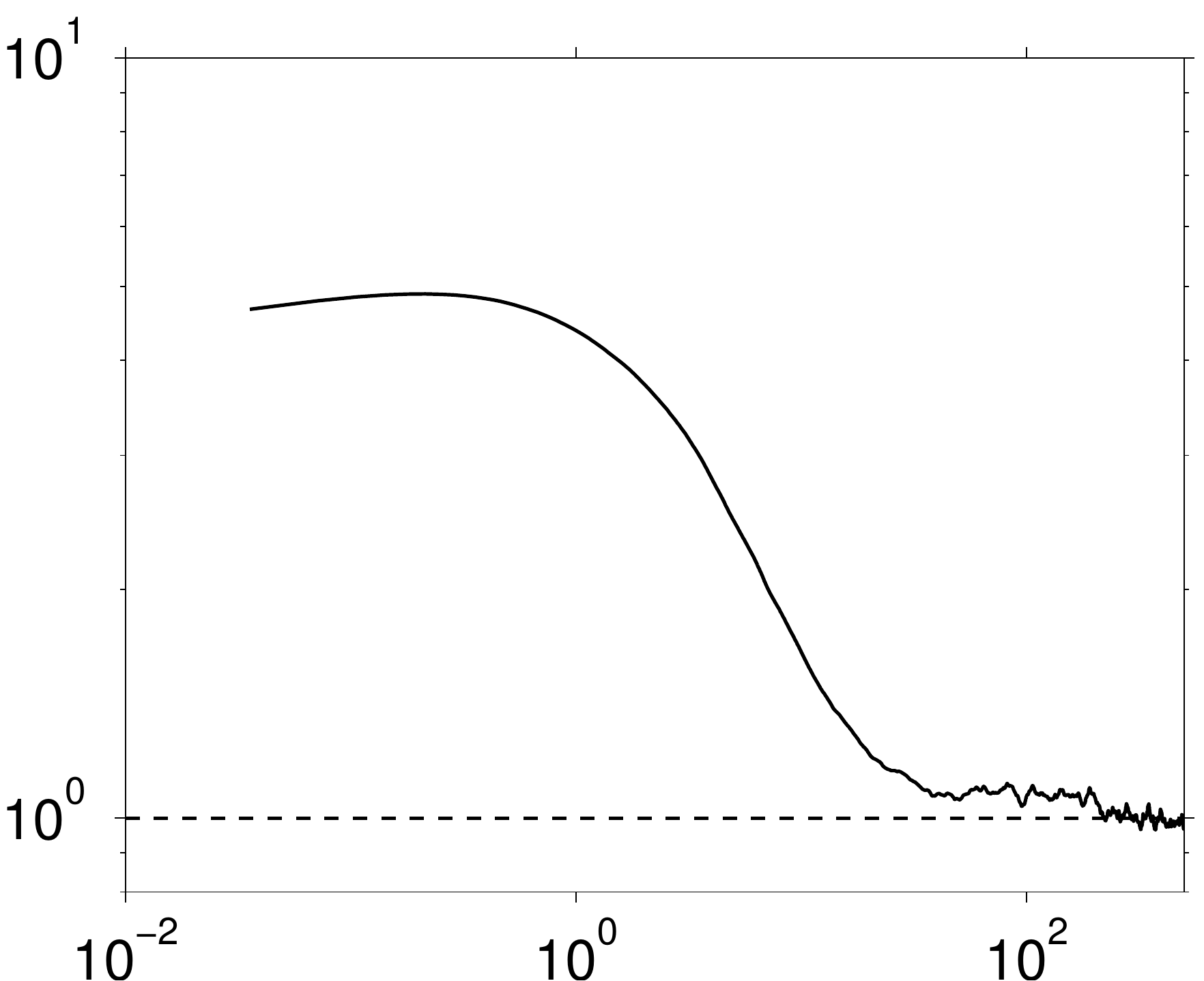}
          \centerline{\small 
            $t\, q_{\rm f}/h_{\rm f}^2$}
        \end{minipage}
  }{
        \caption{$(a)$ 
          Time evolution of the particle flow rate $q_p$ defined in 
          \eqref{instantaneous-particle-flux} for case BL08,
          normalized by the mean value in the statistically stationary
          regime.
          $(b)$ shows the same data in logarithmic scale. 
        \protect\label{fig:particle_flux_evolution}
        } 
      }
\end{figure}

As an initial condition for each fluid-solid simulation a sediment bed
composed of quasi-randomly packed particles is generated. 
For this purpose, a DEM simulation was performed under the conditions
of each simulation (i.e.\ regarding domain size, number of particles,
particle diameter, particle density, value of gravitational
acceleration and contact model parameters) but ignoring hydrodynamic
forces, i.e.\ considering dry granular flow with gravity.
Each DNS-DEM simulation is then started by ramping up the flow rate
from zero to its prescribed value 
over a short time interval
of the order of $10^{-2}$ bulk time units $T_b=\hfluid^2/\qfmean$
(where \hfluid\ is the fluid height defined in
\S~\ref{subsubsec:determination-of-fluid-height}).  
After an initial transient, a statistically stationary state
of the particle-fluid system is reached with -- upon average -- 
constant bed height and constant mean particle flux.
%
The time evolution of the instantaneous particle flux 
$q_p$ (cf.\ equation \eqref{instantaneous-particle-flux} and
figure \ref{fig:particle_flux_evolution}, discussed below)
is used to determine the statistically stationary state interval 
which is used to compute statistics.
%
\subsubsection{Reference experimental data}
\label{subsubsec:reference-experimetal-data}
A detailed set of experimental data  
in a similar configuration studied by \citet{Aussillous2013} is
available for validation purposes. 
In the experiment the transport of spherical particles
in a rectangular channel in the laminar regime is considered. 
Two combinations of fluid and particle properties were investigated by
the authors with particle-to-fluid density ratios $\dratio=1.11$ and 
$2.10$ as well as Galileo number values $\Ga=0.36$ and
$0.38$. 
A pump is used to impose a constant flow rate, which is varied 
such that the bulk Reynolds number ranges from $\Reb=0.27$ up to 
$1.12$. 
Note that these values are two orders of magnitude smaller than the
\revision{Reynold}{Reynolds} 
number values in the present simulations. 
However, the main control parameter of the problem (the Shields
number) covers a similar range ($\shields\approx0.2-1$) as the present
simulations. 
Contrary to our simulation setup, the sediment bed
thickness is not kept constant in the experiment. 
First, a given amount of sediment is 
filled into the channel yielding the initial sediment bed height.   
After start-up of the flow, particles are eroded and transported 
out of the test section, resulting in a fluid height which increases 
with time until it reaches its final equilibrium value. 
Before the cessation of particle erosion is reached, statistics are
accumulated at intervals of $5\,s$ (averaged over a small time window of
$0.5\,s$) in which the flow is assumed to be steady.
An index matching technique is utilized in order to perform optical
measurements of both the velocity of the fluid phase and that of the
particle phase, including locations deep inside the sediment bed.
%
\subsubsection{Determination of the fluid height}
\label{subsubsec:determination-of-fluid-height}
In the reference experiments, a flourescent dye is mixed into the fluid
which upon illumination by a laser sheet emits light detectable by a
camera. In the recorded images the particle positions show up as low
intensity (i.e.\ dark) regions. Such an intensity map can then be
averaged over a number of frames and 
converted into a vertical profile of the solid volume fraction; 
by means of thresholding the fluid-bed interface location can be
determined \citep{Lobkovsky2008,Aussillous2013}. 
Here we have adopted the same approach to determine the location of
the interface based upon our DNS-DEM simulation data. 
In this manner we can directly compare our results with the
experimental counterpart.  

In this spirit we first define a solid phase indicator function
$\phi_p(\mathbf{x},t)$  which equals unity for a point $\mathbf{x}$
that is located inside any particle at time $t$ and zero otherwise  
(cf.\ appendix~\ref{sec-ensemble-averaging}). 
%
%
Next we compute the average of $\phi_p(\mathbf{x},t)$ over time (in
the statistically stationary interval) and over the spanwise
direction, yielding $\langle\phi_p\rangle_{zt}(x,y)$.
An example of this two-dimensional map for case~BL24 is shown in
figure~\ref{fig:particle_bed_grayscale}$(a)$. 
%
Further averaging of $\langle\phi_p\rangle_{zt}(x,y)$ over the
streamwise direction yields the wall-normal profile of the mean solid
volume fraction $\phimeansmooth(y)$, an example of which is given in 
figure~\ref{fig:particle_bed_grayscale}$(b)$.  
Note that the angular brackets without subscripts $\langle\,\rangle$
denote an average over the two homogeneous space directions $x$, $z$
and over time.  
The mean solid volume fraction $\phimeansmooth(y)$ is an alternative
quantity to $\langle \phi_{\rm s}\rangle(y)$
which is 
computed from the number density of
particles in wall-normal averaging bins of finite size (cf.\ 
appendix~\ref{sec-ensemble-averaging}).  
The former approach gives more precise results in cases when there
exists a strong gradient in the solid volume fraction profile as in 
figure~\ref{fig:particle_bed_grayscale}; it is therefore generally
preferred. The latter quantity enters the definition of the particle
flux as defined in equation~(\ref{instantaneous-particle-flux})
below. 
Finally, the fluid-bed interface location is
defined as the wall-normal coordinate \yref\ where the value of the mean
solid volume fraction equals a prescribed threshold value
$\phimeansmooth^{thresh}=0.10$, viz.
\begin{eqnarray}\label{eq:fluid-height-and-bed-height} 
   \yref   &=& y \;\; \vert \; \phimeansmooth(y)
                                =\phimeansmooth^{thresh}\;.
\end{eqnarray}
Consequently, the fluid height is given by 
\begin{equation}
  \hfluid = \Ly - \yref.
  \label{eq:fluid-height}
\end{equation}
For the example of case BL24 the resulting interface location is shown
in figure~\ref{fig:particle_bed_grayscale}. 

It can also be seen in figure~\ref{fig:particle_bed_grayscale}$(b)$
that a reasonable value for the mean solid volume fraction in the bulk
of the sediment bed can be defined as the following average 
\begin{equation}\label{equ-def-mean-solid-vol-fract-in-bulk}
  \solidvolfracbed
  =\frac{1}{y_2-y_1}\,\int_{y_1}^{y_2} \phimeansmooth \,\mbox{d}y
  \,,
\end{equation}
where the interval is delimited by $y_1=3D$ and 
$y_2=6D$. 
The resulting values of $\solidvolfracbed$ vary between $0.43$ and
$0.53$ as reported in table~\ref{tab:physical_parameters}. 
Due to the combined effects of considering exactly mono-dispersed
particles and using a finite force range $\Delta_c$ introduced into
the collision model (cf.\
\S~\ref{subsec:inter-particle-collision-model}),  
these values are somewhat smaller than what is found in experiments. 
\revision{For instance, in the study of \cite{Lobkovsky2008} values in the range 
  of $\solidvolfracbed=0.5-0.65$ are reported. 
  %
  %
}{%
  For instance, in the study of \cite{Aussillous2013} values in the range 
  of $\solidvolfracbed=0.55-0.62$ are measured while
  \cite{Lobkovsky2008} report $\solidvolfracbed=0.5-0.6$.}
Note that the maximum packing fraction of a homogeneously sheared
assembly of frictional spheres measures $0.585$ \citep{boyer:11}.

%% file: bdld_relevant_paramters_table_A.tex
%
%
BL01&375&6.77&13038&9.57&0.16&4900&20.04&0.12&0.44&D4 \\
BL02&375&6.77&13038&13.03&0.21&6670&15.59&0.20&0.43&D4 \\
BL03&375&6.77&13038&16.21&0.27&8300&10.52&0.44&0.43&D4 \\
BL04&375&7.41&10865&20.89&0.30&10695&8.14&0.62&0.43&D5 \\
BL05&375&8.56&8149&9.57&0.16&4900&20.16&0.08&0.44&D4 \\
BL06&375&8.56&8149&13.03&0.21&6670&15.97&0.12&0.44&D4 \\
BL07&375&8.56&8149&16.21&0.27&8300&11.65&0.23&0.44&D4 \\
BL08&375&8.56&8149&18.24&0.30&9341&7.51&0.54&0.43&D4 \\
BL09&133&13.98&5432&6.34&0.21&3246&8.03&0.06&0.44&D6 \\
BL10&266&13.98&5432&6.34&0.21&3246&7.72&0.14&0.43&D6 \\
BL11&333&13.98&5432&6.34&0.21&3246&7.47&0.18&0.43&D6 \\
BL12&300&18.15&7243&6.34&0.21&3246&7.91&0.09&0.44&D6 \\
BL13&400&18.15&7243&6.34&0.21&3246&7.78&0.12&0.44&D6 \\
BL14&500&18.15&7243&6.34&0.21&3246&7.47&0.16&0.43&D6 \\
BL15&111&14.86&8344&14.74&0.30&7246&8.83&0.04&0.48&D7 \\
BL16&222&14.86&8344&14.74&0.30&7246&8.83&0.08&0.48&D7 \\
BL17&333&14.86&8344&14.74&0.30&7246&8.63&0.12&0.48&D7 \\
BL18&444&14.86&8344&14.74&0.30&7246&8.23&0.18&0.47&D7 \\
BL19&200&6.30&16689&34.31&0.35&16864&12.52&0.19&0.47&D8 \\
BL20&267&6.30&16689&34.31&0.35&16864&12.12&0.27&0.47&D8 \\
BL21&333&6.30&16689&34.31&0.35&16864&11.42&0.39&0.47&D8 \\
BL22&400&7.72&11126&34.31&0.35&16864&12.02&0.28&0.48&D8 \\
BL20$^b$&267&6.30&16689&34.31&0.35&16864&11.96&0.28&0.47&D8$^b$ \\
BL21$^b$&333&6.30&16689&34.31&0.35&16864&11.26&0.40&0.47&D8$^b$ \\
BL23&375&7.41&10865&20.89&0.30&10695&13.90&0.21&0.53&D9 \\
BL24&375&8.56&8149&18.24&0.30&9341&12.65&0.19&0.53&D10 \\
%
%

%% file: lam_bdld_tras_results.tex
\subsection{Results}
\label{subsec:laminar-bedload-result}
\subsubsection{Particle flux}
\label{subsec:particle-flux}
%
\begin{figure}
  \figpap{
        %
        \begin{minipage}{2ex}
          \rotatebox{90}
          {\small $\qpmean/\qrefv$}
        \end{minipage}
        \begin{minipage}{.45\linewidth}
          \centerline{\hspace*{8ex}$(a)$}
          \includegraphics[width=\linewidth]
          {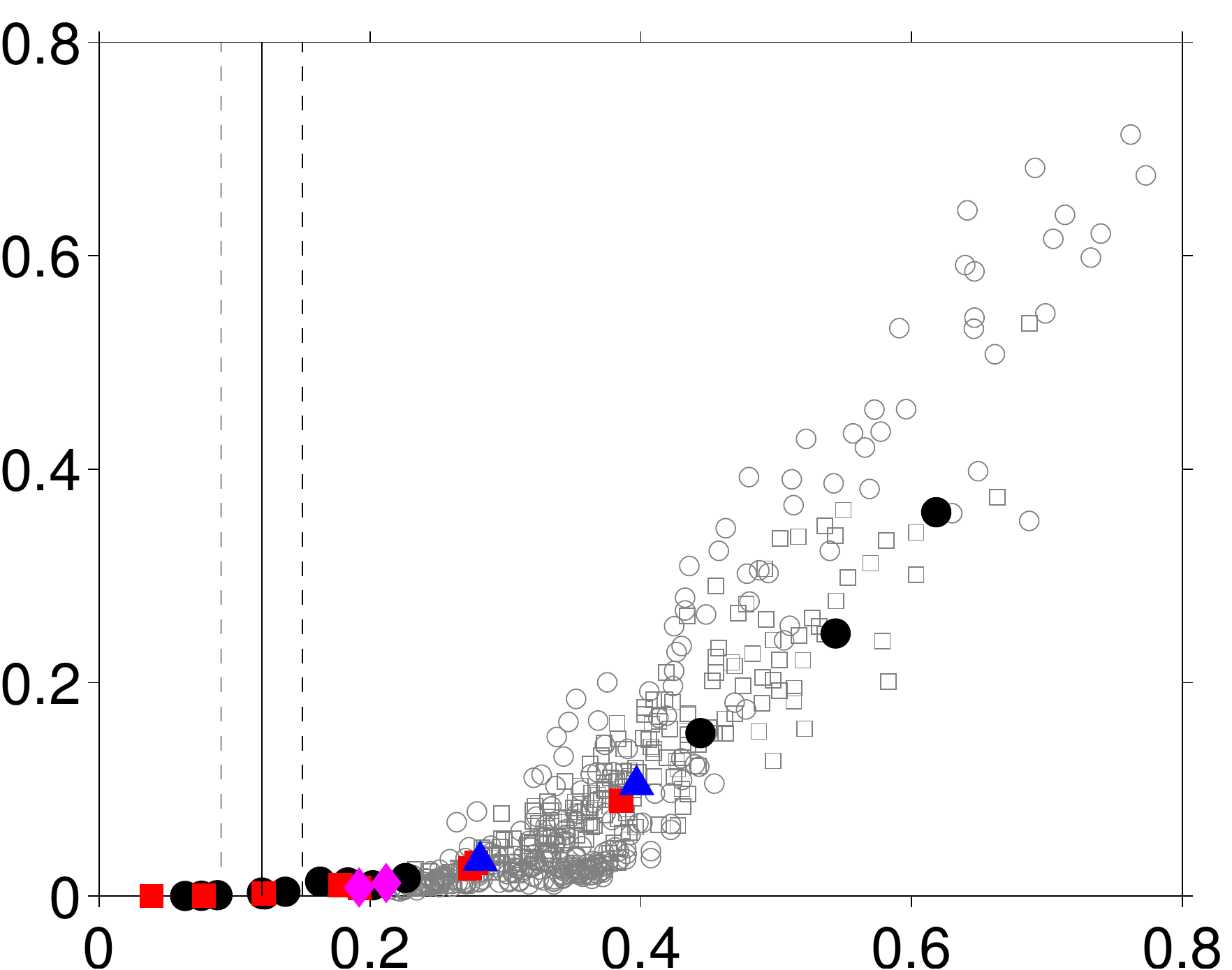}
          \begin{picture}(0,0)(-20,-85)
          \includegraphics[width=0.45\linewidth]
          {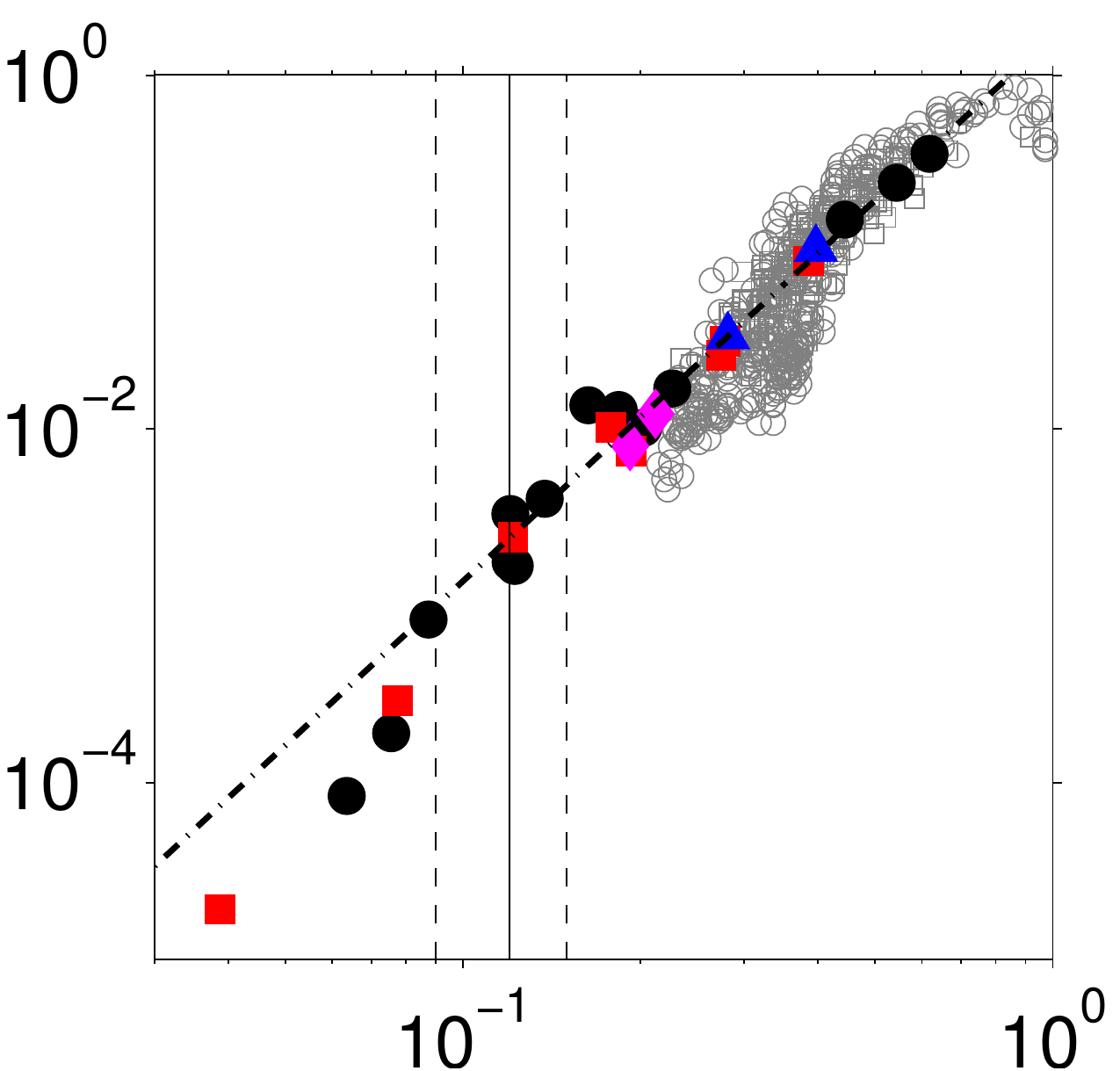}
          \end{picture}
          \centerline{\small \shields}
        \end{minipage}
        \hfill
        \begin{minipage}{2ex}
          \rotatebox{90}
          {\small $\qpmean/\qrefhf$}
        \end{minipage}
        \begin{minipage}{.45\linewidth}
          \centerline{\hspace*{8ex}$(b)$}
          \includegraphics[width=\linewidth]
          {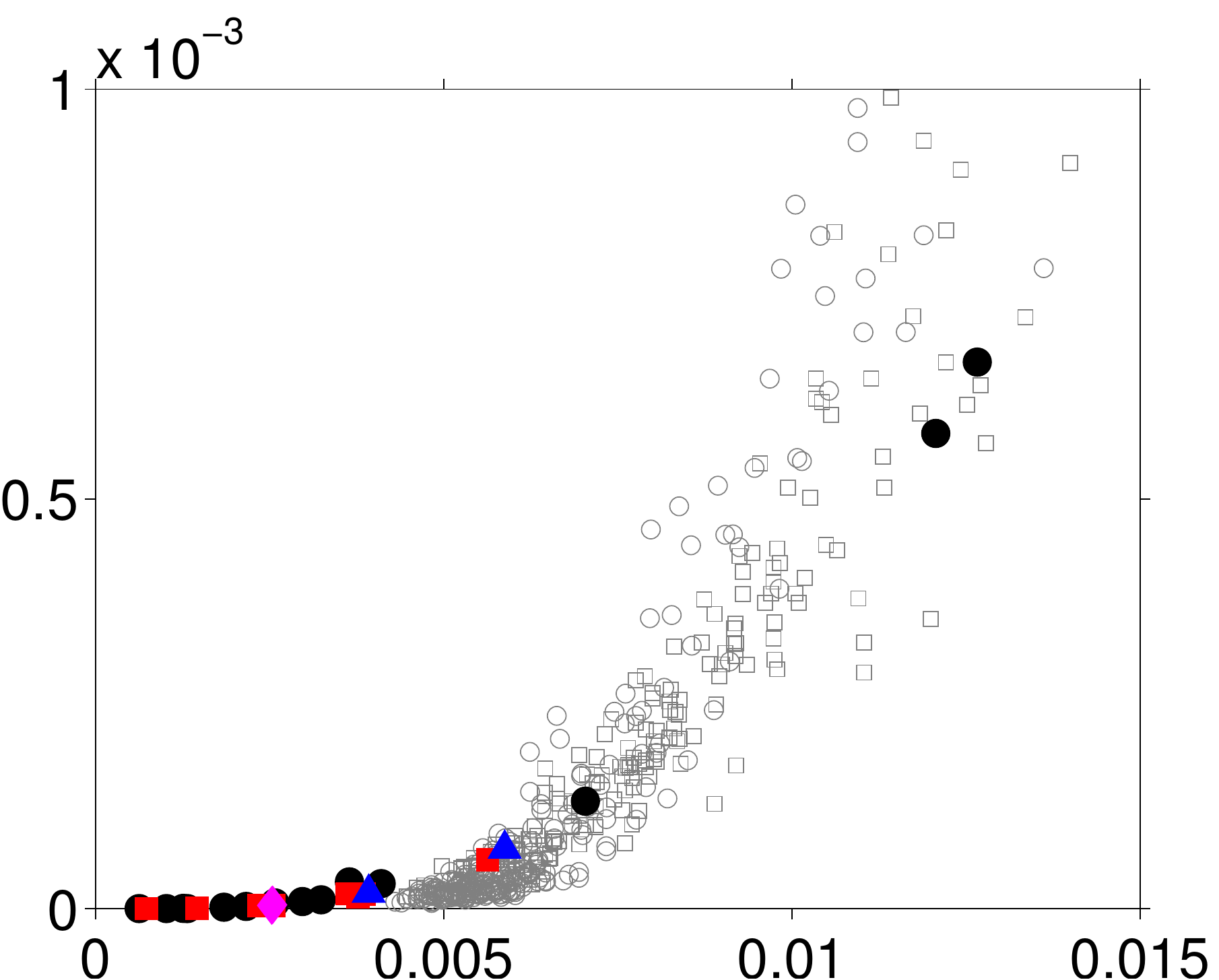}
          \begin{picture}(0,0)(-20,-85)
          \includegraphics[width=0.45\linewidth]
          {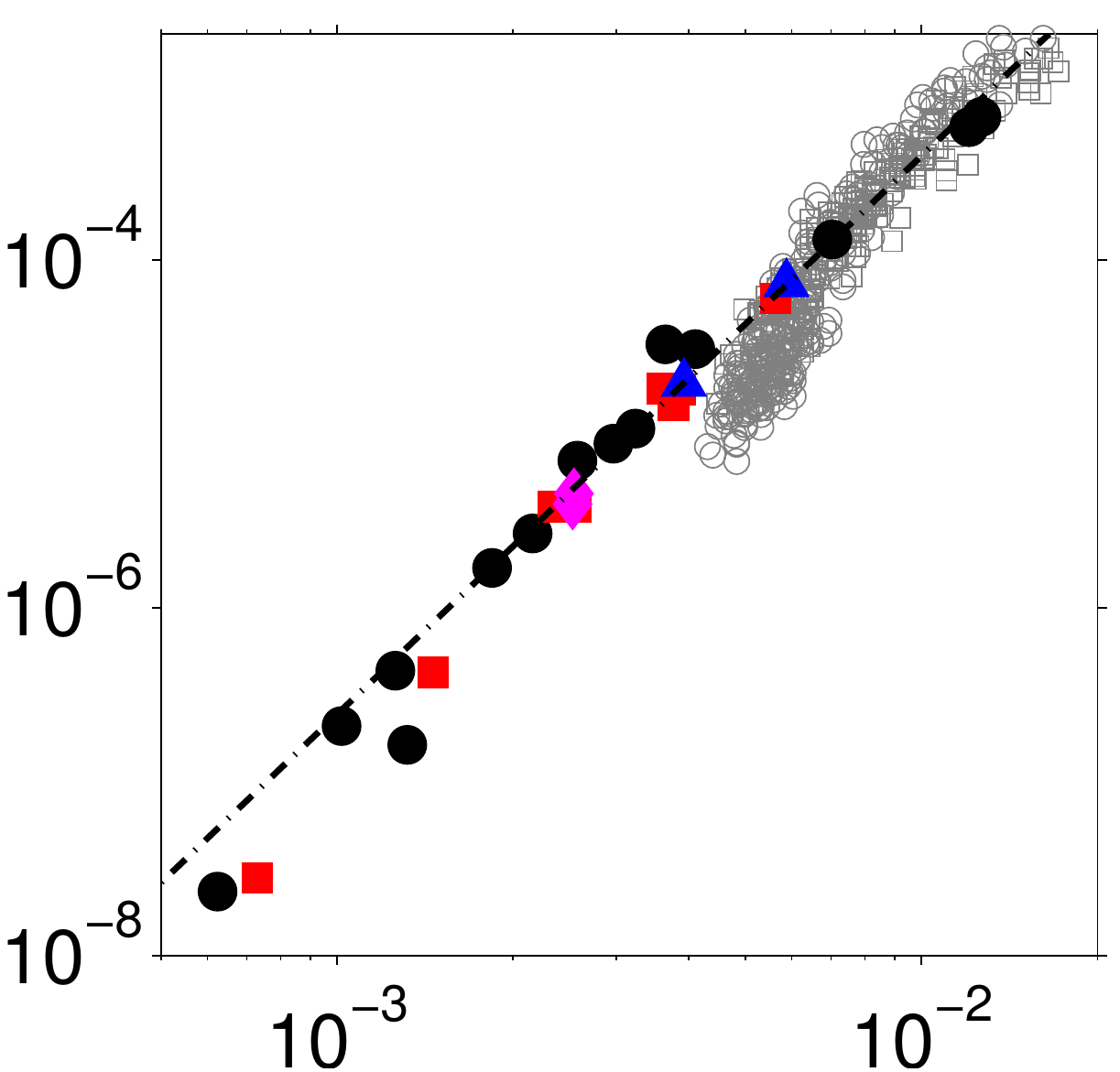}
          \end{picture}
          \centerline{\small \qfstar}
        \end{minipage}
      }{
        \caption{($a$) Mean particle flow rate 
          \qpmean\
          non-dimensionalized by the viscous scaling
          \qrefv\
          plotted as a function of the Shields number \shields.
          The vertical solid line  
          corresponds to the critical Shields number
          for particle erosion $\shields^c~=~0.12~\pm~0.03$
          reported by \citet{Ouriemi2007}, the dashed lines
          indicating the tolerance range. 
          The inset shows the same data plotted in logarithmic scale.
          The black chain-dotted line in the inset corresponds to a power law
          $\qpmean/\qrefv~=~1.6584\shields^{3.08}$. 
          %
          %
          ($b$) The same quantity \qpmean\ as in $(a)$, but 
          normalized by \qrefhf\ and plotted 
          as a function of the non-dimensional fluid flow rate 
          \qfstar. 
          The power law 
          $\qpmean/\qrefhf~=~1055.3(\qfstar)^{3.21}$ is
          indicated by the black chain-dotted line. 
          In both graphs the results from the present
          simulations are indicated by the symbols given in
          table~\ref{tab:numerical_parameters}. 
          Data points marked by (\opencircle,\opensquare) 
          correspond to the experiment of
          \citet{Aussillous2013} 
          (\opencircle, combination ``A'' of materials in their
          table~1; \opensquare, combination ``B''). 
          \protect\label{fig:particle_flux}
        }
        }
\end{figure}
In all the simulations performed in the present work the flow remains
laminar. 
Nevertheless, the individual particle motion is highly unsteady,
exhibiting 
sliding, rolling, reptation, and to a lesser degree saltation, within
and in the vicinity of the mobile layer of the sediment bed. 
It is only occasionally that particles are suspended in the flow over
longer times.  
%
%
Therefore, transport of particles as suspended load is insignificant in
the present configuration, and hereafter 
we will not further distinguish the particle flux due to the different
modes. 

\revision{%
  The volumetric flow rate of the particle phase (per unit span) is
  given by the following expression:
  \begin{equation}
    q_p(t) \;=\; \int_0^{\Ly} \upmean_{xz}(y,t)\phimean_{xz}(y,t){\rm d}y
    \;=\; \frac{\pi D^3}{6 \Lx\Lz}
    \sum_{l=1}^{N_p}u_p^{(l)}(t)
    \,,
    \label{instantaneous-particle-flux}
  \end{equation}
  where $u_p^{(l)}(t)$ is the streamwise component of the instantaneous 
  velocity of the $l$th particle at time $t$, and the operator
  $\langle \cdot \rangle_{xz}$ denotes an instantaneous average
  over the two homogeneous spatial directions (cf.\ appendix
  \ref{sec-ensemble-averaging}). 
  Note that the quantity $q_p$ is termed `grain flux' by
  \cite{Lobkovsky2008}.}{%
  The instantaneous volumetric flow rate of the particle phase (per
  unit span), $q_p(t)$, is given by the sum (over all particles) of the
  streamwise particle velocity times the particle volume,
  divided by the product of the streamwise and spanwise extent of the
  domain, viz.
  \begin{equation}
    q_p(t) 
    \;=\; \frac{\pi D^3}{6 \Lx\Lz}
    \sum_{l=1}^{N_p}u_p^{(l)}(t)
    \,.
    \label{instantaneous-particle-flux}
  \end{equation}
  In (\ref{instantaneous-particle-flux}) 
  $u_p^{(l)}(t)$ denotes the streamwise component of the velocity of
  the $l$th particle at time~$t$.
  Note that the quantity $q_p$ is termed `grain flux' by
  \cite{Lobkovsky2008} 
  who define it as the wall-normal integral over the product between
  the streamwise particle velocity and the solid volume fraction.}
%

%
The time evolution of $q_p$ for case BL08 is shown in figure
\ref{fig:particle_flux_evolution}.  
It can be seen that after initialization of the flow the particle flow
rate decreases from an
initially large value, tending towards a statistically stationary
state with a mean value \qpmean. 
The duration of the transient varies among the different simulations.
Since the focus of the present work is the fully developed  
regime, no attempt was made to find an appropriate scaling for the
transient time. 

Figure~\ref{fig:particle_flux}($a$) shows the mean particle flow rate 
in the statistically stationary regime, 
\qpmean, plotted for all cases as a function of the Shields number
\shields. 
The experimentally determined critical value for the Shields number in
the laminar flow regime, $\shieldscrit=0.12$, together with the
confidence interval $\pm 0.03$ \citep{Ouriemi2007} are indicated
with vertical lines in figure~\ref{fig:particle_flux}($a$). 
One possible reference quantity for the 
particle flow rate is 
the following viscous scale as suggested by \citet{Aussillous2013},  
\begin{equation}
  \qrefv=\frac{(\dratio-1)gD^3}{\nu} = \Ga^2\,\nu
  \,,
  \label{eq:qrefvis}
\end{equation}
which is equivalent to $18\usettling D$, where \usettling\ 
is the Stokes settling velocity of a single particle. 
The quantity defined in (\ref{eq:qrefvis}) is used to
non-dimensionalize \qpmean\ 
in figure~\ref{fig:particle_flux}($a$), while an alternative
normalization will be explored in figure~\ref{fig:particle_flux}($b$).  
It is clearly observable from figure~\ref{fig:particle_flux}($a$) that
the particle flow rate obtained from the present DNS-DEM is negligibly
small for values of $\shields\lesssim0.12$ in good agreement with what
is reported in the literature 
\citep[cf.][]{Charru2004,Loiseleux2005a,Ouriemi2009}. 
Animations of the particle motion (as provided in the 
supplementary material, cf.\ appendix~\ref{sec-suppl-mat}) 
confirm this finding.
%
%
Those supplementary movies also show the well-known 
intermittent erratic movement of a small number of particles for 
Shields number values smaller 
than \shieldscrit, as has been observed experimentally 
\citep[][]{Charru2004}.
On the other hand, in those runs in which the value of the Shields
number is above the critical value, the particles
near the fluid-bed interface are observed to be 
continuously in motion. 
The magnitude of the saturated particle flux \qpmean\ monotonically
increases with increasing value of \shields. This trend is clearly
seen in figure~\ref{fig:particle_flux}($a$). The corresponding 
experimental data of \citet{Aussillous2013}\footnote{%
The particle flow rate defined in (\ref{instantaneous-particle-flux})
is not presented in the paper by \citet{Aussillous2013}. We have
computed it from the data provided by the authors as supplementary
material.}  
is also added to the 
figure, exhibiting a very good agreement with our present DNS-DEM
results. 
Moreover, a fit of the simulation data in
figure~\ref{fig:particle_flux}($a$) for super-critical parameter
points ($\shields > \shieldscrit$) yields the following power law
relation: $\qpmean/\qrefv \propto \shields^{3.08}$. 
\cite{Ouriemi2009} have described the present bedload transport system
by means of 
a two-fluid model which makes use of the Einstein effective
viscosity. They have shown that above the critical value \shieldscrit\
the model predicts a variation of the particle flow rate with the
third power of the Shields number. As can be seen from the fit in 
figure~\ref{fig:particle_flux}($a$), their conclusion is fully
confirmed by the present data.
An alternative scaling of the particle flow rate data is to replace
the particle diameter in (\ref{eq:qrefvis}) with the fluid height, viz.
\begin{equation}
  \qrefhf=\frac{(\dratio-1)g\hfluid^3}{\nu}
  =
  \qrefv\,\left(\frac{\hfluid}{D}\right)^3
  \;.
  \label{eq:qrefhf}
\end{equation}
This scaling was advocated by \citet{Aussillous2013} as more 
consistent with a continuum (two-fluid) approach. 
Figure~\ref{fig:particle_flux}$(b)$ shows that our data for the
particle flow rate as a function of the fluid flow rate, both
normalized with the scale defined in equation (\ref{eq:qrefhf}), 
again matches the experimental data of \citet{Aussillous2013} very
well. 
We also note that the simulation data is very well represented by the
power law 
$\qpmean/\qrefhf~\propto~(\qfstar)^{3.21}$ which is again close to a
cubic variation. In fact it can be shown that a cubic variation of
$\qpmean/\qrefhf$ with $\qfstar$ follows directly from a cubic
variation of $\qpmean/\qrefv$ with $\shields$. 
%
%
%
\begin{figure}
  \figpap{
    %
    \begin{minipage}{2ex}
      \rotatebox{90}
      {\small $\qpmean/\qrefv$}
    \end{minipage}
    \begin{minipage}{.45\linewidth}
      \centerline{$(a)$}
      \includegraphics[width=\linewidth]
      {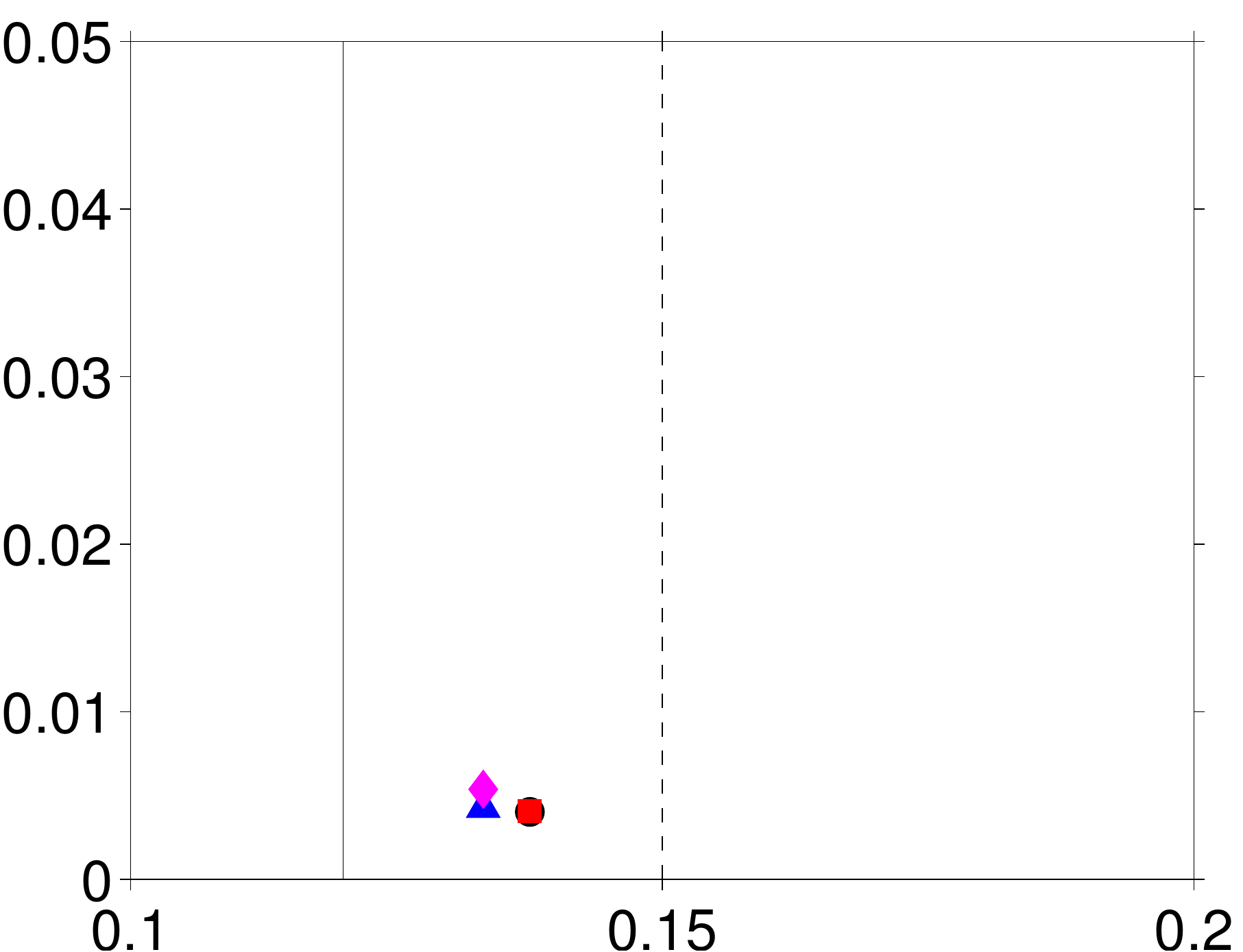}
      \\
      \centerline{\small \shields}
    \end{minipage}        
    \hfill
    \begin{minipage}{2ex}
      \rotatebox{90}
      {\small $\qpmean/\qrefv$}
    \end{minipage}
    \begin{minipage}{.45\linewidth}
      \centerline{$(a)$}
      \includegraphics[width=\linewidth]
      {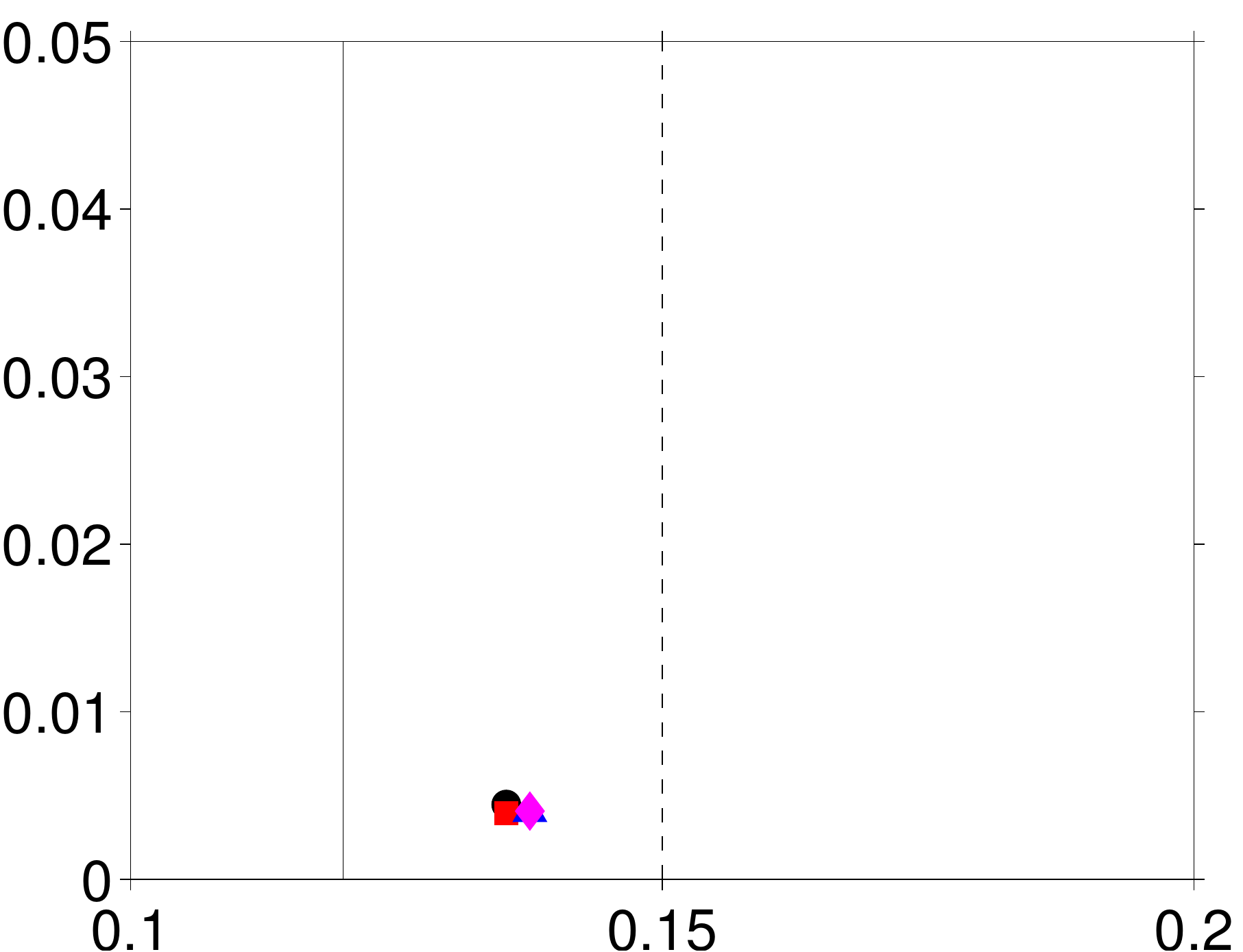}
      \\
      \centerline{\small \shields}
    \end{minipage}        
  }{
    \caption{%
      \revision{}{%
        As figure~\ref{fig:particle_flux}$(a)$, but 
        showing  the effect of changing the parameters of the
        collision model in case BL10 with $Re_b=266$ and $Ga=13.98$
        (cf.\ table~\ref{tab:physical_parameters}). 
        In $(a)$ the dry restitution coefficient
        $\varepsilon_d$ is varied while all remaining physical and
        numerical parameters values are maintained:  
        %
        {\color{black}$\bullet$}, $\varepsilon_d=0.30$; 
        {\color{red}\small$\blacksquare$}, $\varepsilon_d=0.60$; 
        {\color{blue}$\blacktriangle$}, $\varepsilon_d=0.90$; 
        {\color{magenta}\small$\blacklozenge$}, $\varepsilon_d=0.97$. 
        In $(b)$ only the Coulomb friction coefficient $\mu_c$  is varied: 
        {\color{black}$\bullet$}, $\mu_c=0.1$; 
        {\color{red}\small$\blacksquare$}, $\mu_c=0.25$; 
        {\color{blue}$\blacktriangle$}, $\mu_c=0.4$; 
        {\color{magenta}\small$\blacklozenge$}, $\mu_c=0.55$. 
        Note that the present axis scales are different than in
        figure~\ref{fig:particle_flux}$(a)$. 
      }
      \protect\label{fig:particle_flux_epsilon_dry_sensitivity}
    }
  }
\end{figure}
\revision{}{%
  \subsubsection{Sensitivity with respect to collision model parameters}
  \label{subsec:bedload-sensitivity}
  The force range $\Delta_c$ has been varied by a factor of two (from
  $\Delta_c=\Delta x$ to $\Delta_c=2\Delta x$) in
  two simulation cases (BL20 and BL21, where $Re_b=267$ and $333$,
  respectively, and $Ga=6.3$ in both cases, cf.\
  tables~\ref{tab:physical_parameters} and
  \ref{tab:numerical_parameters}). 
  It can be seen from figure~\ref{fig:particle_flux} that the effect
  of modifying this parameter upon the particle flow rate is
  insignificant in the given range. The same conclusion holds for the
  quantities discussed below
  (\S~\ref{subsec:thickness-of-mobile-bed-layer}-\ref{subsec:mean-fluid-and-particle-velocities}). 

  We have likewise tested the influence of the choice of the value for
  the dry restitution coefficient $\efferest_d$ upon the results in
  the present configuration. For this purpose the flow case denoted
  BL10 in table~\ref{tab:physical_parameters} has been repeated three
  times with modified values of the dry restitution coefficient (while
  maintaining all remaining numerical and physical parameters at their
  original
  value). Figure~\ref{fig:particle_flux_epsilon_dry_sensitivity}$(a)$
  shows the resulting mean particle flow rate for the set of 
  coefficients $\efferest_d=0.3$, $0.6$, $0.9$, $0.97$. As can be
  observed, the impact of this parameter variation is very limited.

  Finally, we have repeated the same case BL10 while independently
  changing the value of the Coulomb friction coefficient  in
  the range $\mu_c=0.1\ldots0.55$. The result is shown in
  figure~\ref{fig:particle_flux_epsilon_dry_sensitivity}$(b)$, where
  again a very small effect upon the mean particle flow rate is
  obtained. 
}
%
\begin{figure}
  \figpap{
        %
        \begin{minipage}{2ex}
          \rotatebox{90}
          {\small $\hmobile/D$}
        \end{minipage}
        \begin{minipage}{.45\linewidth}
          \centerline{\hspace*{8ex}$(a)$}
          \includegraphics[width=\linewidth]
          {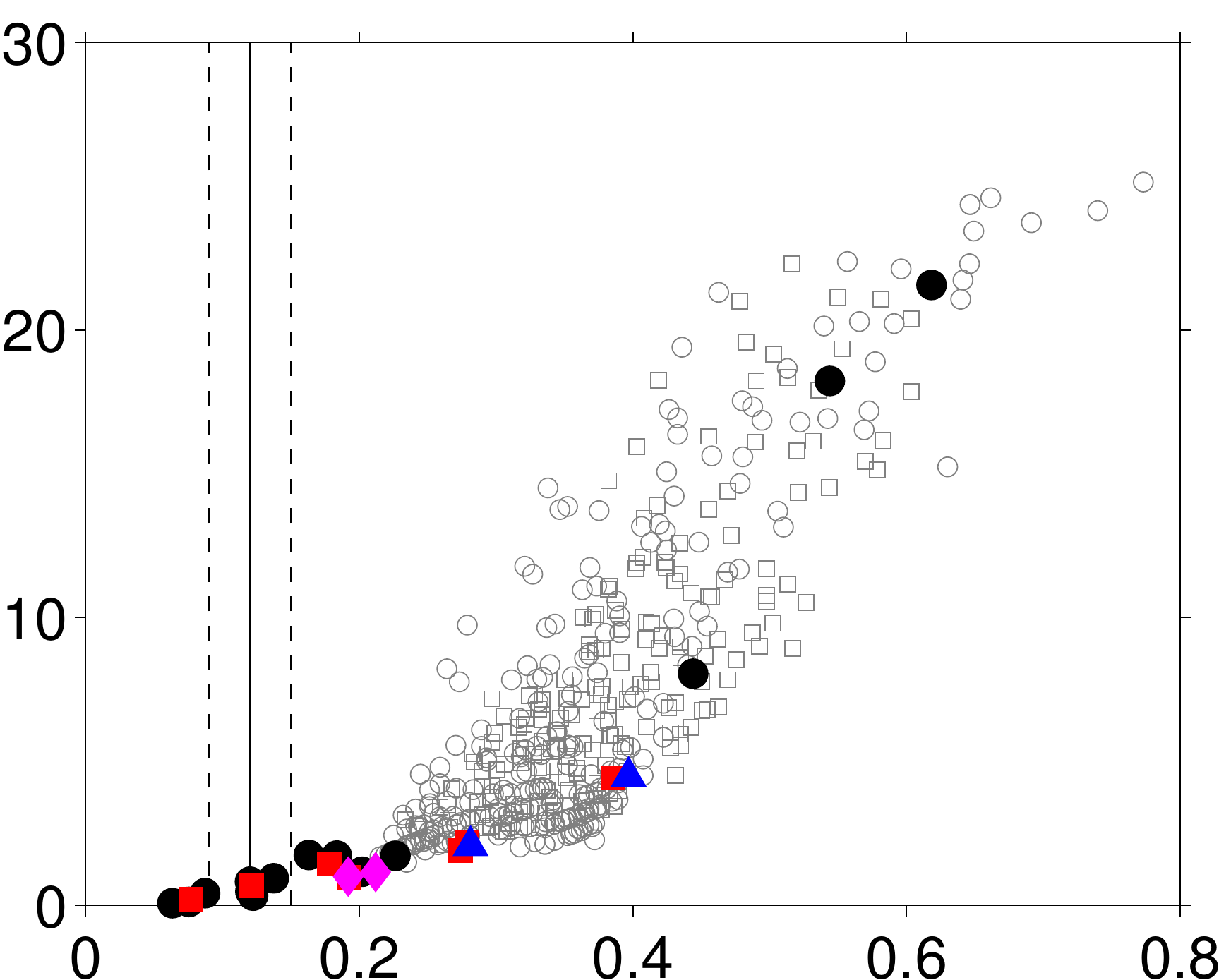}
          \begin{picture}(0,0)(-15,-95)
          \includegraphics[width=0.45\linewidth]
          {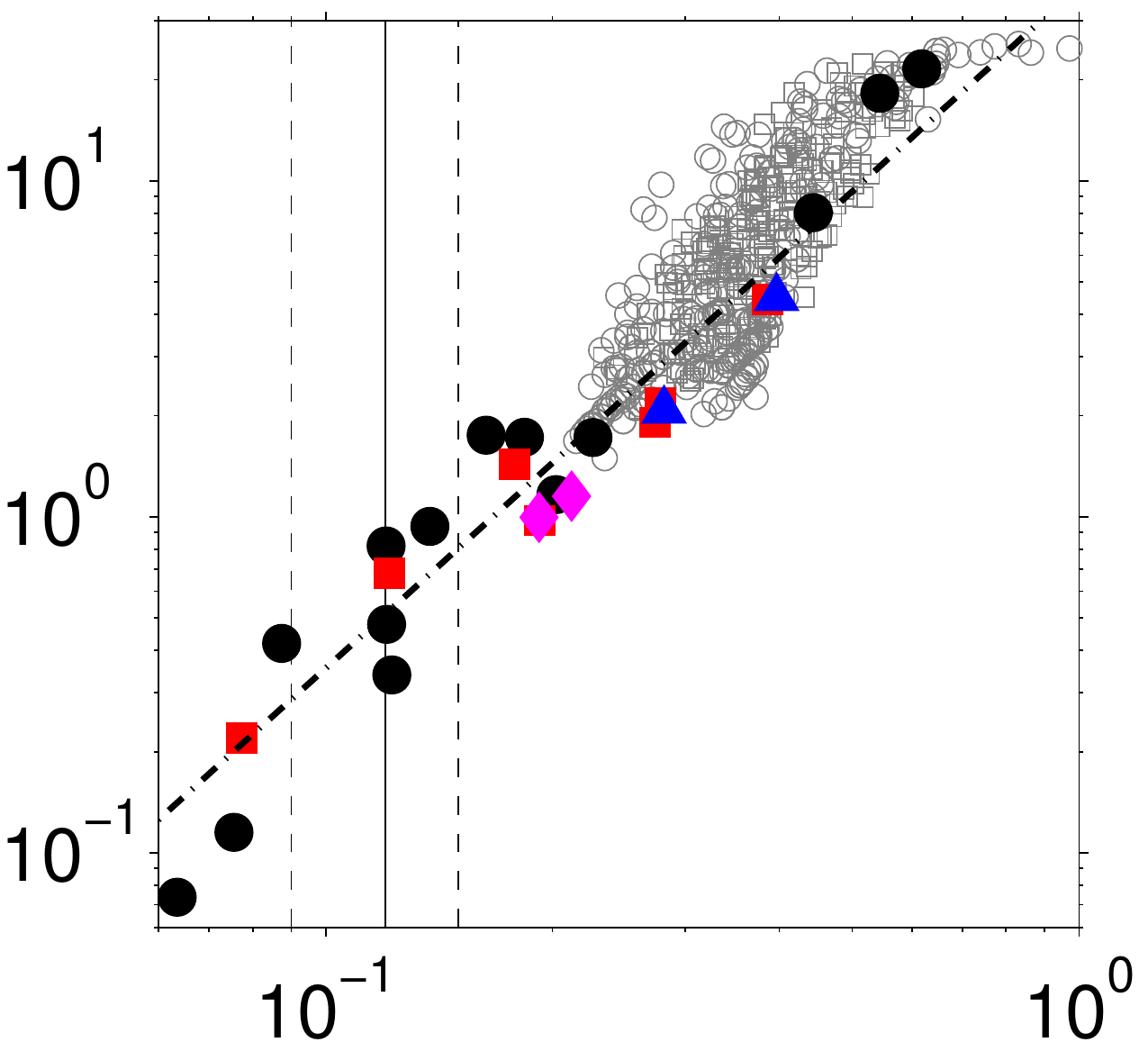}
          \end{picture}
          \centerline{\small \shields}
        \end{minipage}
        \hfill
        \begin{minipage}{2ex}
          \rotatebox{90}
          {\small $\hmobile/\hfluid$}
        \end{minipage}
        \begin{minipage}{.45\linewidth}
          \centerline{\hspace*{8ex}$(b)$}
          \includegraphics[width=\linewidth]
          {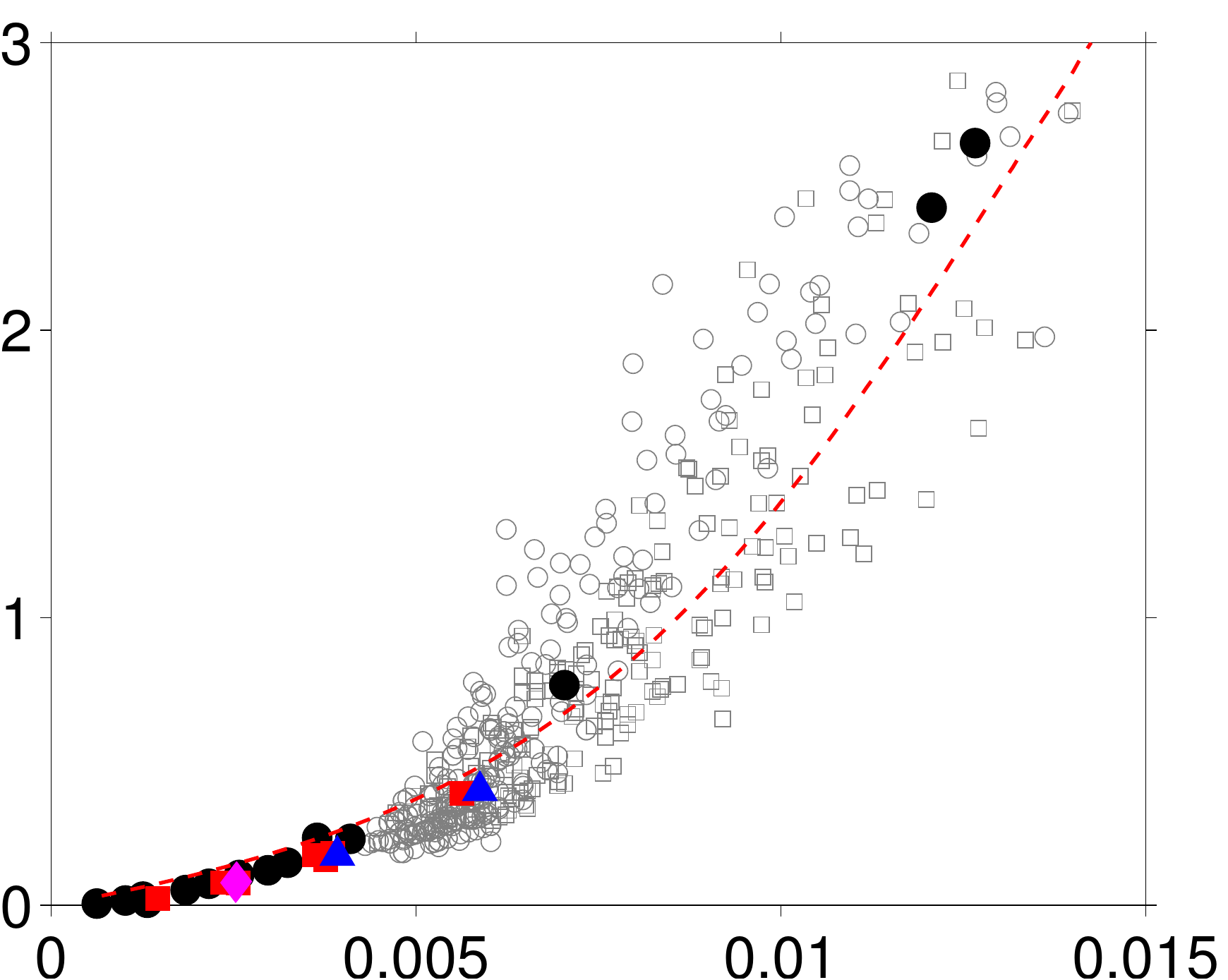}
          \begin{picture}(0,0)(-10,-90)
          \includegraphics[width=0.45\linewidth]
          {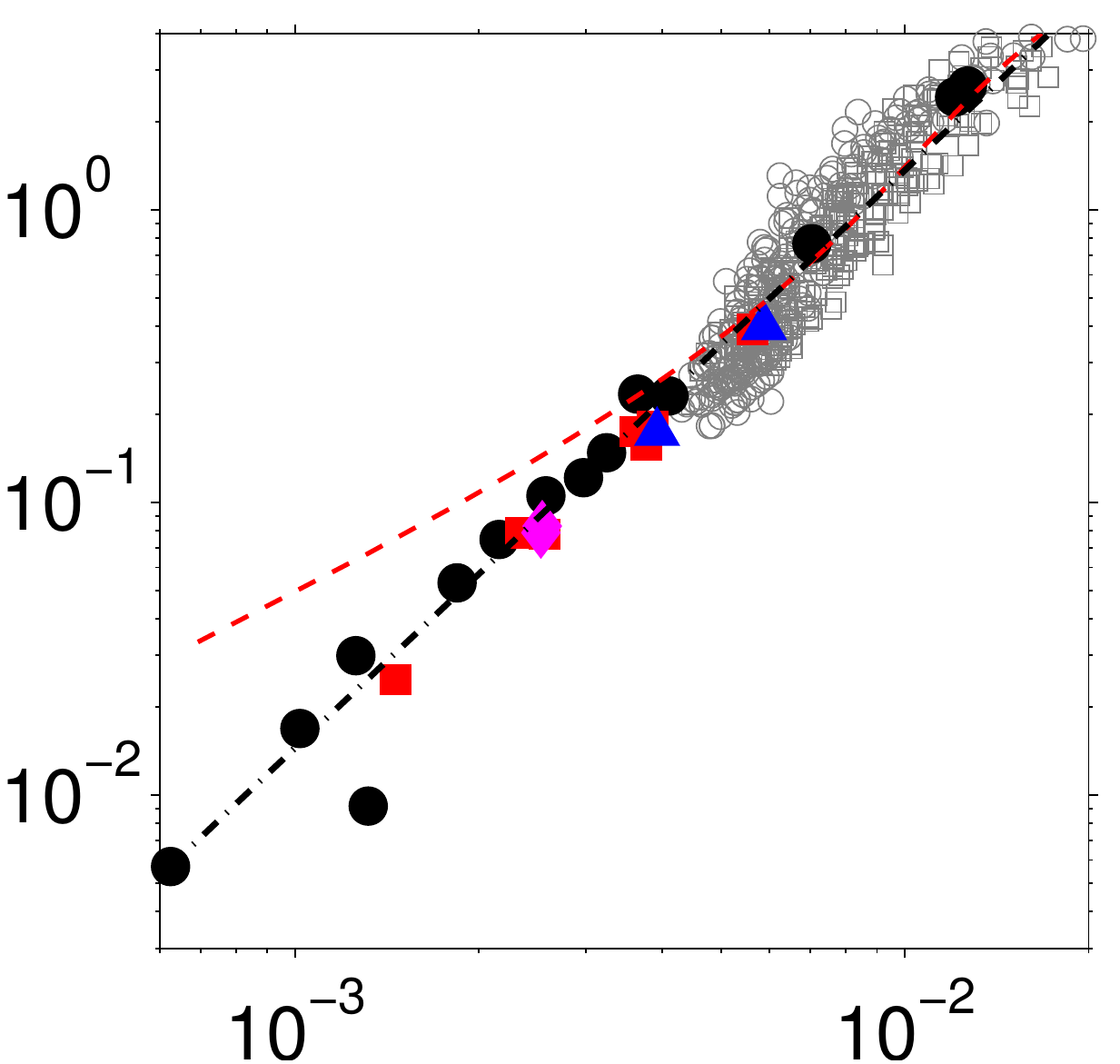}
          \end{picture}
          \centerline{\small  \qfstar}
        \end{minipage}
  }{
        \caption{%
          \revision{
            Thickness of the mobile bed layer \hmobile: 
            ($a$) normalized by the particle diameter $D$ and
            given as a function of the Shields number \shields;  
            ($b$) normalized by 
            the fluid height \hfluid\ and given  
            as a function of the non-dimensional fluid flux
            \qfstar. 
            The insets show the same data in logarithmic scale.
            The chain-dotted lines indicate the following data
            fits: 
            $(a)$ $\hmobile/D = 38.01\,\shields^{2.03}$; 
            $(b)$ $\hmobile/\hfluid = 12096\,(\qfstar)^{1.97}$. 
            Data points marked by (\opencircle,\opensquare) 
            correspond to the experiment of \citet{Aussillous2013}.}{%
            Thickness of the mobile 
            bed layer \hmobile: 
            ($a$) normalized by the particle diameter $D$ and
            given as a function of the Shields number \shields;  
            ($b$) normalized by 
            the fluid height \hfluid\ and given  
            as a function of the non-dimensional fluid flux
            \qfstar. 
            The insets show the same data in logarithmic scale.
            The chain-dotted lines indicate the following data
            fits: 
            $(a)$ $\hmobile/D = 38.01\,\shields^{2.03}$; 
            $(b)$ $\hmobile/\hfluid = 12096\,(\qfstar)^{1.97}$. 
            Data points marked by (\opencircle,\opensquare) 
            correspond to the experiment of \citet{Aussillous2013}.
            %
            In $(b)$ the dashed red line indicates the results
            obtained with a two-fluid model including a shear-rate
            dependent friction coefficient 
            \citep[black dashed line in figure~6$b$
            of][]{Aussillous2013}.  
          }
                \protect\label{fig:thickness-of-mobile-layer}
              }
            }
\end{figure}
\subsubsection{Thickness of the mobile sediment bed}
\label{subsec:thickness-of-mobile-bed-layer}
In practical applications involving bedload transport it is often of
interest to predict the extent of the layer of particles which
exhibits significant streamwise motion. 
Referring to the schematic of the configuration shown in
figure~\ref{fig:bedload-schematic-diagram}, we define the 
thickness of the mobile layer \hmobile\ as the distance between the
fluid-bed interface location ($\yref$) and the location inside the bed
where the mean particle velocity \upmean\ is equal to a prescribed
threshold value $\upmean^{thresh}$. 
In the present work we set this threshold velocity to 
$\upmean^{thresh}=0.005\,\max\ufmean$. 
In terms of the Stokes settling velocity, the chosen value ranges  
from $0.01\usettling$ to $0.09\usettling$ in the different cases which
we have simulated.  
In the study of \cite{Aussillous2013} the mobile layer thickness
\hmobile\ has been determined by a similar thresholding criterion. 

Figure~\ref{fig:thickness-of-mobile-layer}($a$)
shows the computed mobile layer thickness \hmobile\ normalized by the
diameter of the particles as a function of the Shields number; 
figure~\ref{fig:thickness-of-mobile-layer}($b$) shows the same
quantity under the alternative normalization with the fluid height, 
plotted as a function of the non-dimensional fluid flow rate.  
Once again, the DNS data shows very good agreement
with the experimental data despite the sensitivity of this quantity
due to thresholding. 
The data in figure~\ref{fig:thickness-of-mobile-layer} clearly shows
that \hmobile\ monotonically increases with increasing
values of the Shields number and with  the non-dimensional fluid flow rate.
It can be observed that the DNS-DEM data  over the presently investigated
parameter range is fairly well represented by a quadratic law in both
graphs. This result is in contrast to 
\revision{}{some of} 
the available models in the literature. 
%
\cite{Mouilleron2009} have considered the viscous re-suspension
theory of \citet{Leighton1986} in which particles are assumed to
have no inertia and a mass balance between downward 
sedimentation and upward diffusion is considered.
\citet{Ouriemi2009} on the other hand consider a 
continuum description of the problem where
a frictional rheology 
\revision{}{(using a Coulomb model with a constant friction coefficient)} 
is assumed to describe the mobile granular layer. 
In both theoretical approaches \citep{Mouilleron2009,Ouriemi2009} the
authors arrive at a linear variation of the thickness of the mobile
layer with the Shields number.
\revision{}{More recently, \cite{Aussillous2013} have essentially
  revisited the two-fluid modelling approach of \citet{Ouriemi2009},
  but employing more sophisticated closures for the stress tensor of
  the particle phase. In particular, they use a granular frictional
  rheology with a shear-rate dependent friction coefficient. These
  authors' results for the mobile layer thickness obtained with this
  continuum approach are included in
  figure~\ref{fig:thickness-of-mobile-layer}$(b)$. It can be seen that
  the match with our data is very reasonable at larger values of the
  fluid flux $q_f$. For smaller values of $q_f$ the continuum model
  results clearly deviate from the presently observed quadratic
  behavior. This, however, is not much of a surprise since the
  thickness of the mobile layer is very small in that range, and the
  continuum approach might not be appropriate.
}
Please note, however, that the definition of the mobile layer thickness
in the continuum model context is not identical to the definition
which is employed in both the present work and in the experiment
\citep{Aussillous2013}.   

Going back to the present simulation data, it can be seen that at
larger values of the two alternative control parameters our
results are in fact not inconsistent with a 
linear variation of the mobile layer thickness with both Shields
number (figure~\ref{fig:thickness-of-mobile-layer}$a$) and with the
fluid flow rate (figure~\ref{fig:thickness-of-mobile-layer}$b$). 
However, in view of the fact that the data points (both in the
experiment and in the simulation) cover only a limited range of these
control parameters, and considering the scatter of the experimental
data, this issue cannot be settled at the present time. 
%
\begin{figure}
  \figpap{
        %
        \begin{minipage}{2ex}
          \rotatebox{90}
          {\small $(y-\yref)/\hfluid$}
        \end{minipage}
        \begin{minipage}{.45\linewidth}
          \centerline{$(a)$}
          \includegraphics[width=\linewidth]
          {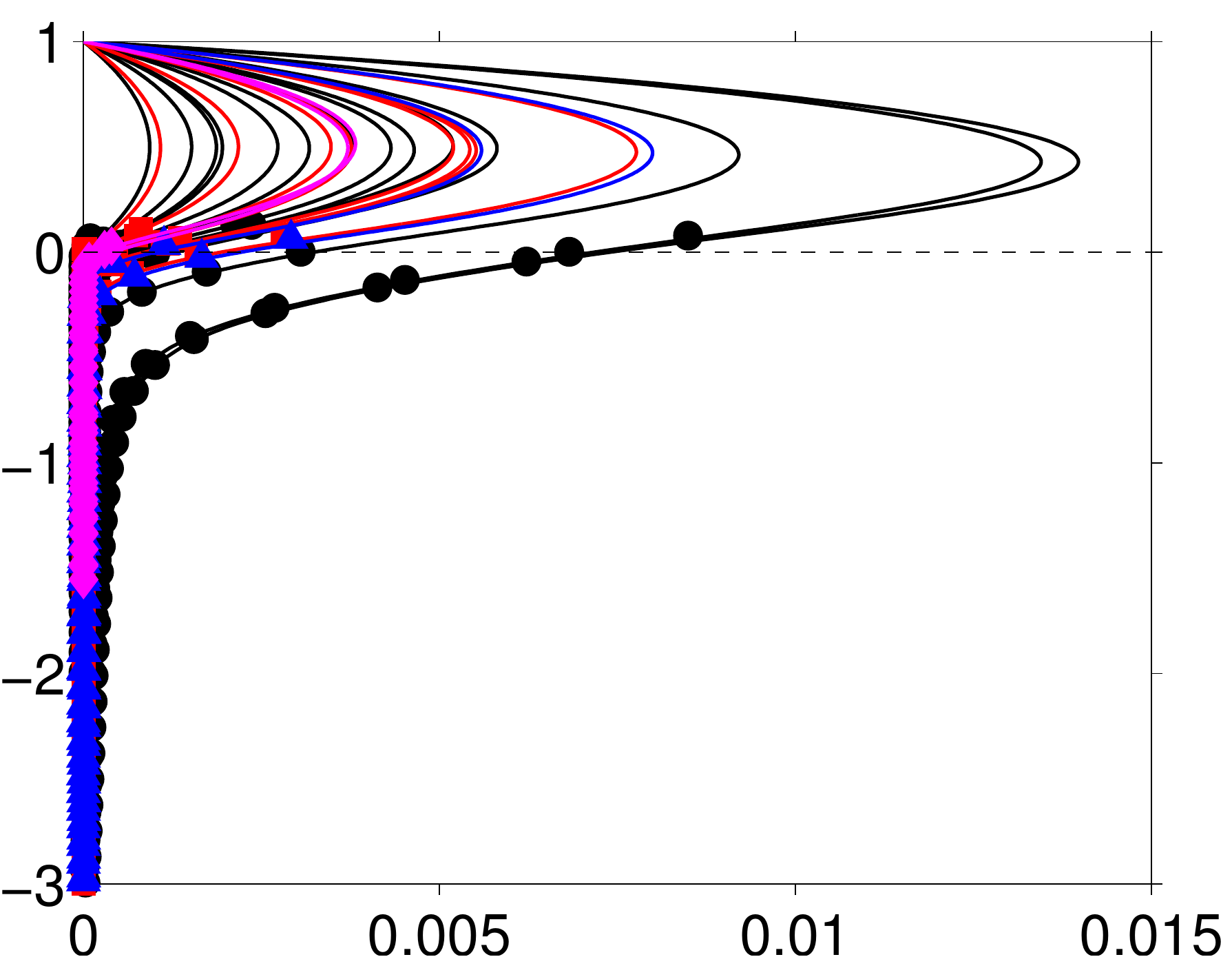}
          \centerline{\small $[\ufmean,\upmean]\hfluid/\qrefhf$}
        \end{minipage}
        \hfill
        \begin{minipage}{2ex}
          \rotatebox{90}
          {\small $(y-\yref+\hmobile)/\hfluid$}
        \end{minipage}
        \begin{minipage}{.45\linewidth}
          \centerline{$(b)$}
          \includegraphics[width=\linewidth]
          {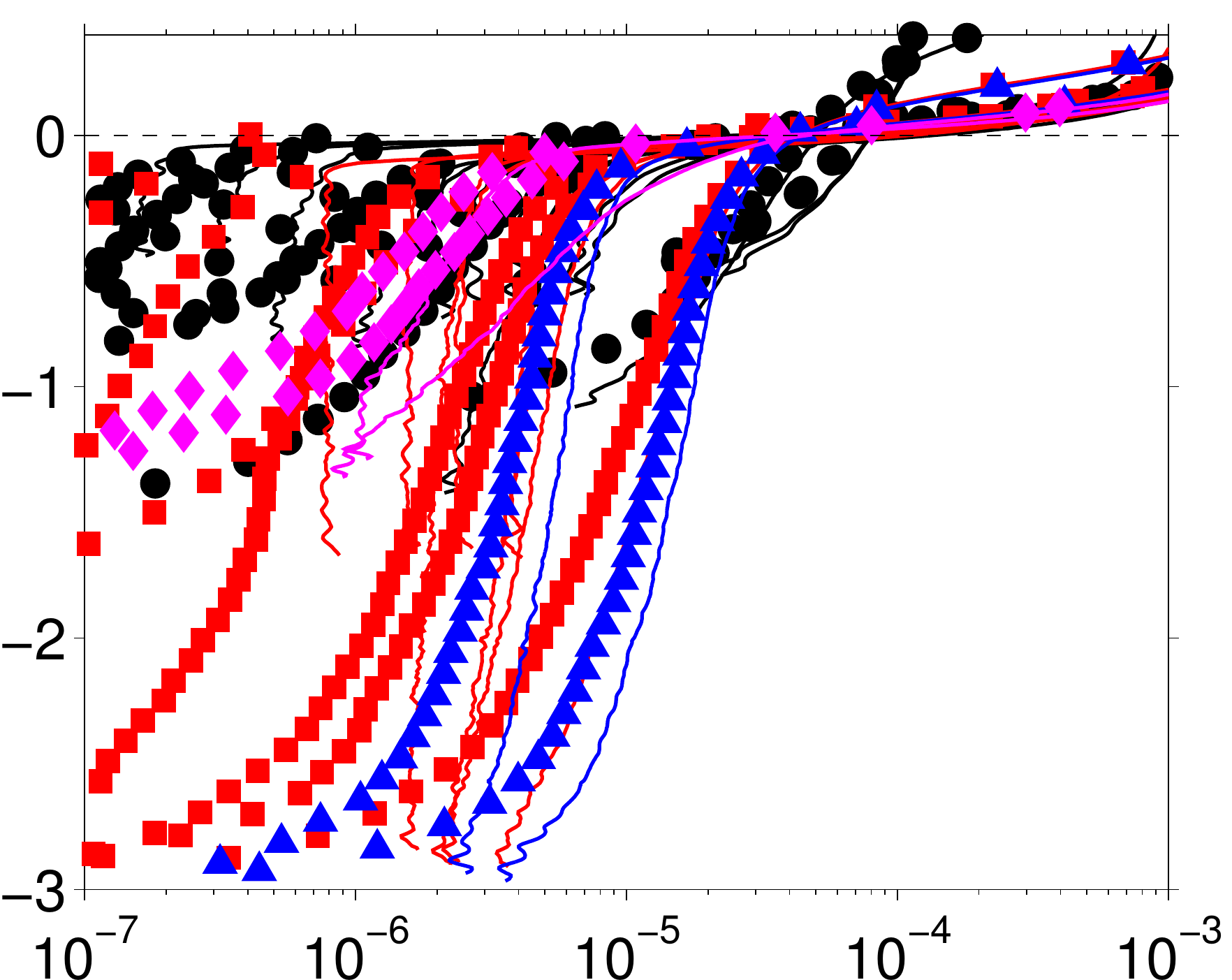}
          \centerline{\small $[\ufmean,\upmean]\hfluid/\qrefhf$}
        \end{minipage}
  }{
        \caption{%
         Wall-normal profiles of the streamwise component of the
         mean fluid velocity \ufmean\ (solid lines) 
         and of the mean particle velocity \upmean\ (symbols); color
         and symbol coding as indicated in
         table~\ref{tab:numerical_parameters}. 
         %
         In $(a)$ the ordinate is shifted to the fluid-bed interface
         location, while in $(b)$ it is shifted to the lower boundary
         of the mobile layer. 
         \protect\label{fig:mean-fluid-and-particle-velocity}
       }
     }
\end{figure}
\subsubsection{Fluid and particle velocities}
\label{subsec:mean-fluid-and-particle-velocities}
The wall-normal profiles of the streamwise component of the 
mean fluid and particle velocities \ufmean\ and \upmean\ for all cases
are shown in figure~\ref{fig:mean-fluid-and-particle-velocity}.  
In this figure 
the length and velocity scales proposed by \citet{Aussillous2013}  
(\hfluid\ and $\qrefhf/\hfluid$, respectively) are used. 
The graph in figure~\ref{fig:mean-fluid-and-particle-velocity}$(a)$ 
features an ordinate which is shifted to the fluid-bed interface
location ($\yref$), while in
figure~\ref{fig:mean-fluid-and-particle-velocity}$(b)$ the location of
the bottom of the mobile layer ($\yref-\hmobile$) is used as the zero
of the ordinate. 
%
%
Similar to what has been observed in the experiments
\citep{Aussillous2013}, the present profiles exhibit three distinct
regions:  
(I) the clear fluid region ($0<y-\yref<\hfluid$), where the 
mean fluid velocity profile is characterized by a near-parabolic
shape, while the solid volume fraction is negligibly small; 
(II) the mobile granular layer ($-\hmobile<y-\yref<0$), where
both the fluid and particles are in motion;
(III) the bottom region ($y < \yref-\hmobile$), where the 
velocities of both phases are vanishingly small. 
It can be observed from
figure~\ref{fig:mean-fluid-and-particle-velocity} that there exists no
significant difference between the mean velocities of the two
phases. 
Note that in the clear fluid region (I) particles are occasionally 
entrained into the bulk flow, reaching larger distances above the
mobile layer. 
However, the mean particle velocity at these points
was not determined with sufficient statistical accuracy, and these
values are, therefore, not shown in
figure~\ref{fig:mean-fluid-and-particle-velocity}.  
%
\begin{figure}
  \figpap{
   \begin{center}
         \begin{minipage}{2ex}
          \rotatebox{90}
          {\small $\tilde{y}/\hufmax$}
        \end{minipage}
        \begin{minipage}{.45\linewidth}
          \centerline{$(a)$}
          \includegraphics[width=\linewidth]
          {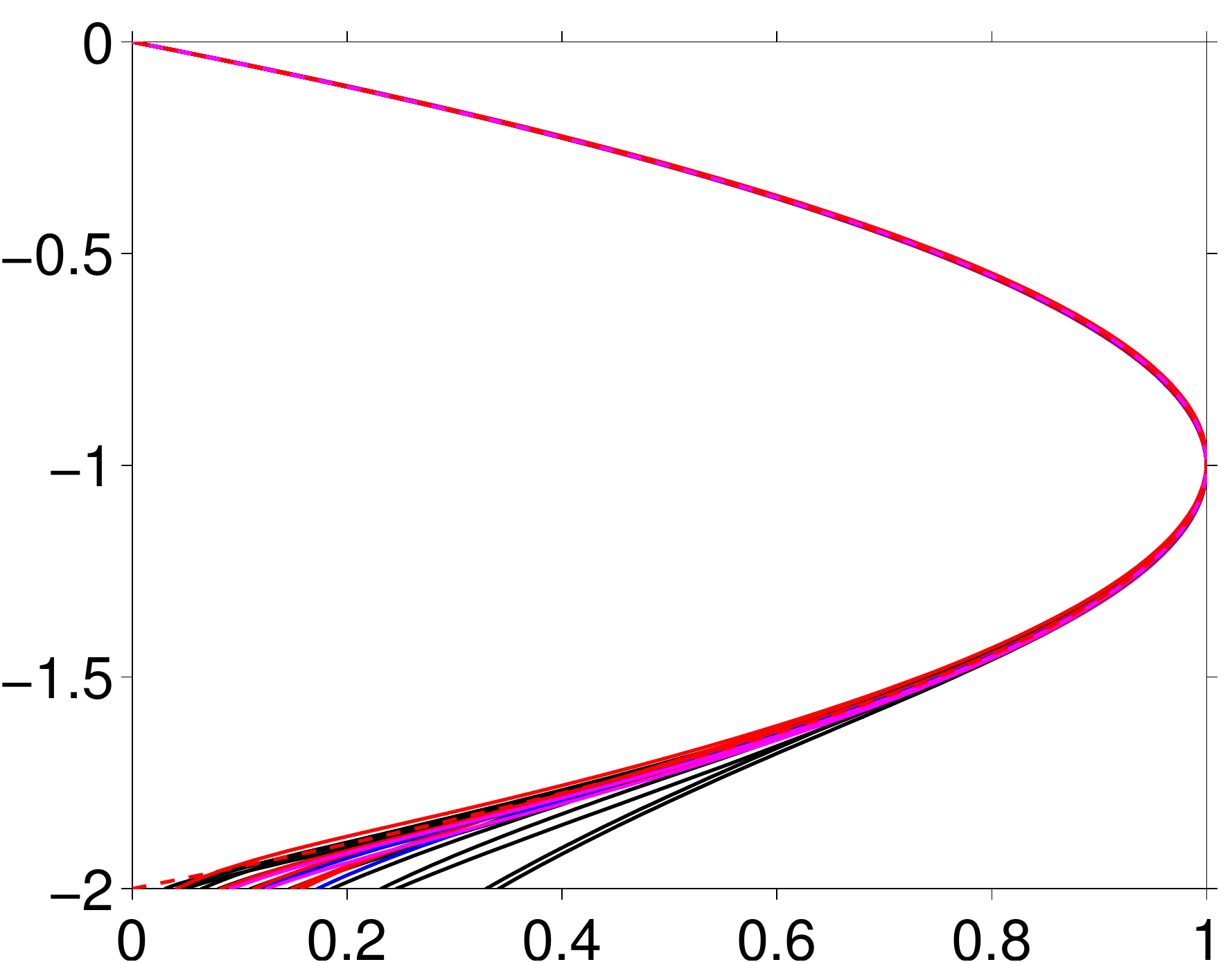}
          \centerline{\small  $\ufmean/\ufmax$}
        \end{minipage}
      \end{center}
         \begin{minipage}{2ex}
          \rotatebox{90}
          {\small $\ufmax\hfluid/\qrefhf$}
        \end{minipage}
        \begin{minipage}{.45\linewidth}
          \centerline{$(b)$}
          \includegraphics[width=\linewidth]
          {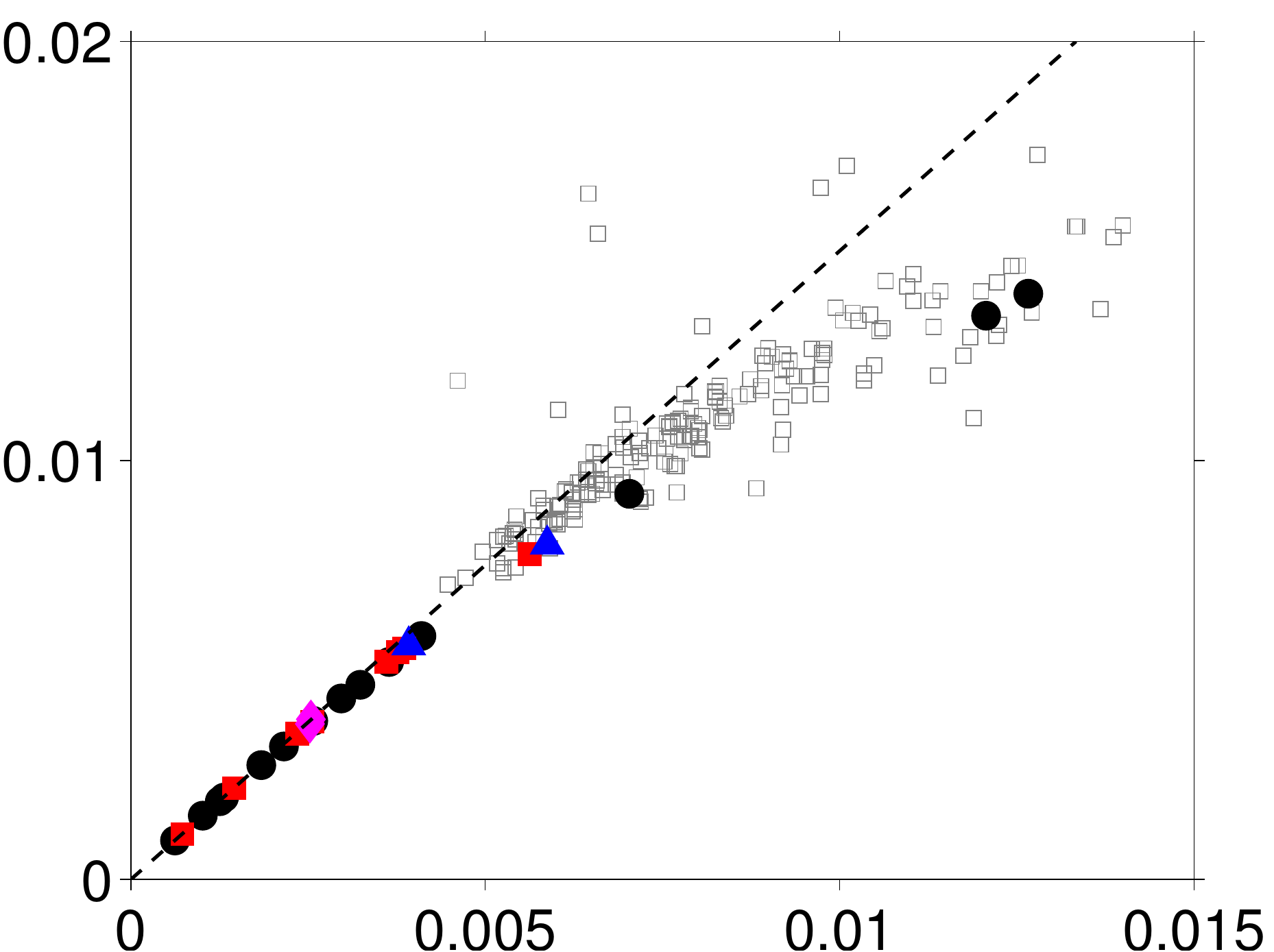}
          \centerline{\small \qfstar}
        \end{minipage}
        \hfill
        \begin{minipage}{2ex}
          \rotatebox{90}
          {\small $\hufmax/\hfluid$}
        \end{minipage}
        \begin{minipage}{.45\linewidth}
          \centerline{$(c)$}
          \includegraphics[width=\linewidth]
          {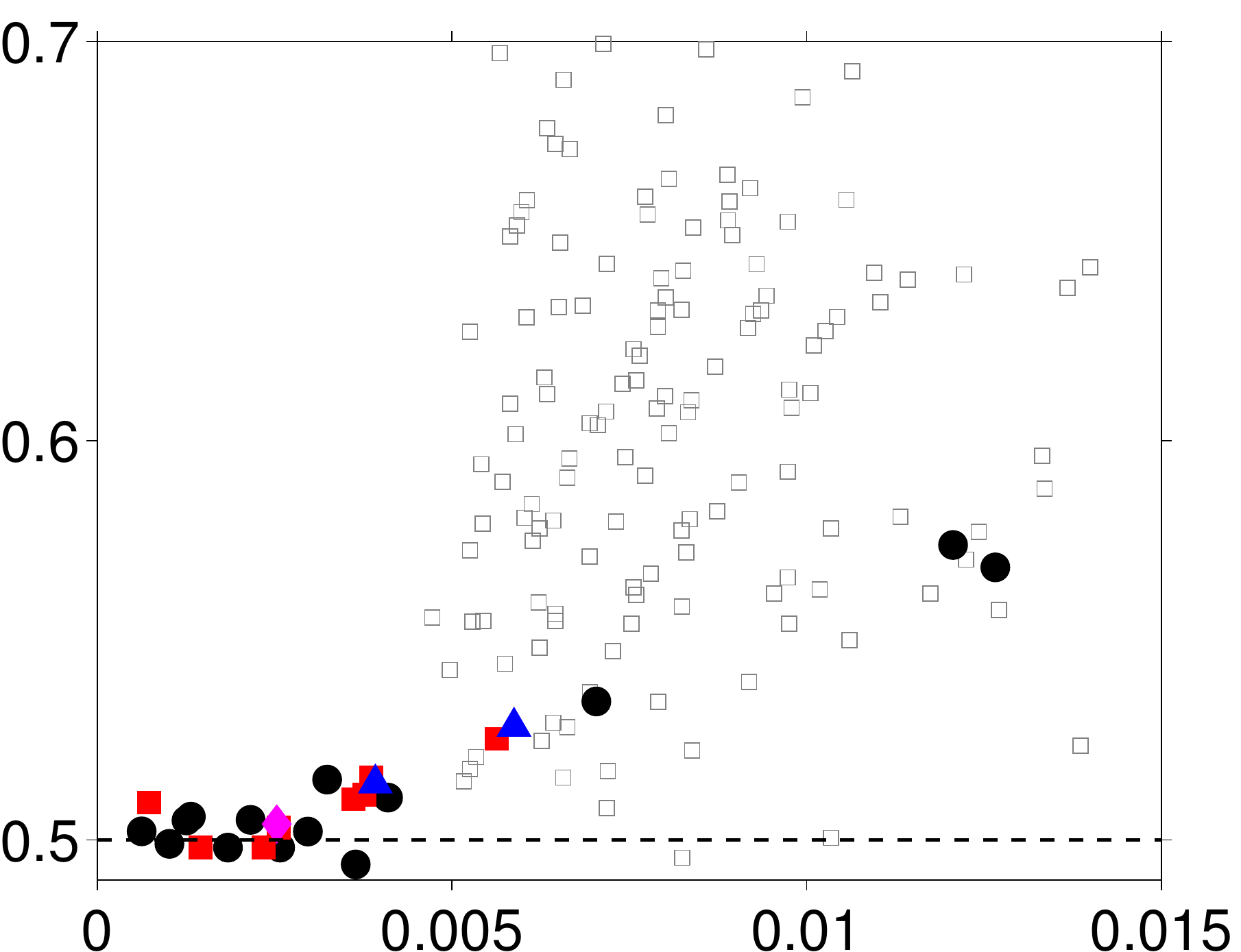}
          \centerline{\small \qfstar}
        \end{minipage}
  }{
        \caption{
                ($a$) Wall-normal profile of the fluid velocity
                \ufmean\, normalized by its  maximum value \ufmax. 
                The vertical axis corresponds to $\tilde{y}=y-L_y$, 
                scaled  by the wall-normal distance
                \hufmax\ 
                between the top upper wall
                and the location of the maximum velocity
                \ufmax. 
                The red dashed curve corresponds
                to the Poiseuille flow parabola
                \eqref{master-parabola}.  
                ($b$) The variation of \ufmax\ 
                as a function of \qfmean. The dashed line
                corresponds to 
                the value for smooth wall Poiseulle flow, 
                i.e.\ $(3/2)\qfmean/\hfluid$. 
                ($c$) The variation of \hufmax as a function of 
                \qfmean\ with the Pouiseuille value
                $\hufmax/\hfluid=1/2$ marked by a 
                dashed line. 
                Data points marked by (\opencircle,\opensquare) 
                correspond to the experiment of \citet{Aussillous2013}. 
                \protect\label{fig:mean-fluid-velocity}
      }
    }
\end{figure}

%
%
%
In the laminar flow regime the fluid velocity profile in the clear
fluid region of the channel would be exactly of parabolic shape if
that region were strictly devoid of particles. Deviations from the
parabola are investigated in figure~\ref{fig:mean-fluid-velocity}. 
For this purpose we normalize the mean particle velocity $\ufmean(y)$
by its respective maximum value $\ufmax = \max\ufmean$, and define a 
coordinate measuring the distance from the top wall,
$\tilde{y}=y-L_y$. 
We also denote by \hufmax\ the distance of the location of \ufmax\
measured from the top wall (cf.\ the schematic diagram in figure 
\ref{fig:bedload-schematic-diagram}). 
Figure~\ref{fig:mean-fluid-velocity}($a$) shows a graph of \ufmean\
under this scaling. It can be seen that all cases collapse upon the
Poiseuille flow parabola,  
\begin{equation}
  \frac{\ufmean}{\ufmax} = -\Big(\frac{\tilde{y}}{\hufmax}\Big)^2
  -2\Big(\frac{\tilde{y}}{\hufmax}\Big)
  \;,
  \label{master-parabola}
\end{equation}  
over a large wall-normal interval. Deviations from the parabolic shape
are noticeable for $y\leq y_0$, which is a case dependent location
under the scaling of figure~\ref{fig:mean-fluid-velocity}($a$). 
These deviations are of the form of larger velocity values than those
given by (\ref{master-parabola}) around the fluid-bed interface. 
%
%
Note that in a smooth wall laminar Poiseuille flow we have 
$\hufmax=\hfluid/2$ and $\qfmean = 2\ufmax\hfluid/3$.
Thus the dependence of the characteristics of the fluid velocity 
profile on the non-dimensional fluid flow rate can be inferred by 
examining the variation of $\ufmax\hfluid/\qrefhf$ and
$\hufmax/\hfluid$ as a function of \qfstar. 
These quantities are shown in
figure~\ref{fig:mean-fluid-velocity}($b$) 
and \ref{fig:mean-fluid-velocity}($c$). 
It is seen that, for values of $\qfmean/\qrefhf$ smaller than
approximately $0.004$, the profiles seem to be adequately 
described by smooth wall Poiseuille flow.
At larger values of  the non-dimensional flow rate, however,
\ufmax\ is observed to 
deviate progressively from the Poiseuille values, exhibiting an
increasingly smaller slope. 
The opposite trend is found for \hufmax, which grows beyond the
Poiseuille flow value with increasing $\qfmean/\qrefhf$. 
Figure~\ref{fig:mean-fluid-velocity}$(b,c)$ also includes
corresponding values computed from the available experimental data of
\citet{Aussillous2013}.  
Despite some scatter in the values from the experiments, a very good
match with the simulation data is obtained for \ufmax. 
On the other hand, the experimental values of
\hufmax\ are highly scattered, making a direct comparison difficult. 

%% file: conclusion.tex
\section{Conclusion}
\label{sec-conclusion}
In the present work we have performed direct numerical simulation of
horizontal channel flow over a thick bed of mobile sediment
particles. 
For this purpose we have employed an existing fluid solver which
features an immersed boundary technique for the efficient and accurate
treatment of the moving fluid-solid interfaces. The algorithm was
coupled with a collision model based upon the soft-sphere approach.  
The forces arising during solid-solid contact are expressed as a
function of the overlap length, with an elastic and a damping normal
force component, as well as a damping tangential force, limited by a 
Coulomb friction law. 
Since the characteristic collision time is typically orders of
magnitude smaller than the time step of the flow solver, the
numerical integration of the equations for the particle motion is
carried out with the aid of a sub-stepping technique, essentially
freezing the hydrodynamic forces acting upon the particles in the
interval between successive flow field updates.  

The collision strategy was validated with respect to the test case of
gravity-driven motion of a single sphere colliding with a horizontally
oriented plane wall in a viscous fluid. Simulations over a range of
collisional Stokes numbers corresponding to roughly three orders of
magnitude have been performed. 
It was found that the present collision strategy works correctly when
coupled to the particulate flow solver, yielding values for the
effective coefficient of restitution in close agreement with reference
data from experimental measurements. 

We have then presented a series of fully-resolved simulations of
bedload transport by laminar flow in a plane channel  
configuration similar to the experiment of \citet{Aussillous2013}. 
Although the Reynolds number in the simulations is two orders of
magnitude larger than in the experiment, the range of values of the
principal control parameters of the problem (either the Shields number
or the non-dimensional flow rate) overlaps significantly between
both approaches, allowing for a direct comparison. 
Our DNS-DEM method was found to provide results which are fully
consistent with the available data from the reference experiment. 
The present simulations yield a cubic variation of the 
particle flow rate (normalized by the square of the Galileo number
times viscosity) with the Shields number, once the threshold value of
the Shields number is exceeded. 
The thickness of the mobile particle layer (normalized with the
particle diameter) obtained from our simulations varies roughly with
the square of the Shields number; when normalized with the height of
the fluid layer, it follows even more closely a quadratic variation
with the normalized fluid flow rate. 
Previous studies using two-fluid models have predicted a linear
dependency of the mobile layer thickness on both parameters
\citep{Mouilleron2009,Ouriemi2009}.  
Unfortunately, the data for this quantity extracted from the
experiment of \citet{Aussillous2013} features a certain amount of
scatter, and, therefore, it does not serve to clearly distinguish
between the two power laws. 
Based upon the DNS-DEM results we were also able to show that the
deviation from a parabolic flow profile in the clear fluid region
above the bed begins to become significant for non-dimensional flow
rates above a value of approximately $0.004$. 

The present work shows that the simple collision model considered
herein is adequate for the purpose of simulating fluid-induced
transport of a dense bed of spherical particles. 
This conclusion opens up the possibility to apply the present
technique to the problem of the formation of sediment patterns (such 
as dunes) in wall-bounded shear flow. 
Another future perspective is to extend the investigation of the
bedload problem itself. It bears a number of open questions, such as
the lack of a precise description of the local balance of forces
acting on the particles as well as a missing analysis of the
Lagrangian aspects of the particle motion, which could be addressed
with the aid of the present methodology.   

%% file: acknowledgement.tex
\section*{Acknowledgments}
%
%
Thanks is due to Pascale Aussillous and \'Elisabeth Guazzelli for
sharing their data in electronic form and for fruitful discussions
throughout this work. 
AGK acknowledges the hospitality of the group ``GEP'' at IUSTI,
Polytech Marseille, during an extended stay. 
This work was supported by the German Research Foundation (DFG) through 
grant UH 242/2-1. The computer resources, technical expertise and 
assistance provided by the staff at LRZ M\"unchen (grant pr58do) 
are thankfully acknowledged.

%% file: appendix_suppl_mat.tex
\section{Supplementary material}
\label{sec-suppl-mat}
Animations of the particle motion in the simulations of 
\S~\ref{sec:erosion-of-granular-bed-sheared-by-laminar-flow} 
can be found in the online version at \href{http://dx.doi.org/10.1016/j.ijmultiphaseflow.2014.08.008}{10.1016/j.ijmultiphaseflow.2014.08.008}. 
The data is also available under the following URL:\\
\href{http://www.ifh.kit.edu/dns_data/particles/bedload}
{\tt http://www.ifh.kit.edu/dns\_data/particles/bedload}.

%% file: appendix_averaging_operations.tex
\section{Averaging operations}
\label{sec-ensemble-averaging}
\subsection{Wall-parallel plane and time averaging}
\label{subsec-wall-parallel-plane-and-time-averaging}
Let us first define an indicator function $\phi_f(\mathbf{x},t)$ 
for the
fluid phase which tells us whether a given point with a position vector
$\mathbf{x}$ lies inside $\Omega_f(t)$, the part of the 
computational domain $\Omega$ which is occupied by the fluid at time $t$, as
follows:
\begin{equation}\label{equ-def-fluid-indicator-fct}
  \phi_f(\mathbf{x},t)
  =
  \left\{
    \begin{array}{lll}
      1&\mbox{if}&
      \mathbf{x}\in\Omega_f(t)
      \\
      0&\mbox{else}&
    \end{array}
  \right.
  \,.
\end{equation}
The solid-phase indicator function $\phi_\mathrm{p}$
is then simply given as the complement of $\phi_\mathrm{f}$, i.e.\
\begin{equation}\label{equ-def-solid-indicator-fct}
  \phi_\mathrm{p}(\mathbf{x},t) = 1 - \phi_\mathrm{f}(\mathbf{x},t).
\end{equation}
Based upon the indicator function $\phi_f$, an instantaneous discrete
counter of fluid sample points in a wall-parallel plane at a given
wall-distance $y_j$ at time $t^m$ is defined as:
\begin{equation}
\label{equ-def-sample-counter-plane-and-time-fluid-only-instant}
  n_{xz}(y_j,t^m)
  =
  \sum_{i=1}^{N_x}
  \sum_{k=1}^{N_z}
  \phi_f(\mathbf{x}_{ijk},t^m)
  \,,
\end{equation}
where $N_x$ and $N_z$ are the number of grid nodes
in the streamwise and spanwise directions, and 
$\mathbf{x}_{ijk}=(x_i,y_j,z_k)^T$ denotes
a discrete grid position.
%
The corresponding counter accounting for $N_t$ time records is defined
as: 
\begin{equation}\label{equ-def-sample-counter-plane-and-time-fluid-only}
  n(y_j)
  =
  \sum_{m=1}^{N_t}
  n_{xz}(y_j,t^m)
  \,.
\end{equation}
Consequently, the ensemble average of a Eulerian quantity
$\boldsymbol{\xi}_f$ of the fluid 
phase over wall-parallel planes (considering only
grid points located in the fluid domain) is defined as:
\begin{eqnarray}
  \langle
  \boldsymbol{\xi}_f
  \rangle_{xz}(y_j,t^m)
   &=&
  \frac{1}{n_{xz}(y_j,t^m)}
  \sum_{i=1}^{N_x}
  \sum_{k=1}^{N_z}
  \phi_f(\mathbf{x}_{ijk},t^m)\,
  \boldsymbol{\xi}_f(\mathbf{x}_{ijk},t^m)
  \label{equ-def-avg-operator-plane-and-time-fluid-only-instant}\\
  \langle
  \boldsymbol{\xi}_f
  \rangle(y_j)
   &=&
  \frac{1}{n(y_j)}
  \sum_{m=1}^{N_t}
  n_{xz}(y_j,t^m)
  \langle
  \boldsymbol{\xi}_f
  \rangle_{xz}(y_j,t^m)
  \,,\label{equ-def-avg-operator-plane-and-time-fluid-only}
\end{eqnarray}
where the operator $\langle\cdot\rangle_{xz}$ indicates an instantaneous
spatial wall-parallel plane average and 
$\langle\cdot\rangle$ indicates an average over space and time.

\subsection{Binned averages over particle-related quantities}
\label{subsec-binned-averages-over-particle-related-quantities}
Concerning Eulerian statistics of (Lagrangian) particle-related
quantities, 
the computational domain was decomposed into discrete wall-parallel
bins of thickness $\Delta h$,   
and averaging was performed over all those particles within each bin.
Similar to (\ref{equ-def-fluid-indicator-fct}), 
we  define an indicator function $\phi_{bin}^{(j)}(y)$ which
tells us whether a given wall-normal position $y$ is located inside or
outside a particular bin with index $j$, viz.
\begin{equation}\label{equ-def-ybin-indicator-fct}
  \phi_{bin}^{(j)}(y)=
  \left\{
    \begin{array}{lll}
      1&\mbox{if}&
      (j-1)\Delta h\leq y<j\Delta h
      \\
      0&\mbox{else}\;.&
    \end{array}
  \right.
\end{equation}
A sample counter for each bin, sampled over an instantaneous  
time $t^m$ as well as sampled over the number of available 
snapshots of the solid phase, $N_t^{(p)}$, is defined as
\begin{eqnarray}
  n_{p_{xz}}^{(j)}(t^m)
  &=&
  \sum_{l=1}^{N_p}
  \phi_{bin}^{(j)}(y_p^{(l)}(t^m))\,,
  \label{equ-def-sample-counter-binned-particles-instant} \\
  n_p^{(j)}
  &=&
  \sum_{m=1}^{N_t^{(p)}}
   n_{p_{xz}}^{(j)}(t^m)
   \,,\label{equ-def-sample-counter-binned-particles}
\end{eqnarray}
respectively.
From the sample counters we can deduce the (instantaneous) average 
solid volume fraction in each bin, viz.
\begin{eqnarray}
  \langle\phi_s\rangle_{xz}(y^{(j)},t^m)
  &=&
  n_{p_{xz}}^{(j)}(t^m)\frac{\pi D^3}{6L_xL_z\Delta h}
  \,,\label{equ-def-avg-solid-volume-frac-instant}\\
  \langle\phi_s\rangle(y^{(j)})
  &=&
  \frac{n_p^{(j)}}{N_t^{(p)}}\,\frac{\pi D^3}{6L_xL_z\Delta h}
  \,.\label{equ-def-avg-solid-volume-frac}
\end{eqnarray}
Alternatively, the solid volume fraction can be deduced from the
indicator function defined in \eqref{equ-def-fluid-indicator-fct}
viz.
\begin{eqnarray}
 \langle \phi_p\rangle_{zt}(x_i,y_j) 
  &=& 
  \frac{1}{N_t^{(p)}N_z}
  \sum_{m=1}^{N_t^{(p)}}
  \sum_{k=1}^{N_z}
   \big(1-\phi_f(\mathbf{x}_{ijk},t^m)\big)
   \,,\label{equ-def-avg-solid-volume-frac-smooth-2d}\\
  \langle \phi_p\rangle(y_j) 
  &=& 
   \frac{1}{N_x}
  \sum_{k=1}^{N_x}
  \langle \phi_p\rangle_{zt}(x_i,y_j) 
  \,,\label{equ-def-avg-solid-volume-frac-smooth}
\end{eqnarray}
where the operator $\langle\cdot\rangle_{zt}$ indicates averaging 
in the spanwise direction and time, and 
$\langle\cdot\rangle$ indicates an average over both homogeneous
directions and time.  
Finally, the binned average  of a
Lagrangian quantity $\boldsymbol{\xi}_p$ is defined as follows:
\begin{eqnarray}\label{equ-def-avg-operator-binned-particles}
  \langle
  \boldsymbol{\xi}_p
  \rangle_{xz}
  (y^{(j)},t^m)
  &=&
  \frac{1}{n_{p_{xz}}^{(j)}(t^m)}
  \sum_{l=1}^{N_p}
  \phi_{bin}^{(j)}(y_p^{(l)}(t^m))\,
  \boldsymbol{\xi}_p^{(l)}(t^m)
  \,,\\
  \langle
  \boldsymbol{\xi}_p
  \rangle
  (y^{(j)})
  &=&
  \frac{1}{n_p^{(j)}}
  \sum_{m=1}^{N_t^{(p)}}
  n_{p_{xz}}^{(j)}(t^m)
  \langle
  \boldsymbol{\xi}_p
  \rangle_{xz}
  (y^{(j)},t^m)
  \,,
\end{eqnarray}
supposing that a finite number of samples has been encountered
($n_{p_{xz}}^{(j)}(t^m)>0$, $n_p^{(j)}>0$).
%
%
%
A bin thickness of $\Delta h = D/4$ was chosen for the evaluation of 
the binned averages, unless otherwise stated.